\documentclass[pra,twocolumn,amsmath,amssymb,superscriptaddress]{revtex4-1}
\usepackage{epsfig,graphicx,graphics,amsmath,amssymb,float}
\usepackage[T1]{fontenc}
\usepackage[latin9]{inputenc}
\usepackage{amsmath}
\usepackage{amssymb}
\usepackage{xcolor}
\usepackage{amscd}
\usepackage{bm}
\usepackage{bbold}
\usepackage{psfrag}
\usepackage{mathrsfs}

\usepackage{bbm} 
\usepackage{chemformula}
\usepackage[caption=false]{subfig}
\usepackage{subfig}
\usepackage{color}
\usepackage[bookmarks=true,colorlinks,linkcolor=blue,urlcolor=blue,citecolor=blue]{hyperref}

\usepackage{dutchcal}
\DeclareMathAlphabet{\matholdcal}{OMS}{cmsy}{m}{n}

\DeclareMathOperator{\Tr}{Tr}

\newcommand{\be}{\begin{equation}}
\newcommand{\ee}{\end{equation}}
\newcommand{\bea}{\begin{eqnarray}}
\newcommand{\eea}{\end{eqnarray}}

\newcommand{\mc}{\matholdcal}

\newcommand{\te}{\text}
\begin{document}

\title{Raman Response of the Charge Density Wave in Cuprate Superconductors}

\author{M. F. Cavalcante}
\affiliation{Departamento de F\'isica, Universidade Federal de Minas Gerais, C. P. 702, 30123-970, Belo Horizonte, MG, Brazil}
\affiliation{Universit\'e Paris-Saclay, CNRS, Laboratoire de Physique des Solides, 91405, Orsay, France}
\author{S. Bag}
\affiliation{Universit\'e Paris-Saclay, CNRS, Laboratoire de Physique des Solides, 91405, Orsay, France}
\author{I. Paul}
\affiliation{Universit\'e Paris Cit\'e, Mat\'eriaux et Ph\'enom\`{e}nes Quantiques,
UMR CNRS 7162, B\^atiment Condorcet, 75205 Paris Cedex 13, France}
\author{A. Sacuto}
\affiliation{Universit\'e Paris Cit\'e, Mat\'eriaux et Ph\'enom\`{e}nes Quantiques,
UMR CNRS 7162, B\^atiment Condorcet, 75205 Paris Cedex 13, France}
\author{M. C. O. Aguiar}
\affiliation{Departamento de F\'isica, Universidade Federal de Minas Gerais, C. P. 702, 30123-970, Belo Horizonte, MG, Brazil}
\author{M. Civelli}
\affiliation{Universit\'e Paris-Saclay, CNRS, Laboratoire de Physique des Solides, 91405, Orsay, France}
\date{\today}

\begin{abstract}
  We study the Raman response, for $B_{1g}$ and $B_{2g}$ light-polarization symmetries, of the charge density wave phase
  appearing in the underdoped region of cuprate superconductors.
  We show that the $B_{2g}$ response provides a distinctive signature of the charge order,
  independently of the details of the electronic structure and from the
  concomitant presence of a pseudogap, in sharp contrast with the behavior of the $B_{1g}$ response.
  This well accounts for the Raman experimental results.
  We then clearly identify a charge density wave energy scale, and show that its doping dependence is
  eventually driven by the monotonic behavior of the pseudogap. This has also been pointed out
  in Raman experiments, and it is suggestive of a pseudogap ruling the multiple energy scales of the exotic phases
  appearing in the cuprate phase diagram.
  

\end{abstract}

\pacs{pacs}

\maketitle
\section{Introduction}

\begin{figure}[t!!]
    \centering
    \includegraphics[scale=0.18]{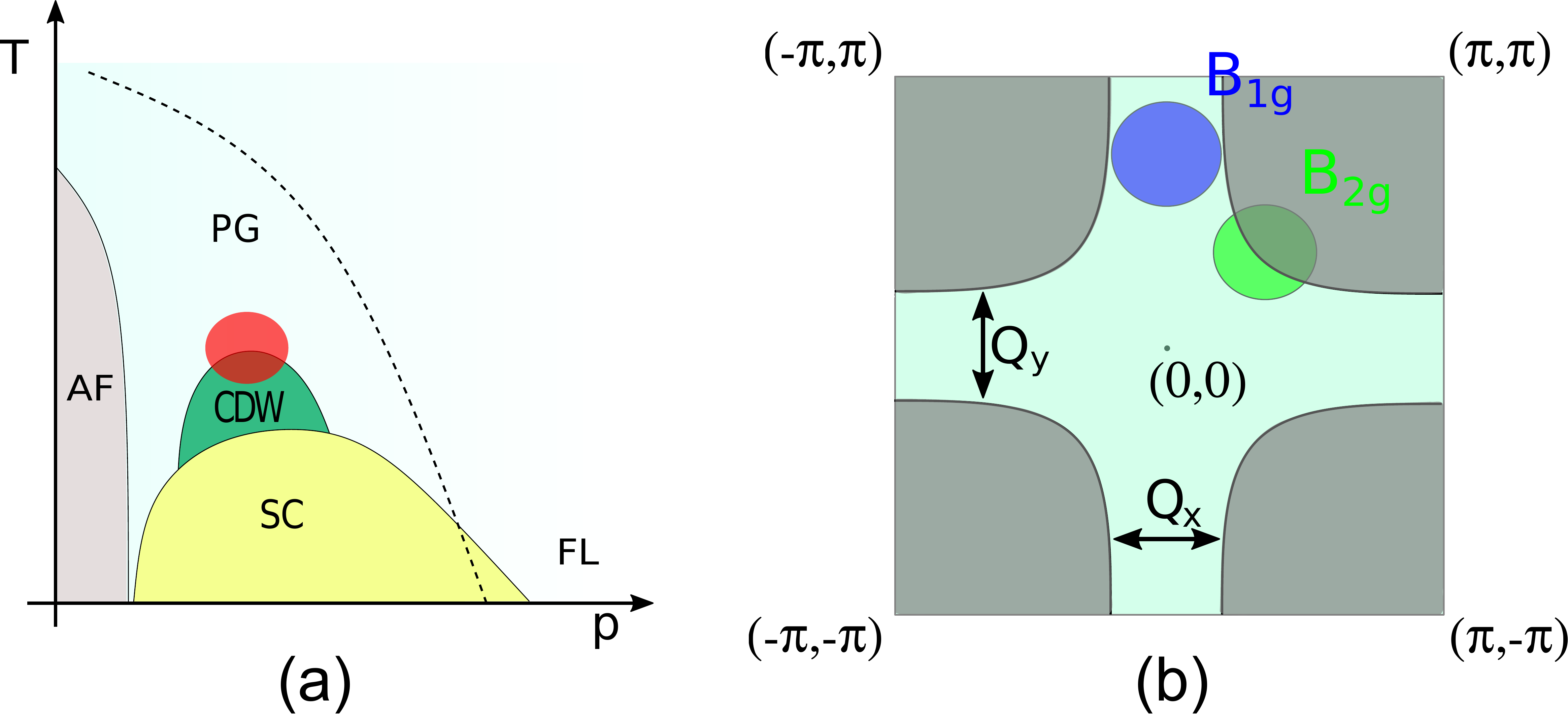}
    \caption{\textbf{(a)} Schematic $T$ vs $p$ phase diagram for the cuprates, displaying
      different quantum phases: anti-ferromagnetism (AF), pseudogap (PG), superconductivity (SC),
      charge density wave (CDW) and Fermi liquid (FL).
      The region highlighted by the red circle is the focus of this study.
      \textbf{(b)} Fermi surface in the overdoped regime of cuprates. The green (blue) circle identify the region that
      is probed by the $B_{2g}$ ($B_{1g}$) Raman response, i.e. the nodal (antinodal) region (see also appendix \ref{vertex_apen}).
      The black arrows mark the nesting points of the CDW wave-vectors $\mathbf Q_x$ and $\mathbf Q_y$.}
    \label{diagram}
\end{figure}

The investigation of key phenomena at the roots of high temperature (HTC) superconductivity (SC) has been at the
center stage of the condensed matter physics since more than thirty years. The parent compound of HTC superconductors
is an unconventional Mott insulator (MI), displaying anti-ferromagnetism (AF) driven by strong electronic correlations.
Upon hole doping, these materials display several distinct phases, as shown on the doping $p$ versus
temperature $T$ diagram [see Fig. \ref{diagram} \textbf{(a)}]. For $p>p_c$, known as the overdoped regime, the system
displays conventional metallic properties, well understood within the framework of the standard theory of metals, the Fermi liquid (FL) theory.
This includes
for instance a large Fermi surface (FS)
[see Fig. \ref{diagram} \textbf{(b)}],
measured e.g. by Angle Resolved Photoemission Spectroscopy (ARPES) \cite{RevModPhys.75.473}.
For $p<p_c$, the underdoped regime, the FL properties are strongly violated. This region is marked by
the presence of the pseudogap (PG) phase \cite{Norman2007}, which
was originally introduced to interpret
an expected loss of spectral weight in spectroscopic responses \cite{PhysRevLett.62.1193,PhysRevLett.63.1700}.
Only a small part of the FS, called the Fermi arcs, is in fact observed by ARPES
\cite{Arpes_FS,Arpes_FS_2}. Like the HTC SC, the PG origin is not well understood.
At intermediate temperatures another exotic phase appears, the charge density wave (CDW) order \cite{Comin}.
Our focus in this work is on this metallic region where the CDW is present.

In the last decade, several X-Rays and STM experiments have confirmed the presence of a CDW phase in the underdoped
regime of several cuprates \cite{Comin}. This phase rises in a wide range of doping and, similar to the SC phase,
is characterized by a dome-like shape in the $T$ vs $p$ diagram [see Fig. \ref{diagram} \textbf{(a)}]. Given the
proximity between these phases, the natural question arises whether and how the CDW phase is related to the
the well-established PG and SC phases. Some experimental and theoretical findings have put into evidence a
competition between them
\cite{Wu2011,LeBoeuf_2012,Ghiringhelli_2012,Wu2013,Huecker_2014,Wu2015,Cyr_Choiniere_2018,PhysRevLett.121.167002,PhysRevB.100.094502,Senechal},
however a non-trivial relationship is believed to take place.

Recent advancements have allowed to measure CDW also within Raman spectroscopy. This technique has allowed
in particular to access for the first time the CDW energy scale \cite{Alain}. These results have provided new clues that
suggest a common microscopic origin of the CDW, SC, PG phases, along the directions of the theoretical proposals in Ref's. \cite{efetov,Fradkin_2015,pepin}. Despite many efforts, though, the formulation of a theory that describes
completely the underdoped regime of cuprates, and in particular the role played by the CDW order, has remained an
open problem.

Our aim with this work is to give a first theoretical contribution to this problem within the Raman spectroscopy
technique. We shall clarify how the CDW can be detected in the Raman response and, in particular, in the B$_{2g}$ light
polarization geometry. We characterize the CDW feature and show that it is universal within the cuprate family even
in the presence of a PG phase.
We then extract the CDW energy scale and compare it with the PG one, providing a description on how the PG can drive the doping-dependent behavior of the CDW,
as observed by the aforementioned Raman experiments \cite{Alain}.

This article is organized in the following form. Sec. \ref{CDW} is divided in three parts.
In the first part, \ref{sub0}, we introduce the tight-binding phenomenological model that describes a
d-wave symmetric bidirectional CDW on the square lattice
and show how to treat the problem in the reduced Brillouin zone (rBZ).
At this stage, we neglect the effects of the PG. In the second part, \ref{sub1},
we calculate the single-particle spectral properties.
In the third part, \ref{sub2}, we discuss the Raman dynamical response, and in particular we
compare two light polarization $B_{1g}$ and $B_{2g}$  symmetries. In subsection \ref{sub2b}
we discuss the commensuration and finite-length of the CDW order.
In the following section, Sec. \ref{CDW+PG}, we investigate the properties of the CDW order in the presence of a PG,
that we introduce by means of the phenomenological Yang-Zhang-Rice-like approach \cite{YZR1,YZR2}.
We prove the robustness of our results even on the presence of the PG, and in particular we extract the CDW and PG
energy scale and establish their mutual interaction. We give our conclusion in Sec. \ref{conclusions}. Finally,
in the Appendix \ref{ap1}, \ref{vertex_apen}, \ref{ap2}, \ref{ap3} and \ref{apen_CDW_domo} we give some further technical details about our calculations.

\section{The CDW order}
\label{CDW}
\subsection{CDW Modeling}
\label{sub0}

Here we take a first order approach and consider a simple tight binding description of the electrons in Cu-O planes of cuprates, adding a CDW interaction. Our goal is to show the pure effects of the formation of the CDW order on the
electronic structure, and in particular on the Raman response. We shall consider the effects of interactions (in particular the PG), which
are indeed relevant in cuprates, in the next section. This approach, which inserts by hand the CDW and the PG, is purely phenomelogical and it is intended to show the effects of such orders on spectroscopic responses. We leave to future developements a more consistent but more involved treatement of the CDW and PG from microscopic theories, which require the implementation of more advanced techniques, like the slave particle approach \cite{slave_cite} or the Dynamical Mean Field theory \cite{DMFT_cite}. This is because, while the CDW can be treated via perturbative self-consistent static mean field theory, the PG is a well-known non-perturbative phenomenon.
%

From the phenomenological view point, we start from the following Hamiltonian on the square lattice (with lattice parameter $a\equiv1$):
\begin{equation}
    H  = \sum_{\mathbf r, \mathbf r'}t_{\mathbf r\mathbf r'}c^{\dagger}_{\mathbf r}c_{\mathbf r'} +   \frac{1}{2}\sum_{\mathbf r, \mathbf a,\mathbf Q}J_{\mathbf Q, \mathbf a}e^{i\mathbf Q\cdot(\mathbf r + \mathbf a/2)}c^{\dagger}_{\mathbf r + \mathbf a}c_{\mathbf r} + \text{H.c.}.
    \label{HCDW}
\end{equation}
This model can be obtained by a mean field decomposition in the particle-hole channel of the $t-J-V$ model; see for example Ref. \cite{Sachdev}. In the above Eq. (\ref{HCDW}) the first term is the kinetic part, with $c^{\dagger}_{\mathbf r}$ the electron creation operator at the site $\mathbf r$. The terms $\mathbf r=\mathbf r'$ are included in the sum, their contribution is the chemical potential $-\mu$. The second term describes the bond density wave between first-neighbor, i.e., $\mathbf a=\pm\hat{x},\pm\hat{y}$, with modulation vector $\mathbf Q$. The modulation amplitude $J_{\mathbf Q, \mathbf a}$ is considered as a fixed parameter in our analysis. The extra term $\mathbf a/2$ in the phase factor
describes a modulation centered at the oxygen atoms. Note that, for sake of convenience, we suppress the spin index since it does not play any role.

The Hamiltonian (\ref{HCDW}) in the $\mathbf k$-space takes the form:
\begin{equation}
    H = \sum_{\mathbf k \in \text{BZ}}\xi_{\mathbf k}c^{\dagger}_{\mathbf k}c_{\mathbf k} + \sum_{\mathbf Q,\mathbf k \in \text{BZ}}\lambda_{\mathbf Q}(\mathbf k) c^{\dagger}_{\mathbf k +\mathbf Q/2}c_{\mathbf k - \mathbf Q/2}  + \text{H.c.},
    \label{HCDW2}
\end{equation}
where $\text{BZ}$ denote the Brillouin zone: $\text{BZ}\equiv \{|k_{x,y}| \leq \frac{\pi}{a}\}$. The free electron dispersion in a general case is given by
\begin{eqnarray}
    \xi_{\mathbf {k}}&=&\sum_{n,\mathbf{a}^{(n)}} t^{(n)} e^{i\mathbf{k} \cdot \mathbf{a}^{(n)}},
    \label{free}
\end{eqnarray}
where $n$ denotes the order of neighbors, i.e., $n=1$ for the first neighbors, $n=2$ for the second neighbors, along with others, and $\mathbf a^{(n)}$ are the corresponding neighbors-vectors on the square lattice. Calculating the first terms of Eq. (\ref{free}) we obtain
\begin{eqnarray}
    \xi_{\mathbf k}&=&-\mu -2t(\cos k_x + \cos k_y) -4t'\cos k_x\cos k_y\nonumber\\
    &&-2t''( \cos 2k_x + \cos 2k_y)+\cdots,
    \label{c-freedis}
\end{eqnarray}
where we have defined $t^{(0)}\equiv-\mu$, $t^{(1)}\equiv t$, $t^{(2)}\equiv t'$, and so on. We fix the energy scale of the system by setting the first neighbor hopping amplitude $t=1$, which typically corresponds to $t\simeq 0.1-0.3$ eV, as obtained for instance
  by fitting ARPES data \cite{He_2011}.

Based on experimental results \cite{d_form}, we consider a $x$-$y$-bond sign alternating
$d$-symmetric CDW interaction, i.e. $J_{\mathbf Q, \pm\hat{x}}=-J_{\mathbf Q, \pm\hat{y}}\equiv \frac{J}{2}$. Then the CDW potential in Eq. (\ref{HCDW2}) can be written as:
\begin{equation}
\lambda_{\mathbf Q}(\mathbf k)=\frac{J}{2}(\cos k_x - \cos k_y).
\label{d-symme}
\end{equation}
We shall consider $\mathbf Q = \mathbf  Q_{x/y}=\frac{2\pi}{n a}\hat{x}/\hat{y}$, i.e., a commensurate modulation of $n$ unitary cells in the $x$ and $y$ directions.
Notice that this assumption changes the translational invariance of the system from $a$ to $n a$.
An immediate consequence of the new translational invariance is the emergence of gaps at the points $\mathbf k$ and $\mathbf k'$ that satisfy $\xi_{\mathbf k} = \xi_{\mathbf k'} \equiv \xi_{\mathbf k \pm \mathbf Q}$, as we shall see below.

\begin{table}
    \centering
    \begin{tabular}{|c|c|c|c|}
        \hline
        Parameters & Allais & Liechtenstein & Schabel\\
        \hline\hline
        $t'$ & $-0.33t$ & $-0.3t$ & $-0.51t$\\
        \hline
        $t''$ & $0.03t$ & $0.2t$ & $0.07t$\\
        \hline
        $t'''$ & 0 & 0 & $-0.05t$\\
        \hline
        $t''''$ & 0 & 0 & $-0.06t$\\
        \hline
        $\mu$ & $-0.808t$ & $-0.766t$ & $-1.069t$\\
        \hline
    \end{tabular}
    \caption{Three band parameter sets and its corresponding chemical potentials at hole doping $p=0.12$ and CDW modulation amplitude $J/2 = 0.2t$. Allais set appeared first in the context of analyzing quantum oscillation frequencies on a reconstructed Fermi surface due to the coexistence of the CDW and SC orders \cite{Allais}. Liechtenstein set was used in Ref. \cite{Liech} to study bilayer cuprates as well as in Ref. \cite{Mattheiss} to calculate the band structure of compound \ch{Ca2CuO2Cl2}. Finally, Schabel set was introduced in Ref. \cite{Schabel} to fit photoemission data on \ch{YBCO}.}
\label{table}
\end{table}

\begin{figure*}[hbt]
    \centering
    \includegraphics[scale=0.08]{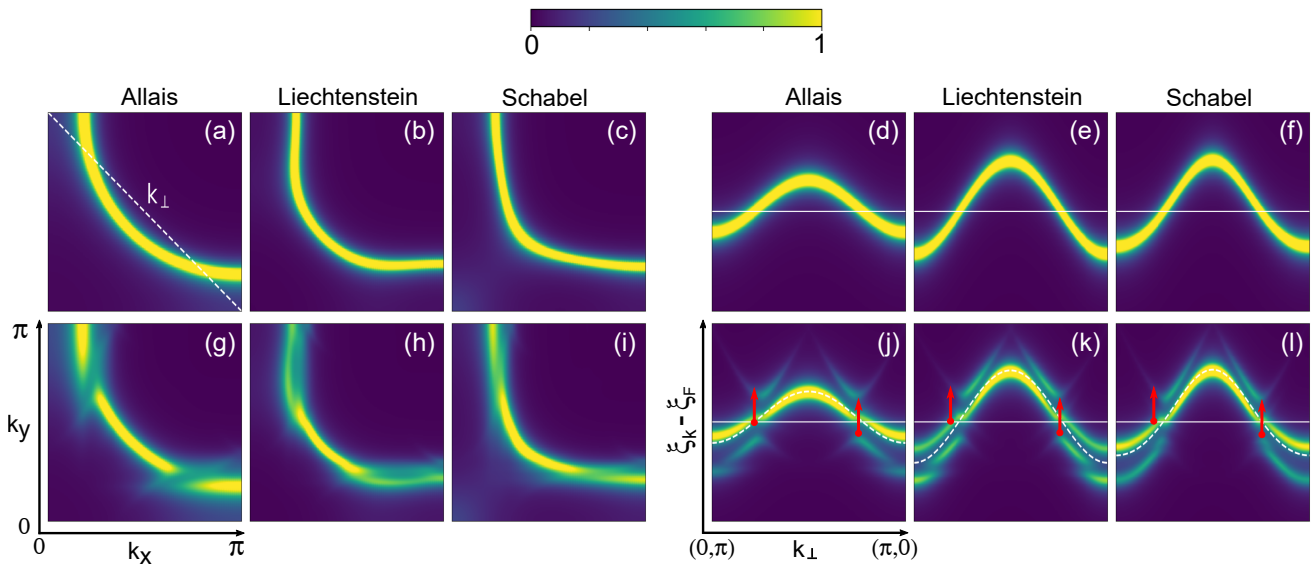}
    \caption{Fermi surface and ARPES dispersion along the path $k_{\perp}$ [indicated in \textbf{(a)}] for the three band parameters set in the FL phase ($J =0$) \textbf{(a-f)} and in the CDW phase \textbf{(g-l)}. The dashed white lines in \textbf{(j-l)} are the free electron band dispersion and the red arrows represent new electronic band transitions between CDW folded band branches.}
    \label{Arpes-CDW}
\end{figure*}

We can rewrite our Hamiltonian in Eq. (\ref{HCDW2}) as a $n^2\times n^2$ matrix (ignoring spin index) \cite{Verret},
by using the relation:
\begin{equation}
    \sum_{\mathbf k \in \text{BZ}}\mathcal g(\mathbf k) = \sum_{\mathbf p \in \text{rBZ}}\sum_{n_{x,y}=0,1,}^{n-1}\mathcal g(\mathbf p + n_x\mathbf Q_x + n_y\mathbf Q_y),
    \label{BZ-to-rBZ}
\end{equation}
for a general function $\mathcal g(\mathbf k)$. The reduced Brillouin zone (rBZ) is defined as rBZ $\equiv \{|p_{x,y}| \leq \frac{\pi}{n}\}$.
Following some experimental results \cite{Comin_2014,Comin}, we will consider here a commensurate modulation with $n=4$.
Then the Hamiltonian (\ref{HCDW2}) becomes:
\begin{equation}
    H = \sum_{\mathbf p \in \text{rBZ}}\Psi^{\dagger}_{\mathbf p}\mc H(\mathbf p)\Psi_{\mathbf p},
    \label{HCDW3}
\end{equation}
where we define the spinors as
\begin{equation}
\Psi^{\dagger}_{\mathbf p}\equiv(c^{\dagger}_{\mathbf p}, c^{\dagger}_{\mathbf p + \mathbf Q_x},c^{\dagger}_{\mathbf p + 2\mathbf Q_x},...,c^{\dagger}_{\mathbf p + 3\mathbf Q_x+3\mathbf Q_y}),
\label{base_rBZ}
\end{equation}
and $\mc H(\mathbf p)$ is a $16\times 16$ matrix (see Appendix \ref{ap1} for its explicit expression).
The corresponding Green's function is given by
\begin{equation}
    \mc G(\mathbf p,\omega) = \frac{1}{\left(\omega + i\eta\right)\mathbb 1 - \mc H(\mathbf p)}.
    \label{Green}
\end{equation}
Here $\mathbb 1$ is the $16\times 16$ identity matrix and
$\eta$ is a pole broadening parameter that can be physically interpreted as the inverse quasiparticle lifetime.
Throughout this work, we fix $\eta=0.2t$.

In order to identify general CDW features of the cuprate phase diagram, we consider three band parameter
sets, previously used to describe different compounds, as detailed in table \ref{table}.

\subsection{Single particle spectrum}
\label{sub1}

From the matrix $\mc G(\mathbf p,\omega)$ in Eq. (\ref{Green}) we can straightforwardly obtain the spectral function,
\begin{equation}
    \mc A(\mathbf p,\omega)=-\frac{1}{\pi}\text{Im}\:\mc G(\mathbf p,\omega),
    \label{espect}
\end{equation}
and, consequently, calculate the Fermi surface $\mc A(\mathbf p,\omega=0)$. Each diagonal element of the matrix
$\mc A(\mathbf p,\omega)$ represents the spectral function in a specific region of the BZ, which is parted in 16 rBZs
by the CDW
translational invariance. However, if we relax the condition $\mathbf p \in \text{rBZ}$ and extend $\mathbf p \in \text{BZ}$ we can work only with the first element $\mc A_{11}(\mathbf p,\omega)$. The FS for the three band parameters sets (keeping $J/2 =0.2t$) is shown in Fig. \ref{Arpes-CDW}. To explicitly identify the effects of the
CDW interaction, we display in Fig. \ref{Arpes-CDW}\textbf{(a-f)} the spectra when the CDW is absent ($J=0$),
which represent a conventional FL. The difference among the three sets of band parameters stands mainly in the degree
of the hole-like
curvature of the band. Upon activation of the CDW interaction [Fig. \ref{Arpes-CDW}\textbf{(g-l)}],
the effect on the FS is
mainly evident at the nesting wave vectors $\mathbf{Q}$ [see Fig. \ref{diagram}\textbf{(b)}],
where a gap opens as pointed out above.

The effect of the CDW 4 $\times$ 4 folding into the rBZ can be however better enlightened by looking at the
energy $\xi_{\mathbf k}- \xi_F$ (where $\xi_F$ is the Fermi energy
setting the zero) vs momentum $\mathbf{k}$ dispersion,
also accessible with ARPES. For clarity, we take the diagonal cut $k_{\perp}$, displayed in Fig. \ref{Arpes-CDW}\textbf{(a)}, and show $\xi_{\mathbf k_{\perp}} -\: \xi_F$, again for sake of comparison for $J=0$ and $J/2=0.2t$, in Fig. \ref{Arpes-CDW}\textbf{(d-f)} and \textbf{(j-l)} respectively. As discussed above, the quasiparticle dispersion is broken at the
Fermi energy at the nesting vector points where the small
gaps open. We can observe that, for energies under and above these Fermi-level regions, folded bands
appear in the spectra due to the CDW folding of the full BZ,
and this has direct consequences on the electronic properties, as we shall see below.

\begin{figure}[t!]
    \centering
    \includegraphics[scale=0.5]{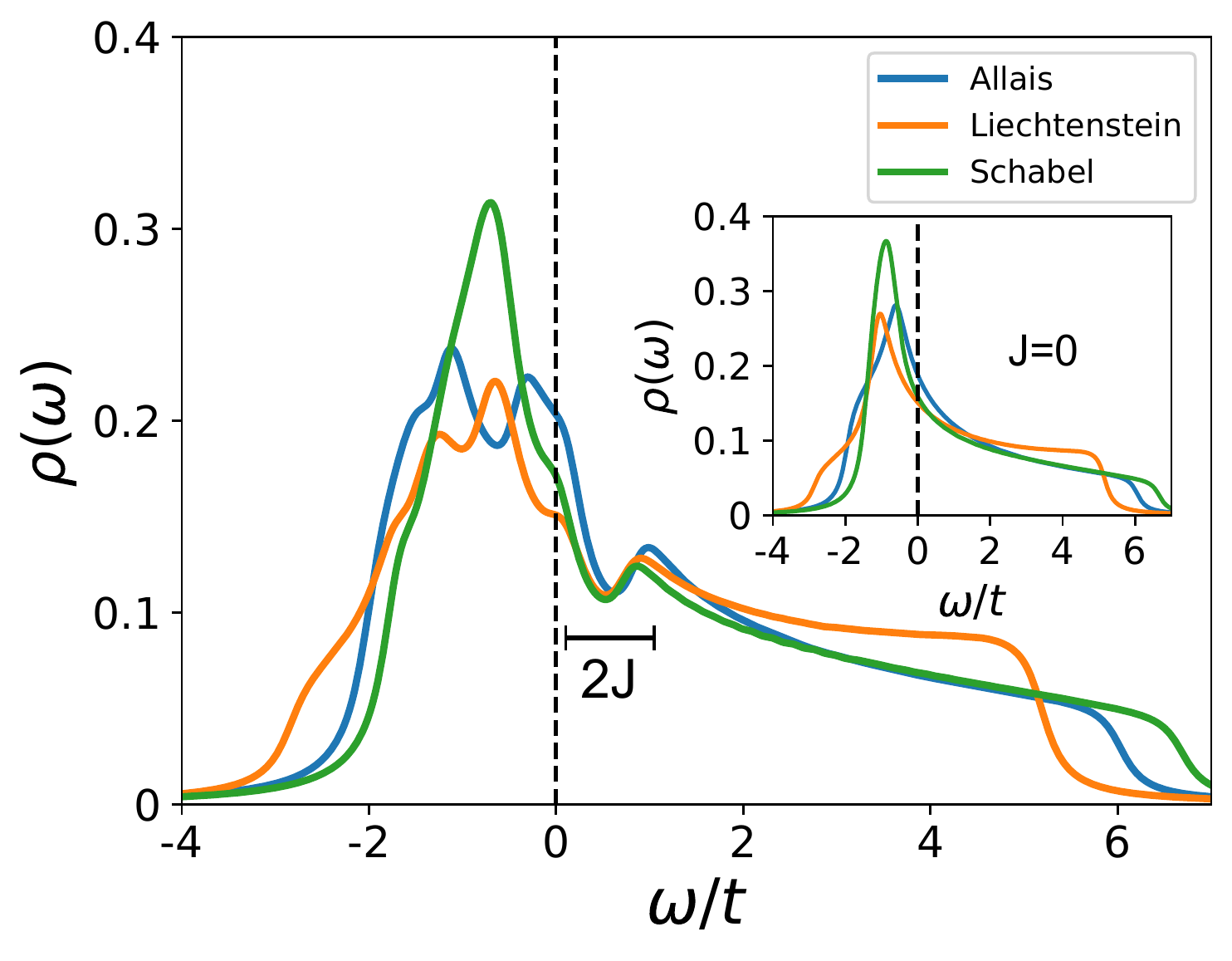}
    \caption{$DOS$ in the CDW phase for the three band parameter sets displayed in table \ref{table}.
      Inset: Same as in the main plot but for the FL phase.
      The dashed black line highlights the Fermi level.
      We observe a dip in the spectral weight at an energy scale $2 J$ which is independent of the band structure details.}
    \label{DOS}
\end{figure}

The effect of the nesting CDW vector and the appearance of the folded band structure can be seen,
for instance, in the local density of states ($DOS$), which is accessible experimentally via scanning tunneling
spectroscopy experiments.
We can calculate $DOS$ directly from the Green's functions:
\begin{equation}
    \rho(\omega) = \frac{1}{\mc N}\sum_{\mathbf p \in \text{BZ} }\mc A_{11}(\mathbf p,\omega),
\end{equation}
where $\mc N$ is the number of states in the BZ. In Fig. \ref{DOS} we display the $DOS$, $\rho(\omega)$,
corresponding to the three parameter sets of table \ref{table}, and the corresponding non-interacting $DOS$ in the inset.
We note a common feature among the three $DOS$. The CDW interaction does not open a full gap around the Fermi level
$\omega=0$, as first pointed out in Ref. \cite{Verret}. It rather changes the slope of the $DOS$ at the Fermi level
and redistributes the spectral weight. The appearance of folded bands is here evidenced by
two features. The first is a band-parameter dependent modulation of the $DOS$, for $\omega< 0$, which can be ascribed
to the folded bands appearing around the antinodal regions [close to $\mathbf{k} \sim (0,\pm\pi), (\pm\pi,0)$]
for $\xi_{\mathbf k} < \xi_F$ [see the row of Fig. \ref{Arpes-CDW}\textbf{(j-l)}].
The second feature is band-parameter independent, and appears as a spectral weight dip, a sort of PG-like feature,
at $\omega >0$ and around $J$. This feature is ascribed to the folded band branches appearing rather close to the nodal regions,
between the CDW nesting points and the quadrant diagonals [close to $\mathbf{k} \sim (\pm\pi/2,\pm\pi/2)$].
On the whole, the CDW effect on the $DOS$ is however rather weak, if compared for example with the effect
of the PG (see section \ref{CDW+PG}).

These results suggest in particular that material independent CDW features should be detectable with probes
that can access at the same time the electronic structure above the Fermi level and close to the nodal
regions. This strongly constrains the range of spectroscopic probes that could be employed.

\subsection{Raman response}
\label{sub2}

A suitable spectroscopic probe that has at the same time access to energy dependent features and
different regions of the BZ is Raman spectroscopy
with different light-polarization symmetries, namely, $B_{2g}$ and $B_{1g}$ \cite{Devereaux}.
The Raman $B_{2g}$ response probes in fact the nodal region of the BZ, while the $B_{1g}$ response probes the
antinodal region [see Fig. \ref{diagram}\textbf{(b)} and Appendix \ref{vertex_apen}].
For our weak CDW potential, we can consider a perturbative expansion of the Raman response $\chi_\nu$ \cite{DMFT,Devereaux}.
As detailed in Appendix \ref{ap2}, at zero temperature, for small momentum transfers between the incident light and the electrons, and ignoring the vertex corrections (which is reasonable for a weak CDW potential)
we obtain \cite{Devereaux}:
\begin{equation}
    \chi_{\nu}(\Omega) = \frac{1}{\mc N}\sum_{\mathbf p \in \text{rBZ}}\int_{-\Omega}^{0}d\omega\Tr\Big[ \Gamma^{\nu}_{\mathbf p}\mc A(\mathbf p,\omega) \Gamma^{\nu}_{\mathbf p}\mc A(\mathbf p,\omega + \Omega) \Big],
    \label{raman}
\end{equation}
where $\Gamma^{\nu}_{\mathbf p}$ is the $16\times 16$ non-renormalized vertex matrix for a given $\nu=B_{1g},B_{2g}$ symmetry and $\Omega$ is the incident photon energy. The physical interpretation of the above equation is immediate: one incident photon with energy $\Omega$ can promote to excited states only the electrons with energy in the range $ \omega \in [-\Omega^+,0]$. This process is described by the first spectral function $\mc A(\mathbf p,\omega)$. By energy conservation, these excited states have energy in the interval $\omega \in [0^+,\Omega]$ and, consequently, are described by $\mc A(\mathbf p,\omega+\Omega)$. The whole scattering process is modulated by the vertex matrix, which fixes specific regions of the BZ.

In the effective mass approximation \cite{Devereaux} and considering  the appropriate light polarization vectors capable to probe the nodal and antinodal regions of the BZ (see appendix \ref{vertex_apen}) \cite{Alain2}, those vertex matrix are given by:
\begin{eqnarray}
    \left[\Gamma^{\text{B2g}}_{\mathbf{p}}\right]_{n,n'} & = & \delta_{n,n'}\frac{\partial^2\xi_n}{\partial p_x p_y}, \label{vertB2g}
     \\
     \left[\Gamma^{\text{B1g}}_{\mathbf{p}}\right]_{n,n'} & = & \delta_{n,n'}\left(\frac{\partial^2\xi_n}{\partial p_x^2} - \frac{\partial^2\xi_n}{\partial p_y^2} \right),
    \label{vertB1g}
\end{eqnarray}
where $\xi_n$ is the n-diagonal element of the $\mc H(\mathbf p)$ (see Appendix \ref{ap1}).
An analysis of the above quantity shows a difference in magnitude of the order of $\sim 10$ (see appendix \ref{vertex_apen}),
then we can already expect a similar
order factor between $\chi_{B_{1g}}$ and $\chi_{B_{2g}}$.

\begin{figure}[t]
{\includegraphics[scale=0.24]{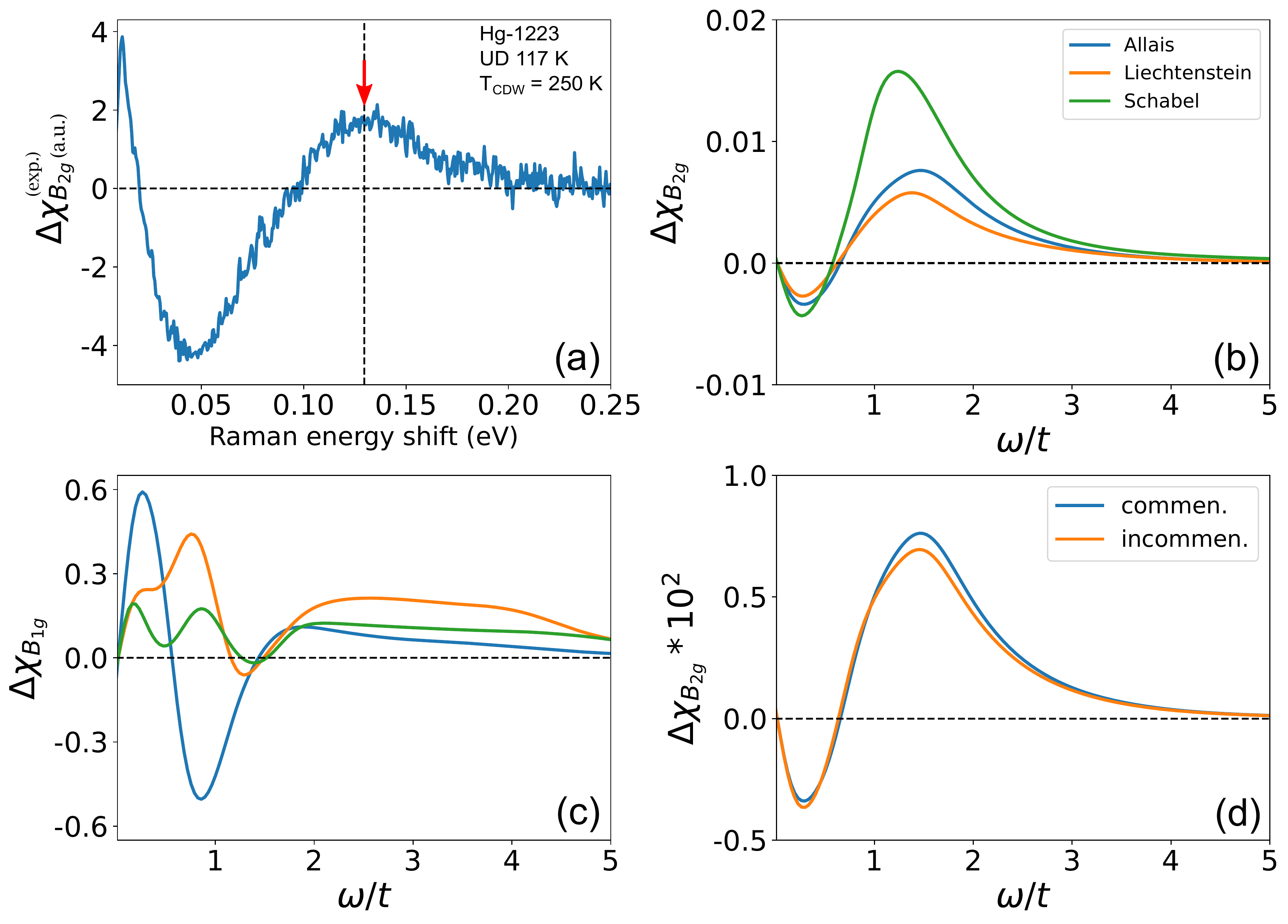} }
\caption{\textbf{(a)} Experimental relative $\Delta\chi^{\te{exp.}}_{B_{2g}}=\chi_{B_{2g}}(117\te{K},\omega)-\chi_{B_{2g}}(290\te{K},\omega)$ for Hg-1223 
  with hole doping $p=0.12$.
  The red arrow indicate the CDW energy scale $E_{\text{CDW}}\approx 0.1295\:\te{eV}\propto 2J$.
  Figure adapted from \cite{Alain}.
  \textbf{(b)} Theoretical relative $B_{2g}$ Raman response for the three band parameter sets
  of the table \ref{table}. \textbf{(c)} relative $B_{1g}$ Raman response (with the same color code).
  \textbf{(d)} The blue line (commen.) corresponds to the Allais curve in panel \textbf{(b)},
  while the orange one is obtained by averaging the responses calculated from the three wave-vectors,
  $\mathbf Q$, $\mathbf Q \pm \delta \mathbf Q$, with $\delta  Q/Q =0.1$.  }
\label{Raman-CDW}
\end{figure}

Following the discussion of the previous section, we expect that the reconstruction of the electronic structure
due to the CDW folding of the BZ affects also the electronic Raman response. As a matter of facts,
in a previous experimental result \cite{Alain},
some of the authors have shown that a characteristic CDW dip-hump feature appears
in the relative $B_{2g}$ Raman response, defined as:
\hbox{$\Delta \chi^{\te{exp.}}_{B_{2g}} = \chi_{B_{2g}}(\omega,T) - \chi^{}_{B_{2g}}(\omega, T= 290\te{K})$}.
Here \hbox{$T= 290\te{K}> T_{CDW} = 250\te{K}$} is higher than the CDW transition temperature $T_{CDW}$
[see Fig. \ref{Raman-CDW}\textbf{(a)}] and close to the PG crossover temperature $T^{\ast}= 325\te{K}$. At this
temperature no CDW feature is present and the PG intensity is negligible.
This operative procedure allows to subtract the {\it normal state} background in the Raman response and put into evidence the features related to the instabilities appearing at lower temperature, like the CDW ones.
In the same publication, by a band-model calculation, similar to the one presented here,
it has been shown that such a dip-hump feature is indeed expected in the theoretical relative $B_{2g}$ Raman response \hbox{$\Delta \chi_{B_{2g}}= \chi_{B_{2g}}(\omega) - \chi^{(0)}_{B_{2g}}(\omega)$}. For sake of simplicity in the choice in the number of free model parameters,
  the calculation is performed at $T=0$. In this case the {\it normal phase} response subtracted,
  $\chi^{(0)}_{B_{2g}}(\omega)$, which has no other instability,
is the Raman response of the FL phase [see Fig. \ref{Raman-CDW}\textbf{(b)}].

The presence of the CDW dip-hump feature in both the experimental and theoretical $\Delta \chi_{B_{2g}}$ supported
  the experimental conclusions. However, a comparison between theory and experiment must be intended only
  at the qualitative level, because of the different temperatures used in the respective analysis.
  Notice, for instance, that the effect of temperature is most evident in the experimental Raman response at small
  energies $\omega <0.025$ eV [see Fig. \ref{Raman-CDW}\textbf{(a)}], where nodal quasiparticles are present.
  One in fact expects that $\chi_{B_{2g}}(\omega,T) \sim \, \tau \,  \omega$ for $\omega \to 0$,
  where $\tau$ is the quasiparticle lifetime at the nodes. As this latter decreases with increasing temperature,
  it originates a positive peak in the $\Delta \chi^{\te{exp.}}_{B_{2g}}$.
  This feature is eventually absent in the $T=0$ theoretical curves [Fig. \ref{Raman-CDW}\textbf{(b)}].
  We remark that low energy quasiparticle are absent at the energies relevant for the CDW features, therefore
  they do not influence our conclusions on the appearance of the dip-hump feature in $\Delta \chi_{B_{2g}}$.

Here we explain the origin of the dip-hump CDW feature, that
arises mainly from the electronic structure in the nodal region. In fact, novel photon-induced Raman excitations
appear between the occupied ($\xi_{\mathbf k}< \xi_F$) and the empty states ($\xi_{\mathbf k}> \xi_F$)
of the cone-like bands induced by the CDW folding in the BZ
[see red arrows in Fig. \ref{Arpes-CDW}\textbf{(j-l)}].
Notice that such a folding is accompanied by a reduction of spectral weight close to the Fermi energy, where
a gap opens at the nesting vector points $\mathbf{Q}$,
and a transfer of this spectral weight to higher energies in correspondence to the cone-like folded bands.
The energy separation between the bottom (in the occupied energy side) and the top (in the empty energy side) of
  the folded cone-like bands is of the order of $2 J$. The transitions between these two bands extend to energies
  higher than $2 J$, hence the CDW feature is expected to appear at energies $\geq 2 J$. This is what indeed is observed
  in Fig. \ref{Raman-CDW}\textbf{(b)}. The maximal intensity of this feature depends on where the maximal spectral
  intensity is located on the bands, as seen on Fig. \ref{Arpes-CDW}\textbf{(j-l)}. The maximum of the CDW hump is
  therefore also located at energies $\geq 2J$, but it is still proportional to $2 J$. Following the experimental
  procedure, we shall identify the maximum of the hump as the CDW energy scale (see following section \ref{DC}).

According to the observations on the $DOS$ of the previous section,
as this CDW dip-hump feature mainly derives from properties of the nodal region it
should be rather universal among different materials. Here we show that this is the case by displaying
$\Delta \chi_{B_{2g}}$ for the different band parameters [see Fig. \ref{Raman-CDW}\textbf{(b)}],
which we introduced in table \ref{table}.
These results are in agreement with the experimental observations of Ref. \cite{Alain},
where such dip-hump feature in $\Delta \chi_{B_{2g}}$
is universal among different members of the cuprate family, including  Hg-based, Y-based and Bi-based compounds,
though with rather different amplitudes depending on the compound disorder level.

In contrast to the $B_{2g}$ Raman response, the $B_{1g}$ one, which is sensitive to Raman transitions in the antinodal
region, has a sharply different behavior. In the previous section, we already pointed out that in the antinodal region
folded bands appear in the occupied side ($\xi_{\mathbf k} < \xi_F$), which strongly depend on the compound band-parameters. We observed that the occupied side ($\omega< 0$) of the $DOS$ (Fig. \ref{DOS}) reflects this material dependence
of the electronic structure. It is then not surprising that we cannot identify a general CDW feature in the $B_{1g}$,
like the dip-hump for the $B_{2g}$, but rather a strong material-dependent response. This is clearly shown
by \hbox{$\Delta \chi_{B_{1g}}= \chi_{B_{1g}}(\omega) - \chi^{(0)}_{B_{1g}}(\omega)$}
in Fig. \ref{Raman-CDW}\textbf{(c)}.
Therefore  $B_{1g}$ is not the appropriate response to clearly identify universal CDW features.
These results settle the longstanding open question on why it has not been possible to
detect CDW order within the $B_{1g}$ Raman response \cite{Alain,Li2019_it}, in spite this being much stronger than the $B_{2g}$ one.
To this one should also add the effects of the PG,
which manifests itself mainly in the $B_{1g}$ geometry.
The PG should further interfere with the CDW, making the identification of the latter even more problematic
within the $B_{1g}$ light-polarization geometry. We shall further discuss the
interaction between CDW and PG in the following section.

\subsubsection{Commensuration and finite-length CDW order}
\label{sub2b}

Experimentally, it is now fairly well-established that the CDW phase has an incommensurate ordering wave vector~\cite{Ghiringhelli_2012,Chang_2012,Huecker_2014,Incom1}, though locally, on a range of several unit cells, the commensuration is preserved
\cite{doi:10.1073/pnas.1614247113,Vinograd2021}.
Incommensuration, in principle, implies complete loss of Bloch periodicity, and therefore the unit cell
of the system is thermodynamically large. Since such a situation cannot be modeled in any calculation, the
standard practice is to truncate the unit cell by a finite size. This is equivalent to imposing an approximate
commensurate order, which revives Bloch periodicity.
In our calculation, we assume a commensurate ordering
vector $Q_{x,y}= \pi/2$, which is close to the actual wave vector found in experiments.
Thus, the Hamiltonian in our calculation is a 16$\times$16 matrix of Eq. (\ref{HCDW3}).
Conceptually, the above approximation does not
affect our main point that the CDW order results in a dip-hump feature in the $B_{2g}$ Raman
response. This is because, irrespective of whether the ordering vector is commensurate or not, the loss of spectral weight in the CDW phase is due to gap opening around the nesting
points in the wave vector space that satisfy $\xi_{\mathbf{k}}= \xi_{\mathbf{k} \pm \mathbf{Q}}$.
This leads to the dip in the Raman response.
Likewise, irrespective of whether $\mathbf{Q}$ is commensurate or not, inter-band transitions
between folded bands give rise to the hump feature.
Furthermore, the leading contribution to
the dip-hump comes from the hybridization between the states at
$\mathbf{k}$ and $\mathbf{k} + \mathbf{Q}$
in the CDW phase. Any hybridization contribution between the states
$\mathbf{k}$ and $\mathbf{k} + n \mathbf{Q}$,
with $n > 1$, is relatively small by a factor of $(J/W)^{n-1}$, where $W$ is the bandwidth, and therefore
can be neglected for sufficiently large $n$.

A second well-known important issue is that, in the absence of a stabilizing magnetic field,
the CDW order is
short range. Therefore, it is legitimate to ask why our model, which is appropriate for a
true long-range density wave order, is applicable. Eventually, this issue is also related to the
question whether Raman responses are able to detect such short-range order.
Such a short range order can be modeled by modifying Eq. (\ref{HCDW2}) by a CDW potential
which is slowly varying in space such that the hybridization is of the form
\begin{align}
  \sum_{\mathbf Q(\mathbf{r}),\mathbf k \in \text{BZ}}
  \frac{J(\mathbf{r})}{2}(\cos k_x - \cos k_y)
  c^{\dagger}_{\mathbf k +\mathbf Q(\mathbf{r})/2}c_{\mathbf k - \mathbf Q(\mathbf{r})/2}
& & \nonumber \\
   + \text{H.c.}, & &
    \label{HCDWR}
\end{align}
where we consider spatial variation of both the energy scale $J$ and the ordering vector
$\mathbf{Q}$. As we discussed above, the CDW feature in the Raman response is around the
energy scale $2J$. Consequently, if the spatial variation is mostly in the ordering vector
$\mathbf{Q(r)}$, which seem to be the case for the cuprates~\cite{Comin},
we expect that the Raman response is not affected too much by the short range
nature of the order. This is because Raman response is an energy-resolved but momentum
averaged probe, and therefore the dip-hump is not blurred by variations in $\mathbf{Q(r)}$, provided
$J$ remains uniform over different patches. This is in contrast to momentum-resolved
probes such as e.g. ARPES, which depends on well-defined momentum-states and are therefore
sensitive to the variation of the ordering wave-vector. As a proof of principle, in
Fig. \ref{Raman-CDW}\textbf{(d)} we plotted $\Delta \chi_{B_{2g}}(\omega)$ obtained by averaging
over the Raman responses
obtained by three different ordering vectors $\mathbf{Q}$ and $\mathbf{Q} \pm \delta\mathbf{Q}$.
We took $\delta Q/Q = 0.1$, which
is consistent with the worst-case-scenario of the broadening of the x-rays lines
used in e.g. Ref. \cite{Comin}.
As expected, we
find that the energy scale at which the dip-hump feature appears does not change significantly
with 10\%
variation of the ordering wavevector, and therefore the feature itself is
not blurred by the spatial average.
While this computation is sufficient to demonstrate that Raman response is robust to spatial
variations of the ordering wavevector, for more accurate quantitative comparison with
experiments (which is not the goal here)
one needs to average the Raman signal with respect to a Gaussian distribution of ordering wavevectors.

\section{CDW and PG coexistence}
\label{CDW+PG}

We shall now introduce the effects of the PG, and show how this can impact the properties of the CDW phase that
we described in the previous sections. This will allow us to draw definitive conclusions on the detection
of the CDW in cuprates within Raman spectroscopy experiments.

\subsection{PG modeling}

PG is introduced here by using the phenomenological ansatz inspired by the proposal of
Yang, Zhang and Rice (YZR model \cite{YZR1}),
and which has proved successful in grasping many experimental features of cuprates \cite{YZR2}.
According to the YZR model, the Green's function develops a pole singularity in the self-energy and assumes the form
at low $\omega$:
\begin{equation}
G_{\text{PG}}(\mathbf k,\omega) \sim \frac{1}{\omega - \xi^{\te{c}}_{\mathbf k} - \frac{|\Delta^{\text{PG}}_{\mathbf k}|^2}{\omega - \xi^{\te f}_{\mathbf k}}}.
\label{gYRZ}
\end{equation}
Here $\xi^{\te c}_{\mathbf k}$ is the free electron dispersion [see Eq. (\ref{c-freedis})], $\xi^{\te f}_{\mathbf k}=2t\left(\cos k_x + \cos k_y\right)$ is a strong scattering diamond-like line which crosses the perfect nesting diagonals in the BZ quadrants.
The origin of this scattering has been strongly debated. It has been ascribed to short-ranged AF order \cite{YZR2}
(which could survive at large doping despite the long-ranged AF correlation dies) or other more exotic mechanisms, like a
not-yet well identified hidden order, or Mott physics. Notice that a pole-like form in the
self-energy had been previously predicted and widely analyzed by Cluster Dynamical Mean Field Theory studies
\cite{PhysRevB.73.165114,PhysRevB.74.125110, PhysRevLett.102.056404}
within the framework of the microscopic two-dimensional Hubbard Model. Following the YZR formulation,
we adopt here a $d$-symmetric PG by setting the self-energy pole strength
$\Delta^{\text{PG}}_{\mathbf k} = \Delta\left( \cos k_x - \cos k_y\right)$.
In order to simplify our analysis, we set the same $t$ in $\xi^{\te f}_{\mathbf k}$ and in $\xi^{\te c}_{\mathbf k}$.
We fix here the doping $p=0.12$. This allows us to disregard the doping dependence of the band parameters originally
considered by YZR. In Sec \ref{DC} we shall re-discuss doping dependencies within the contest of the interplay between CDW and PG.

\begin{figure}[]
    \centering
    \includegraphics[scale=0.4]{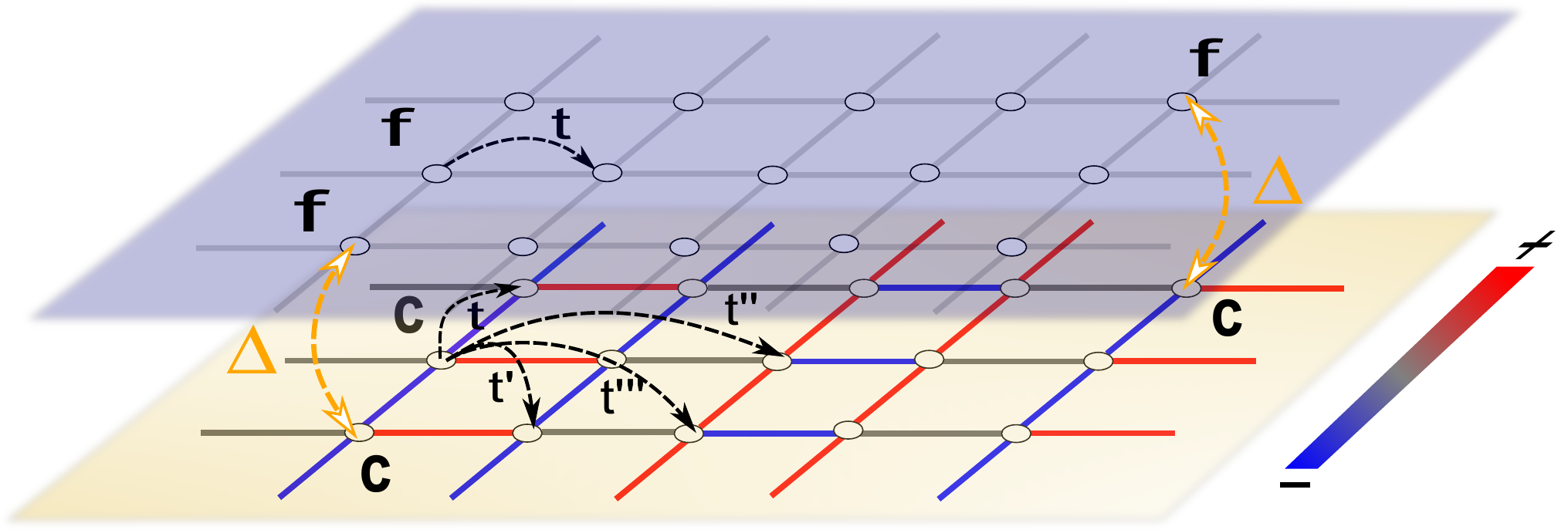}
    \caption{Schematic representation of the model in Eq.'s (\ref{HPG}) and (\ref{H_CDW-PG}).
      The $f$ fermions live in the top abstract layer and hybridize with the $c$ fermions.
      The strength of the hybridization
      potential $\Delta$ has a $d$-wave symmetric form.
      The color map on the bonds of the physical layer represents a CDW with wave-vector
      $\mathbf Q_x=\frac{2\pi}{4a}\hat{x}$.}
    \label{f-fermions}
\end{figure}
The Green's function (\ref{gYRZ}) can be obtained from a two-band theory that describes the coupling between two
different kinds of fermions $c^{\dagger}_{\mathbf k}$ and $f^{\dagger}_{\mathbf k}$ \cite{hidden},
\begin{equation}
    H_{\text{PG}} = \sum_{\mathbf k \in \text{BZ}}\psi^{\dagger}_{\mathbf k}\mc H_{\text{PG}}(\mathbf k)\psi_{\mathbf k},
    \label{HPG}
\end{equation}
where $\psi^{\dagger}_{\mathbf k} = (c^{\dagger}_{\mathbf k},f^{\dagger}_{\mathbf k})$ and
\begin{equation}
\mc H_{\text{PG}}(\mathbf k) = \begin{pmatrix}
\xi^{\te c}_{\mathbf k} & \Delta^{\text{PG}^*}_{\mathbf k}\\
\Delta^{\text{PG}}_{\mathbf k} & \xi^{\te f}_{\mathbf k}
\end{pmatrix}.
\end{equation}
We can describe the $f$ fermions as living in an abstract layer and hybridizing with the physical $c$ fermions;
see Fig. \ref{f-fermions}. To obtain the Green's function in Eq. (\ref{gYRZ}) we need to integrate out the
$f$ fermions \cite{hidden}. This is easily obtained within the path-integral formalism \cite{coleman2015introduction}
by considering an effective action $\mc S_{\text{eff}}[\bar{c},c]$ for the $c$ fermions only (see Appendix \ref{ap3}):
\begin{equation}
    \exp\left(-\mc S_{\text{eff}}[\bar{c},c] \right) \propto \int\mc D(\bar f,f)\exp\left( -\mc S[\bar{c},c,\bar{f},f]\right),
    \label{efaction}
\end{equation}
where $\mc S[\bar{c},c,\bar{f},f]$ is the action corresponding to the Hamiltonian $H_{\text{PG}}$ in Eq. (\ref{HPG}).


\subsection{PG and CDW modeling}

Within our formalism, we can consider now at the same time the CDW and the PG,
by proposing the following Hamiltonian in the $\mathbf k$-space:
\begin{eqnarray}
    \mathcal H&=&\sum_{\mathbf k \in \text{BZ}}\xi^{\te c}_{\mathbf k}c^{\dagger}_{\mathbf k}c_{\mathbf k} + \sum_{\mathbf Q,\mathbf k \in \text{BZ}}\lambda_{\mathbf Q}(\mathbf k) c^{\dagger}_{\mathbf k +\mathbf Q/2}c_{\mathbf k - \mathbf Q/2}  + \text{H.c}\nonumber\\
    &&+\sum_{\mathbf k \in \text{BZ}}\xi^{\te f}_{\mathbf k}f^{\dagger}_{\mathbf k}f_{\mathbf k} + \sum_{\mathbf k \in \text{BZ}}\Delta^{\text{PG}}_{\mathbf k} c^{\dagger}_{\mathbf k}f_{\mathbf k}  + \text{H.c}.
    \label{H_CDW-PG}
\end{eqnarray}
\begin{figure}[]
    \centering
    \includegraphics[scale=0.48]{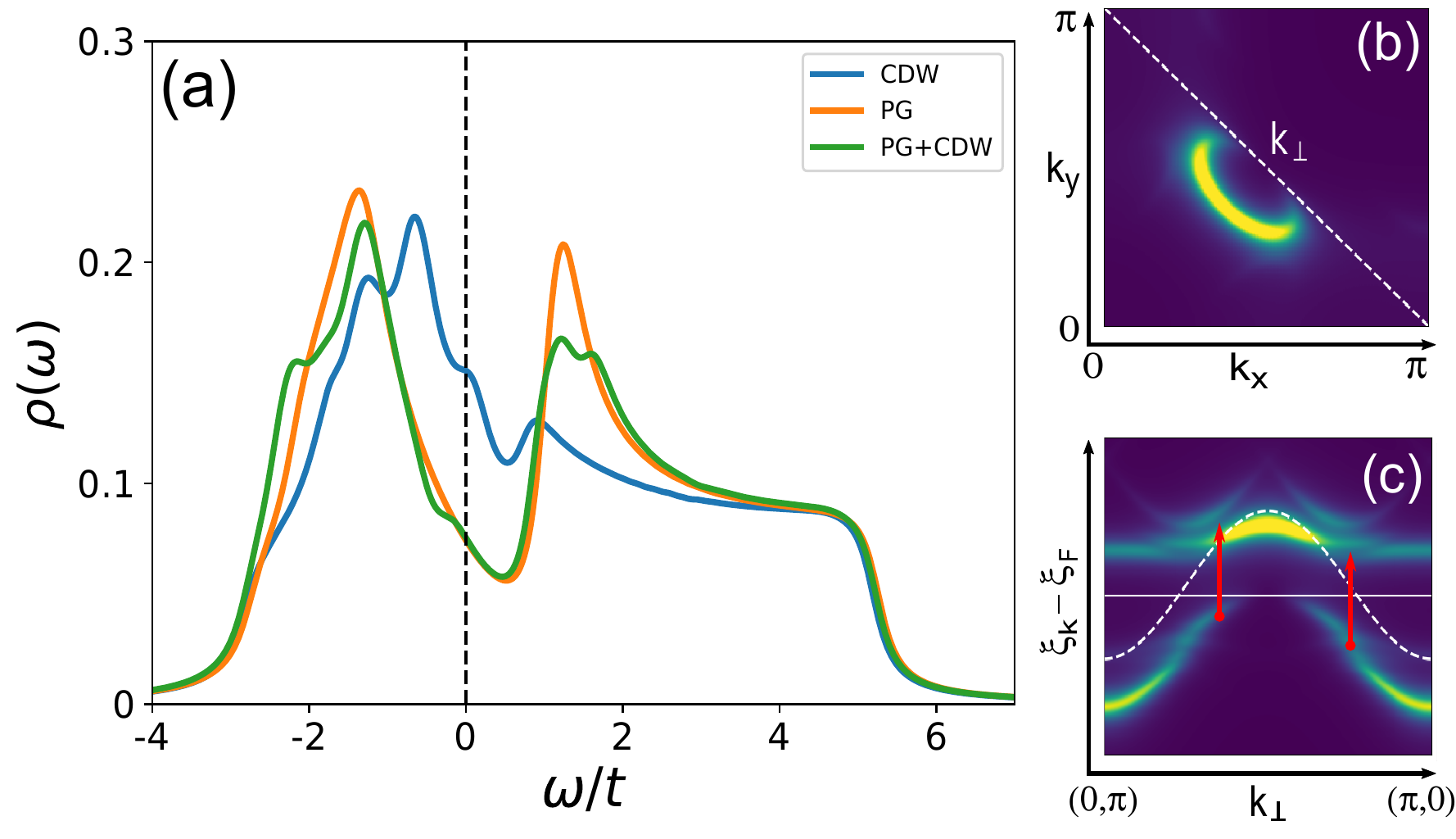}
    \caption{
      \textbf{(a)} Comparison between the $DOS$ of the CDW, PG and CDW+PG phases.
      \textbf{(b)} FS for the CDW+PG phase (with the same color map code adopted in Fig. \ref{Arpes-CDW}).
      \textbf{(c)} Corresponding ARPES dispersion along the path $k_{\perp}$. The dashed white line is the free
      electron dispersion and the red arrows are indicating new electronics transitions. Liechtenstein parameter
      set is adopted here. }
    \label{DOS_PG_CDW}
\end{figure}
We need again to integrate out the auxiliary $f$ fermions in order to obtain the corresponding $c$ fermions effective
action. Letting the details of this integration for Appendix \ref{ap3}, we obtain:
\begin{equation}
    \mc S_{\text{eff}}[\bar \Psi,\Psi]\:\:=-\sum_{\omega_n,\mathbf p \in \text{rBZ}}\bar \Psi_{n\mathbf p}\mathbf G^{-1}(\mathbf p,i\omega_n)\Psi_{n\mathbf p},
    \label{coe_action}
\end{equation}
where $\bar \Psi_{n\mathbf p}$ is the $n$-Matsubara component of the spinor $\bar \Psi_{\mathbf p}$ (see Eq. (\ref{base_rBZ})),
\begin{equation}
   \mathbf G^{-1}(\mathbf p,i\omega_n) = i\omega_n\mathbb 1 - \mc H(\mathbf p) - \frac{|\Lambda_{\mathbf p}|^2}{i\omega_n\mathbb 1 - \mc H^{\te f}_{\mathbf p}},
   \label{Green_CDW_PG}
\end{equation}
with
\begin{equation}
\Lambda_{\mathbf p} = \begin{pmatrix}
\Delta^{\text{PG}}_{\mathbf p} & 0 & \cdots & 0\\
0 & \Delta^{\text{PG}}_{\mathbf p+ \mathbf Q_x} & \cdots & 0\\
\vdots & \vdots & \ddots & \vdots\\
0 & 0 & \cdots & \Delta^{\text{PG}}_{\mathbf p+3\mathbf Q_x + 3\mathbf Q_y}
\end{pmatrix},
\label{PG_gap_matrix}
\end{equation}
and a similar definition for $\mc H^{\te f}_{\mathbf p}$ but with $\Delta^{\text{PG}}_{j}\to\xi^{\te f}_{j}$.

Like for the CDW-only case, we can relax the condition $\mathbf p\in \text{rBZ}$ and work only with the first element of the $\mathbf G(\mathbf p,\omega)$. Thus, the $c$ fermions spectral function is given by
\begin{equation}
    \mathcal A(\mathbf k, \omega) = -\frac{1}{\pi}\text{Im}\:\mathbf G_{11}(\mathbf k,\omega),
    \label{c-spectral}
\end{equation}
where $\mathbf k \in \text{BZ}$. From the Green's function, we can calculate all single particle quantities. As
we stressed in the previous section \ref{CDW}, the change of the band parameters does not produce too much differences
in determining the CDW feature in the Raman $B_{2g}$ response.
For sake of convenience, we shall consider here the Liechtenstein set, $J/2 =0.2t$,
$\Delta=0.8t$ and a hole-doping $p=0.12$, which corresponds to a chemical potential $\mu =-0.804t$. 

\subsection{CDW and PG spectrum}

In Fig. \ref{DOS_PG_CDW}\textbf{(a)} we show the $DOS$, $\rho(\omega)$, with both CDW and PG (green line).
For sake of comparison, we also show the same $DOS$ when only the CDW order [Eq. (\ref{HCDW3})] and only
the PG [Eq. (\ref{HPG})] are present (blue and orange lines, respectively).
We note the characteristic PG spectral weight loss around the Fermi level $\omega=0$,
and that the CDW spectral weight dip is sub-leading (see Fig. \ref{DOS}).
The characteristic modulations due to the bands folding
away from $\omega=0$ remain well visible, but are also a rather small feature compared to the large
PG spectral depression. These results are in line with the several observations on the $DOS$ of cuprates via
STM \cite{RevModPhys.79.353, JPSJ.81.011005},
where the PG features are now well established, while the features related to the CDW could be
revealed only with a more involved real-space map analysis \cite{science.1066974}. Notice though that
while the presence of the CDW could finally be detected in the $DOS$, it has not been
possible to extract from these measures a clear CDW energy scale.

It is also useful to observe the spectral density in momentum space $\mathcal A(\mathbf{k},\omega)$,
which can be directly accessed by ARPES measurements. In Fig. \ref{DOS_PG_CDW}\textbf{(b)} we plot a
spectral intensity cut $\mathcal A(\mathbf{k},\omega=0)$ at the Fermi level. This panel should be compared
with Fig. \ref{Arpes-CDW}\textbf{(g-i)}. Again the PG dominates the spectra, and we recover the typical Fermi arc,
which is a well known feature in 
underdoped cuprates \cite{YZR1,Arpes_FS_2}, and which the CDW order alone is not able
to reproduce (unless considered at artificially intense values, not observed in cuprates). An energy vs momentum
cut along the BZ diagonal $k_{\perp}$ is displayed in  Fig. \ref{DOS_PG_CDW}\textbf{(c)}.
Again, this panel should be compared with the spectral plot along the same cut for only the CDW case in
Fig. \ref{Arpes-CDW}\textbf{(j-l)}. By looking at the occupied side $\xi_{\mathbf k} < \xi_F$, we can observe that the
dominant PG opens a wide gap close to the antinodal regions, shifting down all the bands, that are like compressed.
The resulting spectra are less dispersive, and, the fact that is more interesting for our discussion, the
CDW folded bands are much less visible than in the case where only the CDW is present. The resulting effect is
the appearance of waterfall-like features, i.e. abrupt vertical jumps in $\xi_{\mathbf{k}}$ vs $\mathbf{k}$  band dispersion,
in going from $\mathbf{k}= (0,\pi) \to (\pi/2,\pi/2)$ and approaching the nodal regions. Waterfall features in
cuprate spectra are well known \cite{Lanzara2001-he, PhysRevB.71.214513, PhysRevLett.98.067004, PhysRevLett.98.167003, PhysRevLett.98.147001, PhysRevB.75.174506},
and ascribed to phonon-electron or electron-electron
correlations. Our result shows that these features may be related to the CDW appearance as well. The subleading order
of the CDW effects compared to the PG ones makes the detection of CDW within ARPES highly nontrivial. An involved
analysis of the bands folding, along the lines of Ref. \cite{Hashimoto2010-qa} may be required, possibly considering precise band structure details.

This result suggests an answer to the open question on why it has been difficult to detect the
CDW order within ARPES up to now, differently from X-rays and STM measurements. Notice however that the CDW
folded bands remain clearly visible in the unoccupied side $\xi_{\mathbf k} > \xi_F$ of the spectra. We predict then that
future advancements in ARPES techniques that could allow accessing the unoccupied states, like laser-pumping
photoemission (on the lines of reference \cite{science.1217423}) or inverse photoemission (see e.g. \cite{Yang2008-zv, RevModPhys.93.025006}), could finally reveal the CDW folded band structure.

We now turn to the Raman response, which is obtained by using Eq. (\ref{raman}). We shall show that thanks to
the momentum space selectivity and the access to unoccupied states, this probe is still able to capture the
CDW even in the presence of a PG, as pointed out in experiments.

\begin{figure}
    \centering
    \includegraphics[scale=0.24]{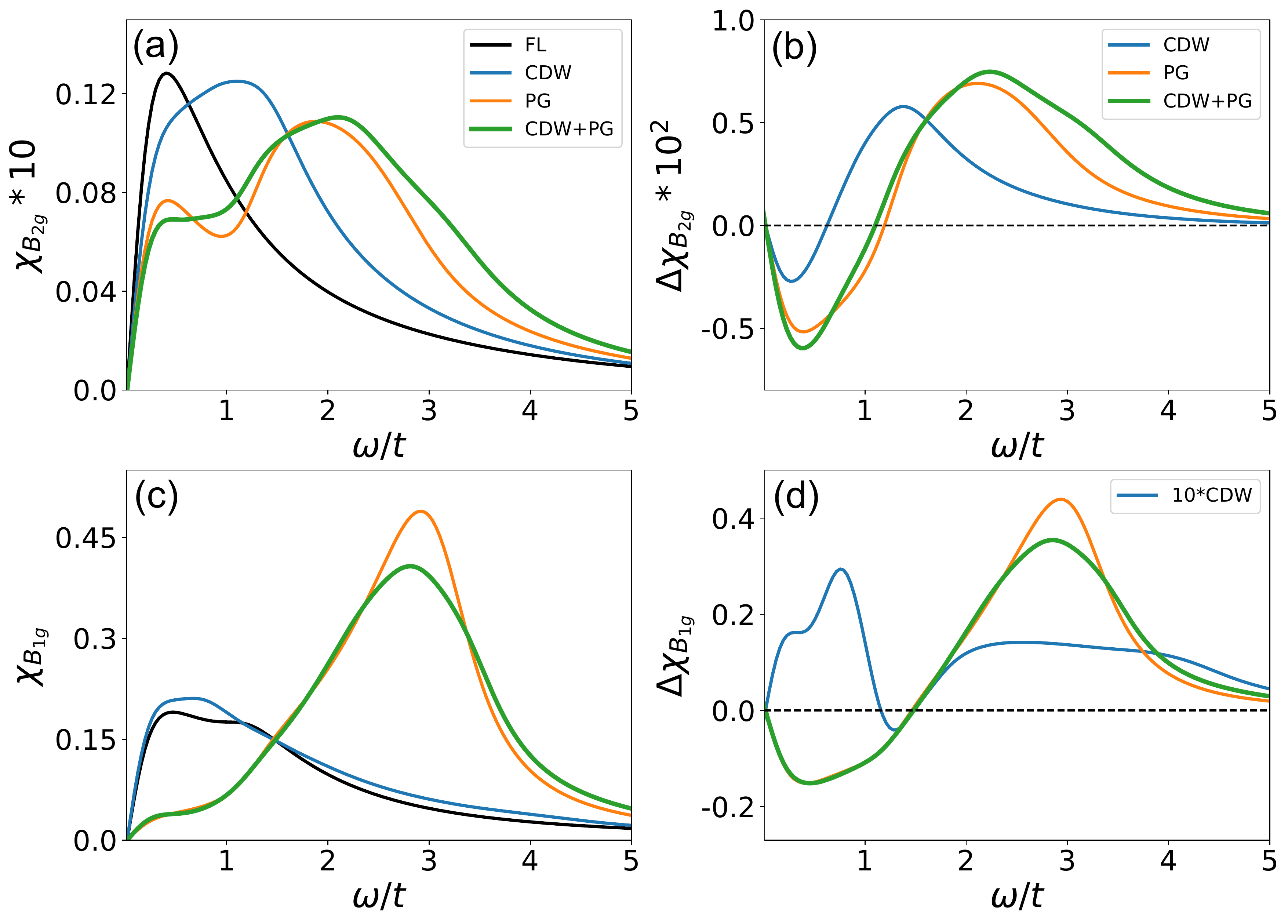}
    \caption{$B_{2g}$ and $B_{1g}$ Raman responses for the different phases, FL, PG, CDW and PG+CDW.
      \textbf{(a)} Absolute $B_{2g}$ Raman response. \textbf{(b)} Relative $B_{2g}$ Raman response.
      \textbf{(c)} Absolute $B_{1g}$ Raman response. \textbf{(d)} Relative $B_{1g}$ Raman response. Here the CDW response is
      10 times weaker than the PG one. }
    \label{Raman_CDW_PG}
\end{figure}

In Fig. \ref{Raman_CDW_PG}\textbf{(a)} we show the $B_{2g}$ response obtained with both the CDW and the PG. For sake of comparison, we also plot the same Raman response with only the PG, CDW, and the simple FL. In the case of the  $B_{2g}$ response,
the CDW and the PG are of comparable intensity, and they ``sum-up'' in the relative  Raman response $\Delta \chi_{B_{2g}}$
to produce a dip-hump feature which has {\it both} contributions,
as clearly shown in Fig. \ref{Raman_CDW_PG}\textbf{(b)}. The $B_{2g}$ response is
still in agreement with the experimental $B_{2g}$ feature [displayed in Fig. \ref{Raman-CDW}\textbf{(a)}], and this shows that
this the Raman  $B_{2g}$ response is indeed capable to detect the CDW even in presence of the PG.

This behavior must be contrasted with the one of the $B_{1g}$ response which is almost completely dominated by
the PG, displaying little contribution from the CDW, see Fig. \ref{Raman_CDW_PG}\textbf{(c-d)}. Moreover, as we discussed in Fig.
\ref{Raman-CDW}\textbf{(c)}, the shape of the relative $\Delta \chi_{B_{1g}}$ response strongly depends on the details of the
band parameters. CDW cannot be well detected in this Raman polarization channel, as pointed out in the
recent experimental work of Ref. \cite{Alain}.


\subsection{CDW and PG energy scales}
\label{DC}

The relation between the CDW and the PG displayed in the Raman $B_{2g}$ response has been strongly debated
\cite{Alain,pepin}. In particular, the experimental results have pointed out that the
Raman $B_{2g}$ response allows extracting the CDW energy scale $E_{CDW}$, defined as the energy of
the maximum of the hump in $\Delta \chi_{B_{2g}}$ [Fig. \ref{Raman-CDW}\textbf{(b)}]. This procedure is similar to the
one well established for the PG energy scale $E_{PG}$, which is at its time
defined as the energy of the maximum in the Raman $\Delta \chi_{B_{1g}}$.
The remarkable experimental fact is that  $E_{CDW}(p)$, as a function of doping $p$, follows the trend
of  $E_{PG}(p)$, rather than the one of $T_{CDW}(p)$, the CDW transition temperature, as it would be naturally
expected from a standard mean-field behavior. Our result about the interaction between CDW and PG, revealed by
our modeling of the $B_{2g}$ response [Fig. \ref{Raman_CDW_PG}\textbf{(b)}],
suggests indeed that the $E_{CDW}(p)$ may be ruled by the PG.

In order to show this, we introduce a doping dependence in our model. We shall neglect at the first order approximation
the doping dependence of band parameters coming from correlation effects (see Ref.\cite{YZR1} for a more
detailed modeling of this), and focus on the doping dependence of the PG,
$\Delta$, which we extract from
Raman experiments \cite{Collapse}:
\begin{equation}
    \Delta(p) = \, \theta(0.2-p)\, (1-2p)t,
    \label{delta_dependence}
\end{equation}
where $\theta(x)$ is the Heaviside step function.
This $\Delta(p)$ form simulates the sudden collapse of the PG phase around $p=0.2$, as it was observed in
experiments \cite{Collapse}.
Always within a first order approximation, we also neglect the doping dependence of the CDW phase considering
\begin{equation}
  J(p) \approx \text{const.}= 0.4t.
    \label{J_dependece}
\end{equation}
With this choice, any possible doping dependence of energy scale must take its roots from the PG one.
Notice that in our model there is not explicit interaction between the PG and CDW. This interaction is
however implicit via the coupling $\Delta$ of the auxiliary pseudogap $f$ fermions and the coupling $J$ of the
CDW interaction with the bare $c$ electron band [see Eq.'s (\ref{H_CDW-PG}) and (\ref{d-symme})].

\begin{figure}[t]
    \centering
    \includegraphics[width=8.7cm,height=3.5cm]{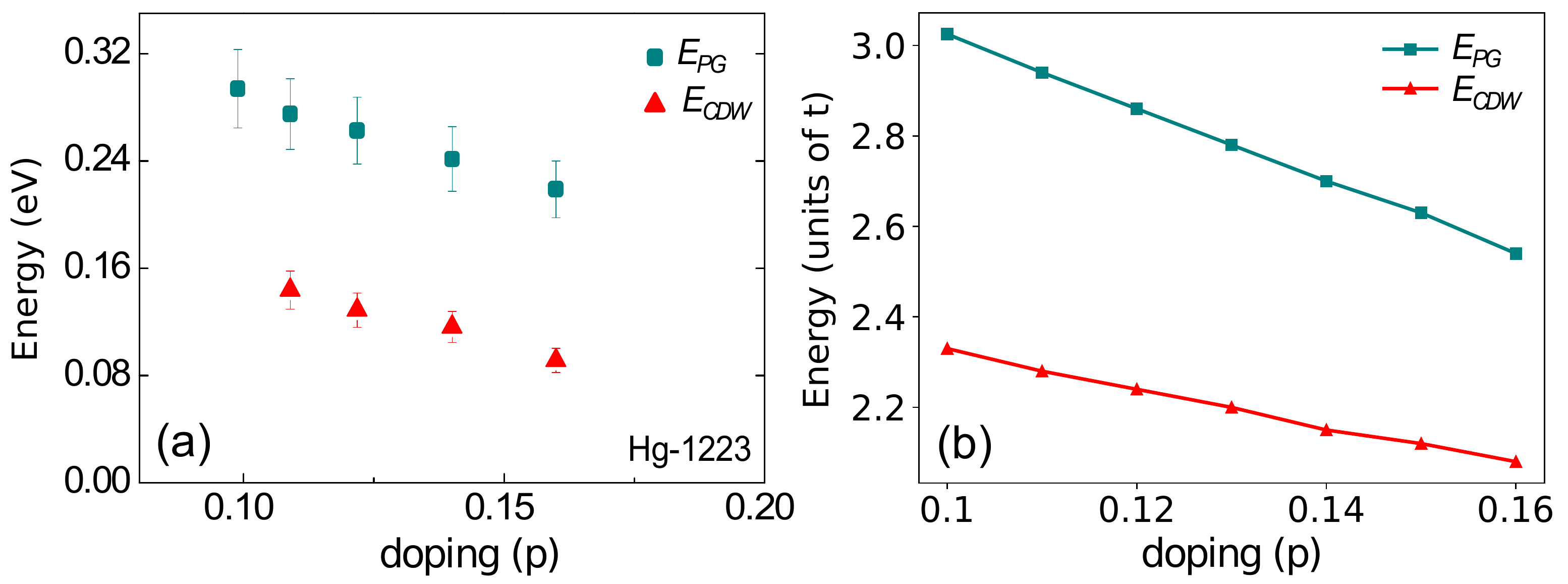}
    \caption{$E_{CDW}(p)$ and $E_{PG}(p)$ energy scales extracted from the $B_{2g}$ and $B_{1g}$ Raman responses, respectively. In \textbf{(a)} is shown the experimental results adapted from \cite{Alain}, while in \textbf{(b)} is shown the theoretical results obtained from our CDW-PG coexistence model given in Eq. (\ref{Green_CDW_PG}). }
    \label{doping_energy}
\end{figure}

We can calculate the Raman response for different symmetries and doping levels
and extract the CDW $E_{CDW}$ and PG $E_{PG}$ energy scales from the $B_{2g}$ and $B_{1g}$ Raman responses
as the frequency $\omega_{\nu}$ where the maximum of the hump following the dip is located, like in
the Raman experiments \cite{Alain,PhysRevB.96.094525} [see Fig. \ref{doping_energy}\textbf{(a)}].
For comparison, we display the theoretical $E_{CDW}$ (red line) and $E_{PG}$ (green line) in Fig. \ref{doping_energy} \textbf{(b)}.
As a matter of facts, the CDW energy scale $E_{CDW}$ definitively follows the PG one $E_{PG}$, in close resemblance
with the aforementioned experiments [see Fig. \ref{doping_energy} \textbf{(a)}] \cite{Alain}, proving that indeed the PG can be the driving force dominating the
CDW energy scale too.

A last point that we would like to call attention to is the role played by the CDW order on the energy scales shown in Fig \ref{doping_energy}\textbf{(b)}. A first look at Fig. \ref{Raman_CDW_PG}\textbf{(b)} may lead the reader to think that the charger order has no influence on the energy scales $E_{CDW}$ and $E_{PG}$. However, given the implicit coupling between $\Delta$ and $J$ in our model [see Eq. (\ref{Green_CDW_PG})], both energy scales carry a $J$-dependence. To show that, in appendix \ref{apen_CDW_domo} we display the doping dependent energy scales for the case $J(p)\propto T_{CDW}(p)$. From this result we see the above mentioned fact.

\section{Conclusions}
\label{conclusions}

We study the charge density wave in cuprate high temperature superconductors, focusing
on its detection within Raman spectroscopy. To this purpose we adopt a microscopic tight binding approach to describe
the Cu-O planes of cuprates, and consider terms that describe both a bi-modal $x$-$y$ CDW (which mimics experiments \cite{Comin}) and the pseudogap phase within a Yang-Zhang-Rice-like approach \cite{YZR1}. Our main result is to show that the CDW produces a distinctive dip-hump feature in the Raman the $B_{2g}$ light polarized response, consistent with what observed in Raman experiments \cite{Alain}. This feature is universal and only weakly dependent on the details of the band parameters of the different cuprate compounds. We interpret these findings in terms of
novel electronic Raman excitations which involve electronic branches folded into the Brillouin Zone reduced by the 4 lattice steps CDW periodization in the $x$ and $y$ directions. These electronic branches are most relevant close to the nodal regions of momentum space, and hence produce a signature in the $B_{2g}$ geometry which is most sensitive to that region. In sharp contrast,
the $B_{1g}$ Raman response is less sensitive to the electronic changes brought about by the CDW order and most sensitive to the details of the band dispersion. The CDW signature in this light polarization is therefore strongly compound dependent. By adding the PG interaction, we show that the  $B_{2g}$ CDW
feature survives, validating its interpretation within Raman experiments on cuprates.

We observe that in general the effects of the CDW order are subleading with respect to the PG ones. PG interaction is overwhelming in the  $B_{1g}$ response, which remains the most adapted one to detect the PG. This latter also dominates the spectral weight depression around the Fermi level in the density of states, though subleading CDW distinctive features can be identified \cite{RevModPhys.79.353}.
PG has more impact on ARPES, where the CDW order detection remains debated. This is in part due to the fact that CDW folded bands are most visible in the unoccupied side of the spectra, which is not easily accessible by ARPES. CDW features in the occupied side, located closer to the antinodes, may produce water-falls structures that are blurred by the short-ranged nature of the CDW order in these materials (in the absence of strong magnetic fields).

Finally, we show that the dominant role of the PG may also be at the origin of the monotonic behavior of the CDW energy scale with doping, that, alike the superconducting gap, does not follow the dome-like shape of its critical temperature. This latter behavior is what would be expected in a conventional mean-field description of a broken symmetry. This is suggestive of the role of the pseudogap in ruling the energy scales on the complex and exotic phase diagram of cuprate high temperature superconductors \cite{efetov,Fradkin_2015,pepin}.
Possible future work should address the competition between the CDW and the PG starting from microscopic theories that treat these two phenomena on the same footsteps, like e.g. Hubbard-like Hamiltonians (see e.g. reference \cite{Senechal}). The charger order should be also studied within the superconductor phase where other non-trivial symmetry broken orders, like pair density wave \cite{PDW}, could emerge.


\acknowledgements
M. F. Cavalcante would like to thank Vitor A. M. Lima for his help with the numerical part of this work. We thank Marc-Henri Julien for useful discussions. We acknowledge financial support from French Agence Nationale de la Recherche (ANR) grant ANR-19-CE30-0019-01 (Neptun). This work is supported by FAPEMIG, CNPq (in particular through INCT- IQ 465469/2014-0), and CAPES (in particular through program CAPES-COFECUB-0899/2018).

\appendix


\section{$16\times 16$ CDW matrix Hamiltonian}
\label{ap1}
In this appendix, we write down the $16\times 16$ CDW matrix Hamiltonian in Eq. (\ref{HCDW3}).

The d-symmetric, four periodic and bidirectional charge modulation matrix in Eq. (\ref{HCDW3}) is given by
\begin{equation}
\mc H(\mathbf p) = \begin{pmatrix}
\textbf{A}_0 & \textbf{B}_\frac{1}{2} & \textbf{0} & \textbf{B}_\frac{7}{2}\\
\textbf{B}_\frac{1}{2} & \textbf{A}_{1} & \textbf{B}_\frac{3}{2} & \textbf{0}\\
\textbf{0} & \textbf{B}_\frac{3}{2} & \textbf{A}_{2} & \textbf{B}_\frac{5}{2}\\
\textbf{B}_\frac{7}{2} & \textbf{0} & \textbf{B}_\frac{5}{2} & \textbf{A}_{3}
\end{pmatrix},
\end{equation}
where
\begin{equation}
\small
\textbf{A}_n = \begin{pmatrix}
\xi_{\mathbf p + n\mathbf Q_y} & \lambda_{\frac{1}{2},n} & 0 & \lambda_{\frac{7}{2},n}\\
\lambda_{\frac{1}{2},n} & \xi_{\mathbf p + n\mathbf Q_y + \mathbf Q_x } & \lambda_{\frac{3}{2},n} & 0\\
0 & \lambda_{\frac{3}{2},n} & \xi_{\mathbf p + n\mathbf Q_y + 2\mathbf Q_x} & \lambda_{\frac{5}{2},n}\\
\lambda_{\frac{7}{2},n} & 0 & \lambda_{\frac{5}{2},n} & \xi_{\mathbf p + n\mathbf Q_y + 3\mathbf Q_x}
\end{pmatrix},
\end{equation}
and
\begin{equation}
\textbf{B}_{n} = \begin{pmatrix}
\lambda_{0,n} & 0 & 0 & 0\\
0 & \lambda_{1,n} & 0 & 0\\
0 & 0 & \lambda_{2,n} & 0\\
0 & 0 & 0 & \lambda_{3,n}
\end{pmatrix},
\end{equation}
with $\textbf{0}$ denoting the $4\times 4$ null matrix, $\lambda_{m,n} \equiv \lambda_{\mathbf Q}(\mathbf p + m\mathbf Q_x + n\mathbf Q_y)$, $\xi_{\mathbf p}$ was defined in Eq. (\ref{free}) and $\mathbf Q_{x,y}$ just below Eq. (\ref{d-symme}).
\newpage
\section{Vertex function}
\label{vertex_apen}

In this appendix, we show the BZ regions probed by the $B_{2g}$ and $B_{1g}$ light polarization.

\begin{figure}[h]
    \centering
    \includegraphics[scale=0.28]{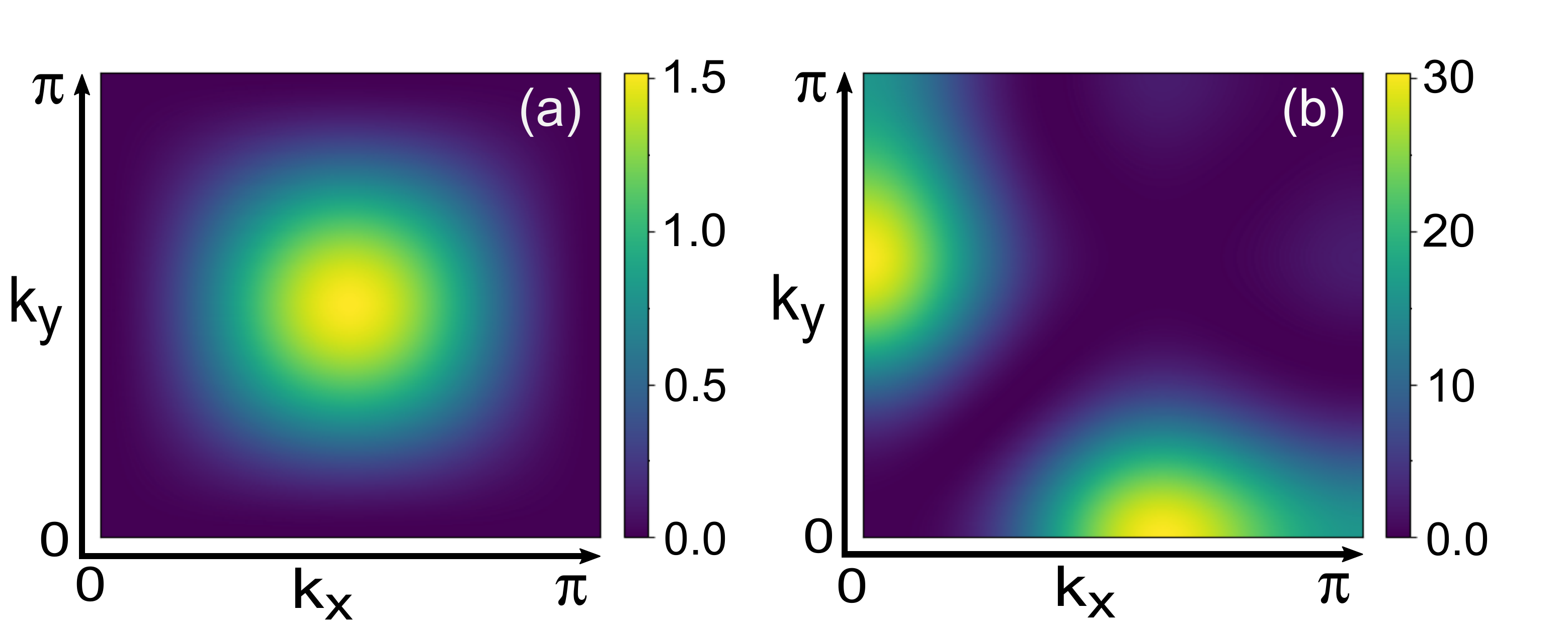}
    \caption{Square of vertex function (Eq.'s (\ref{ape_vertex_b2g}) and \ref{ape_vertex_b1g}) for the $B_{2g}$ \textbf{(a)} and the $B_{1g}$ \textbf{(b)} channels.}
    \label{fig_vertex}
\end{figure}

For a simple single-band model, the vertex matrix in Eq.'s (\ref{vertB2g}) and (\ref{vertB1g}) become scalars. Besides, assuming Liechtenstein band parameters set (see table \ref{table}), we have explicitly
\begin{equation}
    \Gamma^{\te{B2g}}_{\mathbf k}=-4t' \sin k_x \sin k_y,
    \label{ape_vertex_b2g}
\end{equation}
\begin{equation}
    \Gamma^{\te{B1g}}_{\mathbf k}=2\left[ t\left(\cos k_x -\cos k_y\right)+4t''\left(\cos 2k_x - \cos 2k_y\right)\right].
    \label{ape_vertex_b1g}
\end{equation}
In Fig. \ref{fig_vertex} we plot the square of each vertex function above in the first quadrant of the BZ. We can see that the $B_{2g}$ vertex probes the nodal region of the BZ whereas the $B_{1g}$ one the antinodal region.

\section{Derivation of Raman response}
\label{ap2}
In this appendix, we derive the Raman responses given in Eq. (\ref{raman}). The following calculation can also be applied to others dynamical responses, such as optical conductivity.

In the Matsubara frequency axis we have the general expression for the response function \cite{coleman2015introduction}
\begin{equation}
    \Tilde{\chi}(\mathbf q, i\Omega_m) = \int_{0}^{\beta} d\tau e^{i\Omega_m\tau}\Tilde{\chi}(\mathbf q,\tau),
\end{equation}
where $\Tilde{\chi}(\mathbf q,\tau) = - \langle T_{\tau} \mathcal O(\mathbf q,\tau) \mathcal O(-\mathbf q,0) \rangle$, with $T_{\tau}$ the complex time $\tau$ ordering operator, is the source perturbation. For the Raman response, we have \cite{Devereaux}
\begin{equation}
    \mathcal O(\mathbf q) \equiv \sum_{n,\mathbf p }\Gamma_{n\mathbf p}\Psi^{\dagger}_{n\mathbf p + \mathbf q}\Psi_{n\mathbf p},
    \label{O-raman}
\end{equation}
where $n$ corresponds to the index of the basis in which the Hamiltonian (\ref{HCDW3}) is written and $\Gamma_{n\mathbf p}$ are the vertex matrix elements (see Eq.'s (\ref{vertB2g}) and (\ref{vertB1g})), which give the scattering strength. One important point to note is that $\mathbf q$ and $\Omega_m$ are bosonic quantities. Without loss of generality, in Eq. (\ref{O-raman}) and the upcoming derivation we suppress the spin index. From Eq. (\ref{O-raman}) we see that the Raman response measures ``effective density fluctuation'' in the system \cite{Devereaux}.

Considering the ladder approximation \cite{DMFT,Devereaux,coleman2015introduction},  we obtain
\small
\begin{eqnarray}
    \Tilde{\chi}(\mathbf q,i\Omega_m)&=-&\sum_{n,n',\mathbf p}\Gamma_{n\mathbf p}\Gamma_{n'\mathbf p+\mathbf q}\int_{0}^{\beta}d\tau e^{i\Omega_m \tau}\mc G_{n'n}(\mathbf p +\mathbf q,-\tau)\nonumber\\
    &&\times \mc G_{nn'}(\mathbf p,\tau) + \text{(vertex correc.)},
    \label{expansion}
\end{eqnarray}
where $\mc G_{nn'}(\mathbf p,\tau) = - \langle T_{\tau}\Psi_{n\mathbf p}(\tau)\Psi^{\dagger}_{n'\mathbf p}(0)\rangle$ is the interacting Green's function and $+\text{(vertex correc.)}$ means the vertex corrections. For $n\neq n'$ this Green's function is finite due to the ``inter-band'' excitation terms present in the CDW Hamiltonian of Eq. (\ref{HCDW3}). Performing a Fourier transformation in the product of the two Green's function with opposite time argument and using the standard relation \cite{coleman2015introduction}
\begin{equation}
    \mc G_{nn'}(\mathbf p,i\Omega_m)=\int_{-\infty}^{+\infty}d\omega' \frac{\mc A_{nn'}(\mathbf p,\omega')}{i\Omega_m - \omega'},
\end{equation}
where $\mc A_{nn'}(\mathbf p,\omega')$ is the spectral function, we obtain (ignoring the vertex corrections hereafter)
\begin{eqnarray}
    \Tilde{\chi}(\mathbf q\to 0,i\Omega_m)&=-&\frac{1}{\beta}\sum_{i\nu}\sum_{n,n',\mathbf p}\Gamma_{n\mathbf p}\Gamma_{n'\mathbf p}\int_{-\infty}^{+\infty}d\omega'd\omega''\nonumber\\
    &&\times\frac{\mc A_{nn'}(\mathbf p,\omega')}{i(\nu+\Omega_m)- \omega'}\frac{\mc A_{n'n}(\mathbf p,\omega'')}{i\nu- \omega''},
\end{eqnarray}
where $\nu$ is a fermionic Matsubara frequency. Summing over the Matsubara frequencies \cite{coleman2015introduction},
\begin{eqnarray}
    \sum_{\nu}\frac{1}{i\nu - y_1}\frac{1}{i\nu - y_2}&=&\frac{\beta}{y_1-y_2}\left[ \frac{1}{e^{\beta y_1}+1} -\frac{1}{e^{\beta y_2}+1}\right],\nonumber\\
\end{eqnarray}
performing the analytic continuation, $i\Omega_m \to \Omega + i\eta$, and taking the limit of zero temperature we obtain for $\text{Im}\Tilde{\chi}(\mathbf q\to 0,\Omega) \equiv \chi(\Omega)$:
\begin{eqnarray}
    \chi(\Omega)&=&\sum_{n,n',\mathbf p}\Gamma_{n\mathbf p}\Gamma_{n'\mathbf p}\int_{-\Omega}^{0}d\omega'\mc A_{nn'}(\mathbf p,\omega')\nonumber\\
    &&\times \mc A_{n'n}(\mathbf p,\omega'+\Omega)\\
    &=&\sum_{\mathbf p \in \text{rBZ}}\int_{-\Omega}^{0}d\omega'\Tr[\Gamma_{\mathbf p}\mc A(\mathbf p,\omega')\Gamma_{\mathbf p}\mc A(\mathbf p,\omega'+\Omega)].\nonumber\\
\end{eqnarray}
Finally, after normalizing the above expression with respect to the number of states in the BZ, $\mc N$, we obtain Eq. (\ref{raman}).
\section{Integrating out the $f$ fermions}
\label{ap3}
In this appendix, we give details about the procedure to integrate out the auxiliary $f$ fermions in the Hamiltonian (\ref{H_CDW-PG}).

In the path-integral formalism the partition function corresponding to the Hamiltonian in Eq. (\ref{H_CDW-PG}) is \cite{coleman2015introduction}
\begin{equation}
    \mc Z = \int \mc D(\bar{c},c)\mc D(\bar{f},f)e^{-\mc S[\bar{c},c,\bar{f},f]}.
    \label{part_funct}
\end{equation}
Here $\mc D(\bar{\alpha},\alpha)=\prod_{\omega_n}\prod_{\mathbf k}d\bar{\alpha}_{\mathbf kn} d\alpha_{\mathbf kn}$, where $\alpha=c,f$ (and $\bar\alpha$) is a Grassmann's variable, and $\omega_n$ are the fermionic Matsubara frequencies. In the imaginary-time axis, the action is
\begin{equation}
    \mc S[\bar{c},c,\bar{f},f] = \int_0^\beta d\tau \sum_{\mathbf k\in \text{BZ}}\left(\bar{c}_{\mathbf k}\partial_{\tau}c_{\mathbf k} + \bar{f}_{\mathbf k}\partial_{\tau}f_{\mathbf k}\right) + \mathcal H,
\end{equation}
where the time dependence of the quantities was omitted. Similar to the spinor $\Psi_{\mathbf p}$ in Eq. (\ref{base_rBZ}), if we define
\begin{equation}
    \bar\Upsilon_{\mathbf p} = \left(\bar f_{\mathbf p}, \bar f_{\mathbf p + \mathbf Q_x},\bar f_{\mathbf p + 2\mathbf Q_x},...,\bar f_{\mathbf p + 3\mathbf Q_x+3\mathbf Q_y}\right),
\end{equation}
we can rewrite the above action as
\begin{eqnarray}
    \mc S[\bar{c},c,\bar{f},f]&=& \sum_{\mathbf p\in \text{rBZ}}\int_0^\beta d\tau \: \Big\{\bar\Psi_{\mathbf p}\left(\partial_{\tau} + \mc H({\mathbf p}) \right)\Psi_{\mathbf p}\nonumber\\
    &&+\bar\Upsilon_{\mathbf p}(\partial_{\tau} + \mc H^{\text{f}}_\mathbf p )\Upsilon_{\mathbf p} + \bar\Psi_{\mathbf p}\Lambda_{\mathbf p}\Upsilon_{\mathbf p} + \text{H.c.}\Big\},\nonumber\\
\end{eqnarray}
where we use the notation $\partial_{\tau} \equiv \partial_{\tau}\mathbb 1$ and $\mc H^{\text{f}}_{\mathbf p}$ and $\Lambda_{\mathbf p}$ are defined in the main text (see Eq. (\ref{PG_gap_matrix})). Taking the Fourier transform and performing the integral over time, we obtain
\small
\begin{eqnarray}
    \mc S[\bar{c},c,\bar{f},f]&=&\sum_{\omega_n,\mathbf p}\Big\{ \bar\Psi_{n\mathbf p}\left( -i\omega_n\mathbb 1 + \mc H({\mathbf p}) \right)\Psi_{n\mathbf p}\nonumber\\
    && +\bar\Upsilon_{n\mathbf p}( -i\omega_n\mathbb 1 + \mc H^\text{f}_{\mathbf p} )\Upsilon_{n\mathbf p} + \bar\Psi_{n\mathbf p}\Lambda_{\mathbf p}\Upsilon_{n\mathbf p} + \text{H.c.} \Big\}.\nonumber\\
\end{eqnarray}
Defining new variables
\begin{equation}
    \bar\Theta_{n\mathbf p} = \bar\Psi_{n\mathbf p}\frac{\Lambda_{\mathbf p}}{\mc H^{\text{f}}_{\mathbf p}-i\omega_n\mathbb 1} + \bar\Upsilon_{n\mathbf p},
\end{equation}
\begin{equation}
    \Theta_{n\mathbf p} = \frac{\Lambda^{*}_{\mathbf p}}{\mc H^{\text{f}}_{\mathbf p}-i\omega_n\mathbb 1 }\Psi_{n\mathbf p} + \Upsilon_{n\mathbf p},
\end{equation}
the partition function in Eq. (\ref{part_funct}) becomes
\begin{equation}
    \mc Z = \int \mc D(\bar \Psi,\Psi)\mc D(\bar \Theta, \Theta)e^{-\mc S[\bar{\Psi},\Psi,\bar{\Theta},\Theta]},
\end{equation}
with the transformed action
\begin{eqnarray}
    \mc S[\bar{\Psi},\Psi,\bar{\Theta},\Theta]&=&-\sum_{\omega_n,\mathbf p}\Big\{\bar\Psi_{n\mathbf p}\mathbf G^{-1}(\mathbf p,i\omega_n)\Psi_{n\mathbf p}\nonumber\\
    &&- \bar\Theta_{n\mathbf p}(-i\omega_n\mathbb 1 + \mc H^{\text{f}}_{\mathbf p}  )\Theta_{n\mathbf p}\Big\},
\end{eqnarray}
where $\mathbf G^{-1}(\mathbf p,i\omega_n)$ is defined in the main text (see Eq. (\ref{Green_CDW_PG})). Since the $\Theta$-part of the above action is free-like, we can directly perform its path integration to obtain $\mc Z_{\Theta}$. Thus
\begin{equation}
    \mc Z = \mc Z_{\Theta}\int \mc D(\bar \Psi,\Psi)e^{-\mc S_{\text{eff}}[\bar{\Psi},\Psi]},
\end{equation}
where  $\mc S_{\text{eff}}[\bar{\Psi},\Psi]$ is the effective action of $c$ fermions given in Eq. (\ref{coe_action}).

\section{Doping-dependent CDW order}
\label{apen_CDW_domo}

We show here how a doping dependence of the CDW phase influences the results on the energy scales displayed in Fig. \ref{doping_energy}\textbf{(b)}.

\begin{figure}[h!]
    \centering
    \includegraphics[scale=0.45]{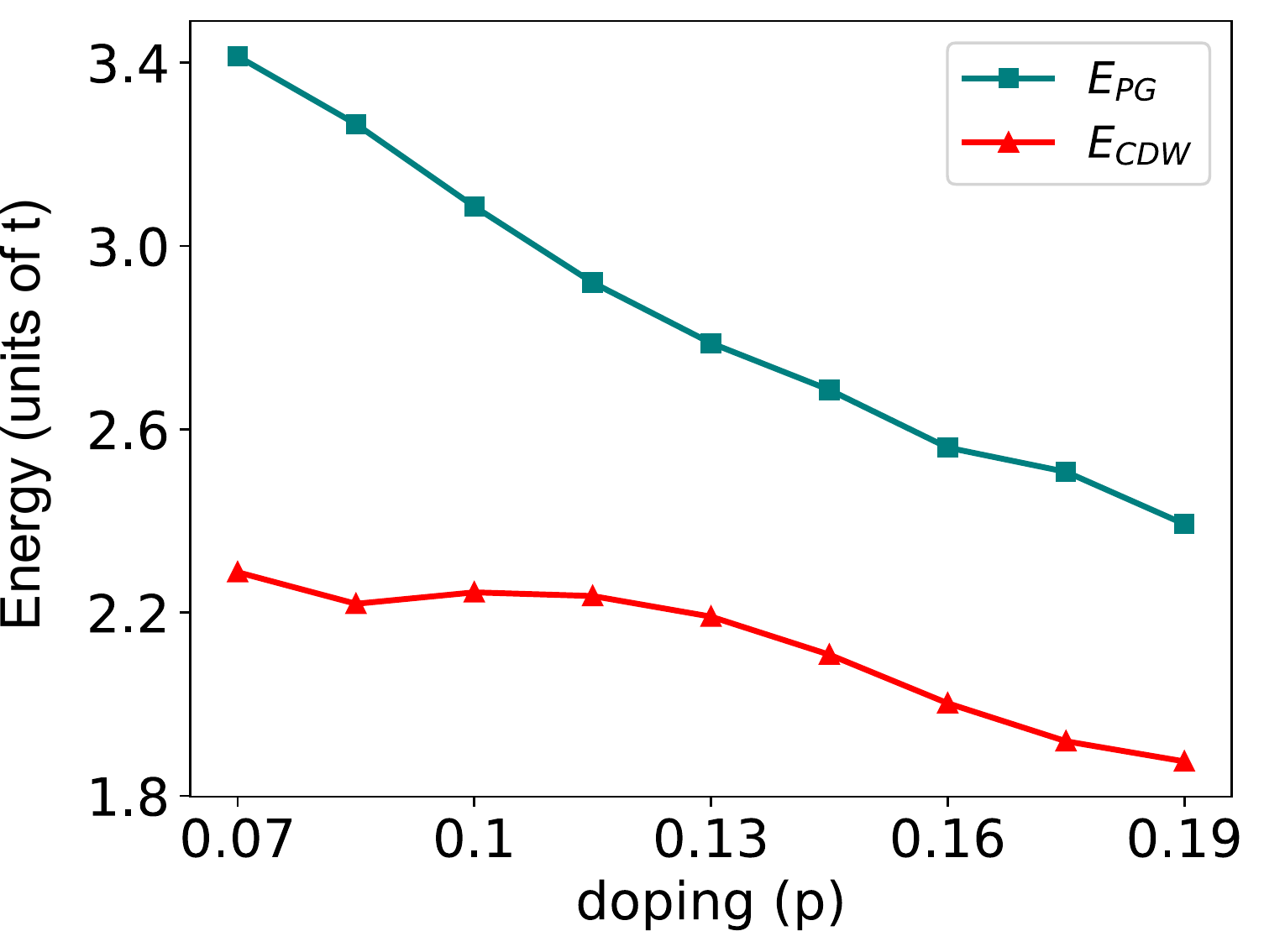}
    \caption{$E_{CDW}(p)$ and $E_{PG}(p)$ energy scales extracted from the $B_{2g}$ and $B_{1g}$ Raman responses, respectively, and considering that the CDW phase has an explicit doping dependence given by Eq. (\ref{J_doping}). }
    \label{omega_doping_domo}
\end{figure}

Within a general meanfield framework, we assume that $J(p)$ follows the dome-like doping dependency of $T_{CDW}(p)$ [see Fig. \ref{diagram}\textbf{(a)}]:
\begin{equation}
    J(p) = 0.4t\sqrt{1-\left(\frac{p-0.13}{0.04}\right)^2},
    \label{J_doping}
\end{equation}
where $J^{\te{max}}=J(p=0.13)=0.4t$ and $J(p)$ goes to zero at $p=0.13 \pm 0.04$. 
We perform again the Raman response calculations for different doping levels and extract the energy scales. The results are displayed in Fig. \ref{omega_doping_domo}. We can see that the doping dependency of the $J$ changes
\textit{both} energy scales, as a consequence of the implicit interaction between the PG and CDW phases in Eq. (\ref{Green_CDW_PG}). This shows that, though the CDW order is sub-dominant with respect to the PG, its effects could be a priori visible on the energy scales.

\bibliography{references.bib}

\begin{thebibliography}{63}%
\makeatletter
\providecommand \@ifxundefined [1]{%
 \@ifx{#1\undefined}
}%
\providecommand \@ifnum [1]{%
 \ifnum #1\expandafter \@firstoftwo
 \else \expandafter \@secondoftwo
 \fi
}%
\providecommand \@ifx [1]{%
 \ifx #1\expandafter \@firstoftwo
 \else \expandafter \@secondoftwo
 \fi
}%
\providecommand \natexlab [1]{#1}%
\providecommand \enquote  [1]{``#1''}%
\providecommand \bibnamefont  [1]{#1}%
\providecommand \bibfnamefont [1]{#1}%
\providecommand \citenamefont [1]{#1}%
\providecommand \href@noop [0]{\@secondoftwo}%
\providecommand \href [0]{\begingroup \@sanitize@url \@href}%
\providecommand \@href[1]{\@@startlink{#1}\@@href}%
\providecommand \@@href[1]{\endgroup#1\@@endlink}%
\providecommand \@sanitize@url [0]{\catcode `\\12\catcode `\$12\catcode `\&12\catcode `\#12\catcode `\^12\catcode `\_12\catcode `\%12\relax}%
\providecommand \@@startlink[1]{}%
\providecommand \@@endlink[0]{}%
\providecommand \url  [0]{\begingroup\@sanitize@url \@url }%
\providecommand \@url [1]{\endgroup\@href {#1}{\urlprefix }}%
\providecommand \urlprefix  [0]{URL }%
\providecommand \Eprint [0]{\href }%
\providecommand \doibase [0]{http://dx.doi.org/}%
\providecommand \selectlanguage [0]{\@gobble}%
\providecommand \bibinfo  [0]{\@secondoftwo}%
\providecommand \bibfield  [0]{\@secondoftwo}%
\providecommand \translation [1]{[#1]}%
\providecommand \BibitemOpen [0]{}%
\providecommand \bibitemStop [0]{}%
\providecommand \bibitemNoStop [0]{.\EOS\space}%
\providecommand \EOS [0]{\spacefactor3000\relax}%
\providecommand \BibitemShut  [1]{\csname bibitem#1\endcsname}%
\let\auto@bib@innerbib\@empty
\bibitem [{\citenamefont {Damascelli}\ \emph {et~al.}(2003)\citenamefont {Damascelli}, \citenamefont {Hussain},\ and\ \citenamefont {Shen}}]{RevModPhys.75.473}%
  \BibitemOpen
  \bibfield  {author} {\bibinfo {author} {\bibfnamefont {A.}~\bibnamefont {Damascelli}}, \bibinfo {author} {\bibfnamefont {Z.}~\bibnamefont {Hussain}}, \ and\ \bibinfo {author} {\bibfnamefont {Z.-X.}\ \bibnamefont {Shen}},\ }\href {\doibase 10.1103/RevModPhys.75.473} {\bibfield  {journal} {\bibinfo  {journal} {Rev. Mod. Phys.}\ }\textbf {\bibinfo {volume} {75}},\ \bibinfo {pages} {473} (\bibinfo {year} {2003})}\BibitemShut {NoStop}%
\bibitem [{\citenamefont {Norman}\ \emph {et~al.}(2005)\citenamefont {Norman}, \citenamefont {Pines},\ and\ \citenamefont {Kallin}}]{Norman2007}%
  \BibitemOpen
  \bibfield  {author} {\bibinfo {author} {\bibfnamefont {M.~R.}\ \bibnamefont {Norman}}, \bibinfo {author} {\bibfnamefont {D.}~\bibnamefont {Pines}}, \ and\ \bibinfo {author} {\bibfnamefont {C.}~\bibnamefont {Kallin}},\ }\href {\doibase 10.1080/00018730500459906} {\bibfield  {journal} {\bibinfo  {journal} {Advances in Physics}\ }\textbf {\bibinfo {volume} {54}},\ \bibinfo {pages} {715} (\bibinfo {year} {2005})},\ \Eprint {http://arxiv.org/abs/https://doi.org/10.1080/00018730500459906} {https://doi.org/10.1080/00018730500459906} \BibitemShut {NoStop}%
\bibitem [{\citenamefont {Warren}\ \emph {et~al.}(1989)\citenamefont {Warren}, \citenamefont {Walstedt}, \citenamefont {Brennert}, \citenamefont {Cava}, \citenamefont {Tycko}, \citenamefont {Bell},\ and\ \citenamefont {Dabbagh}}]{PhysRevLett.62.1193}%
  \BibitemOpen
  \bibfield  {author} {\bibinfo {author} {\bibfnamefont {W.~W.}\ \bibnamefont {Warren}}, \bibinfo {author} {\bibfnamefont {R.~E.}\ \bibnamefont {Walstedt}}, \bibinfo {author} {\bibfnamefont {G.~F.}\ \bibnamefont {Brennert}}, \bibinfo {author} {\bibfnamefont {R.~J.}\ \bibnamefont {Cava}}, \bibinfo {author} {\bibfnamefont {R.}~\bibnamefont {Tycko}}, \bibinfo {author} {\bibfnamefont {R.~F.}\ \bibnamefont {Bell}}, \ and\ \bibinfo {author} {\bibfnamefont {G.}~\bibnamefont {Dabbagh}},\ }\href {\doibase 10.1103/PhysRevLett.62.1193} {\bibfield  {journal} {\bibinfo  {journal} {Phys. Rev. Lett.}\ }\textbf {\bibinfo {volume} {62}},\ \bibinfo {pages} {1193} (\bibinfo {year} {1989})}\BibitemShut {NoStop}%
\bibitem [{\citenamefont {Alloul}\ \emph {et~al.}(1989)\citenamefont {Alloul}, \citenamefont {Ohno},\ and\ \citenamefont {Mendels}}]{PhysRevLett.63.1700}%
  \BibitemOpen
  \bibfield  {author} {\bibinfo {author} {\bibfnamefont {H.}~\bibnamefont {Alloul}}, \bibinfo {author} {\bibfnamefont {T.}~\bibnamefont {Ohno}}, \ and\ \bibinfo {author} {\bibfnamefont {P.}~\bibnamefont {Mendels}},\ }\href {\doibase 10.1103/PhysRevLett.63.1700} {\bibfield  {journal} {\bibinfo  {journal} {Phys. Rev. Lett.}\ }\textbf {\bibinfo {volume} {63}},\ \bibinfo {pages} {1700} (\bibinfo {year} {1989})}\BibitemShut {NoStop}%
\bibitem [{\citenamefont {Shen}\ \emph {et~al.}(2005)\citenamefont {Shen}, \citenamefont {Ronning}, \citenamefont {Lu}, \citenamefont {Baumberger}, \citenamefont {Ingle}, \citenamefont {Lee}, \citenamefont {Meevasana}, \citenamefont {Kohsaka}, \citenamefont {Azuma}, \citenamefont {Takano}, \citenamefont {Takagi},\ and\ \citenamefont {Shen}}]{Arpes_FS}%
  \BibitemOpen
  \bibfield  {author} {\bibinfo {author} {\bibfnamefont {K.~M.}\ \bibnamefont {Shen}}, \bibinfo {author} {\bibfnamefont {F.}~\bibnamefont {Ronning}}, \bibinfo {author} {\bibfnamefont {D.~H.}\ \bibnamefont {Lu}}, \bibinfo {author} {\bibfnamefont {F.}~\bibnamefont {Baumberger}}, \bibinfo {author} {\bibfnamefont {N.~J.~C.}\ \bibnamefont {Ingle}}, \bibinfo {author} {\bibfnamefont {W.~S.}\ \bibnamefont {Lee}}, \bibinfo {author} {\bibfnamefont {W.}~\bibnamefont {Meevasana}}, \bibinfo {author} {\bibfnamefont {Y.}~\bibnamefont {Kohsaka}}, \bibinfo {author} {\bibfnamefont {M.}~\bibnamefont {Azuma}}, \bibinfo {author} {\bibfnamefont {M.}~\bibnamefont {Takano}}, \bibinfo {author} {\bibfnamefont {H.}~\bibnamefont {Takagi}}, \ and\ \bibinfo {author} {\bibfnamefont {Z.-X.}\ \bibnamefont {Shen}},\ }\href {\doibase 10.1126/science.1103627} {\bibfield  {journal} {\bibinfo  {journal} {Science}\ }\textbf {\bibinfo {volume} {307}},\ \bibinfo {pages} {901} (\bibinfo {year} {2005})},\ \Eprint
  {http://arxiv.org/abs/https://www.science.org/doi/pdf/10.1126/science.1103627} {https://www.science.org/doi/pdf/10.1126/science.1103627} \BibitemShut {NoStop}%
\bibitem [{\citenamefont {Fournier}\ \emph {et~al.}(2010)\citenamefont {Fournier}, \citenamefont {Levy}, \citenamefont {Pennec}, \citenamefont {McChesney}, \citenamefont {Bostwick}, \citenamefont {Rotenberg}, \citenamefont {Liang}, \citenamefont {Hardy}, \citenamefont {Bonn}, \citenamefont {Elfimov},\ and\ \citenamefont {Damascelli}}]{Arpes_FS_2}%
  \BibitemOpen
  \bibfield  {author} {\bibinfo {author} {\bibfnamefont {D.}~\bibnamefont {Fournier}}, \bibinfo {author} {\bibfnamefont {G.}~\bibnamefont {Levy}}, \bibinfo {author} {\bibfnamefont {Y.}~\bibnamefont {Pennec}}, \bibinfo {author} {\bibfnamefont {J.~L.}\ \bibnamefont {McChesney}}, \bibinfo {author} {\bibfnamefont {A.}~\bibnamefont {Bostwick}}, \bibinfo {author} {\bibfnamefont {E.}~\bibnamefont {Rotenberg}}, \bibinfo {author} {\bibfnamefont {R.}~\bibnamefont {Liang}}, \bibinfo {author} {\bibfnamefont {W.~N.}\ \bibnamefont {Hardy}}, \bibinfo {author} {\bibfnamefont {D.~A.}\ \bibnamefont {Bonn}}, \bibinfo {author} {\bibfnamefont {I.~S.}\ \bibnamefont {Elfimov}}, \ and\ \bibinfo {author} {\bibfnamefont {A.}~\bibnamefont {Damascelli}},\ }\href {\doibase 10.1038/nphys1763} {\bibfield  {journal} {\bibinfo  {journal} {Nature Physics}\ }\textbf {\bibinfo {volume} {6}},\ \bibinfo {pages} {905} (\bibinfo {year} {2010})}\BibitemShut {NoStop}%
\bibitem [{\citenamefont {Comin}\ and\ \citenamefont {Damascelli}(2016)}]{Comin}%
  \BibitemOpen
  \bibfield  {author} {\bibinfo {author} {\bibfnamefont {R.}~\bibnamefont {Comin}}\ and\ \bibinfo {author} {\bibfnamefont {A.}~\bibnamefont {Damascelli}},\ }\href {\doibase 10.1146/annurev-conmatphys-031115-011401} {\bibfield  {journal} {\bibinfo  {journal} {Annual Review of Condensed Matter Physics}\ }\textbf {\bibinfo {volume} {7}},\ \bibinfo {pages} {369} (\bibinfo {year} {2016})},\ \Eprint {http://arxiv.org/abs/https://doi.org/10.1146/annurev-conmatphys-031115-011401} {https://doi.org/10.1146/annurev-conmatphys-031115-011401} \BibitemShut {NoStop}%
\bibitem [{\citenamefont {Wu}\ \emph {et~al.}(2011)\citenamefont {Wu}, \citenamefont {Mayaffre}, \citenamefont {Kr{\"a}mer}, \citenamefont {Horvati{\'{c}}}, \citenamefont {Berthier}, \citenamefont {Hardy}, \citenamefont {Liang}, \citenamefont {Bonn},\ and\ \citenamefont {Julien}}]{Wu2011}%
  \BibitemOpen
  \bibfield  {author} {\bibinfo {author} {\bibfnamefont {T.}~\bibnamefont {Wu}}, \bibinfo {author} {\bibfnamefont {H.}~\bibnamefont {Mayaffre}}, \bibinfo {author} {\bibfnamefont {S.}~\bibnamefont {Kr{\"a}mer}}, \bibinfo {author} {\bibfnamefont {M.}~\bibnamefont {Horvati{\'{c}}}}, \bibinfo {author} {\bibfnamefont {C.}~\bibnamefont {Berthier}}, \bibinfo {author} {\bibfnamefont {W.~N.}\ \bibnamefont {Hardy}}, \bibinfo {author} {\bibfnamefont {R.}~\bibnamefont {Liang}}, \bibinfo {author} {\bibfnamefont {D.~A.}\ \bibnamefont {Bonn}}, \ and\ \bibinfo {author} {\bibfnamefont {M.-H.}\ \bibnamefont {Julien}},\ }\href {\doibase 10.1038/nature10345} {\bibfield  {journal} {\bibinfo  {journal} {Nature}\ }\textbf {\bibinfo {volume} {477}},\ \bibinfo {pages} {191} (\bibinfo {year} {2011})}\BibitemShut {NoStop}%
\bibitem [{\citenamefont {LeBoeuf}\ \emph {et~al.}(2012)\citenamefont {LeBoeuf}, \citenamefont {Kr\"amer}, \citenamefont {Hardy}, \citenamefont {Liang}, \citenamefont {Bonn},\ and\ \citenamefont {Proust}}]{LeBoeuf_2012}%
  \BibitemOpen
  \bibfield  {author} {\bibinfo {author} {\bibfnamefont {D.}~\bibnamefont {LeBoeuf}}, \bibinfo {author} {\bibfnamefont {S.}~\bibnamefont {Kr\"amer}}, \bibinfo {author} {\bibfnamefont {W.~N.}\ \bibnamefont {Hardy}}, \bibinfo {author} {\bibfnamefont {R.}~\bibnamefont {Liang}}, \bibinfo {author} {\bibfnamefont {D.~A.}\ \bibnamefont {Bonn}}, \ and\ \bibinfo {author} {\bibfnamefont {C.}~\bibnamefont {Proust}},\ }\href {\doibase 10.1038/nphys2502} {\bibfield  {journal} {\bibinfo  {journal} {Nature Physics}\ }\textbf {\bibinfo {volume} {9}},\ \bibinfo {pages} {79} (\bibinfo {year} {2012})}\BibitemShut {NoStop}%
\bibitem [{\citenamefont {Ghiringhelli}\ \emph {et~al.}(2012)\citenamefont {Ghiringhelli}, \citenamefont {Tacon}, \citenamefont {Minola}, \citenamefont {Blanco-Canosa}, \citenamefont {Mazzoli}, \citenamefont {Brookes}, \citenamefont {Luca}, \citenamefont {Frano}, \citenamefont {Hawthorn}, \citenamefont {He}, \citenamefont {Loew}, \citenamefont {Sala}, \citenamefont {Peets}, \citenamefont {Salluzzo}, \citenamefont {Schierle}, \citenamefont {Sutarto}, \citenamefont {Sawatzky}, \citenamefont {Weschke}, \citenamefont {Keimer},\ and\ \citenamefont {Braicovich}}]{Ghiringhelli_2012}%
  \BibitemOpen
  \bibfield  {author} {\bibinfo {author} {\bibfnamefont {G.}~\bibnamefont {Ghiringhelli}}, \bibinfo {author} {\bibfnamefont {M.~L.}\ \bibnamefont {Tacon}}, \bibinfo {author} {\bibfnamefont {M.}~\bibnamefont {Minola}}, \bibinfo {author} {\bibfnamefont {S.}~\bibnamefont {Blanco-Canosa}}, \bibinfo {author} {\bibfnamefont {C.}~\bibnamefont {Mazzoli}}, \bibinfo {author} {\bibfnamefont {N.~B.}\ \bibnamefont {Brookes}}, \bibinfo {author} {\bibfnamefont {G.~M.~D.}\ \bibnamefont {Luca}}, \bibinfo {author} {\bibfnamefont {A.}~\bibnamefont {Frano}}, \bibinfo {author} {\bibfnamefont {D.~G.}\ \bibnamefont {Hawthorn}}, \bibinfo {author} {\bibfnamefont {F.}~\bibnamefont {He}}, \bibinfo {author} {\bibfnamefont {T.}~\bibnamefont {Loew}}, \bibinfo {author} {\bibfnamefont {M.~M.}\ \bibnamefont {Sala}}, \bibinfo {author} {\bibfnamefont {D.~C.}\ \bibnamefont {Peets}}, \bibinfo {author} {\bibfnamefont {M.}~\bibnamefont {Salluzzo}}, \bibinfo {author} {\bibfnamefont {E.}~\bibnamefont {Schierle}}, \bibinfo {author} {\bibfnamefont
  {R.}~\bibnamefont {Sutarto}}, \bibinfo {author} {\bibfnamefont {G.~A.}\ \bibnamefont {Sawatzky}}, \bibinfo {author} {\bibfnamefont {E.}~\bibnamefont {Weschke}}, \bibinfo {author} {\bibfnamefont {B.}~\bibnamefont {Keimer}}, \ and\ \bibinfo {author} {\bibfnamefont {L.}~\bibnamefont {Braicovich}},\ }\href {\doibase 10.1126/science.1223532} {\bibfield  {journal} {\bibinfo  {journal} {Science}\ }\textbf {\bibinfo {volume} {337}},\ \bibinfo {pages} {821} (\bibinfo {year} {2012})}\BibitemShut {NoStop}%
\bibitem [{\citenamefont {Wu}\ \emph {et~al.}(2013)\citenamefont {Wu}, \citenamefont {Mayaffre}, \citenamefont {Kr{\"a}mer}, \citenamefont {Horvati{\'{c}}}, \citenamefont {Berthier}, \citenamefont {Kuhns}, \citenamefont {Reyes}, \citenamefont {Liang}, \citenamefont {Hardy}, \citenamefont {Bonn},\ and\ \citenamefont {Julien}}]{Wu2013}%
  \BibitemOpen
  \bibfield  {author} {\bibinfo {author} {\bibfnamefont {T.}~\bibnamefont {Wu}}, \bibinfo {author} {\bibfnamefont {H.}~\bibnamefont {Mayaffre}}, \bibinfo {author} {\bibfnamefont {S.}~\bibnamefont {Kr{\"a}mer}}, \bibinfo {author} {\bibfnamefont {M.}~\bibnamefont {Horvati{\'{c}}}}, \bibinfo {author} {\bibfnamefont {C.}~\bibnamefont {Berthier}}, \bibinfo {author} {\bibfnamefont {P.~L.}\ \bibnamefont {Kuhns}}, \bibinfo {author} {\bibfnamefont {A.~P.}\ \bibnamefont {Reyes}}, \bibinfo {author} {\bibfnamefont {R.}~\bibnamefont {Liang}}, \bibinfo {author} {\bibfnamefont {W.~N.}\ \bibnamefont {Hardy}}, \bibinfo {author} {\bibfnamefont {D.~A.}\ \bibnamefont {Bonn}}, \ and\ \bibinfo {author} {\bibfnamefont {M.-H.}\ \bibnamefont {Julien}},\ }\href {\doibase 10.1038/ncomms3113} {\bibfield  {journal} {\bibinfo  {journal} {Nature Communications}\ }\textbf {\bibinfo {volume} {4}},\ \bibinfo {pages} {2113} (\bibinfo {year} {2013})}\BibitemShut {NoStop}%
\bibitem [{\citenamefont {H\"ucker}\ \emph {et~al.}(2014)\citenamefont {H\"ucker}, \citenamefont {Christensen}, \citenamefont {Holmes}, \citenamefont {Blackburn}, \citenamefont {Forgan}, \citenamefont {Liang}, \citenamefont {Bonn}, \citenamefont {Hardy}, \citenamefont {Gutowski}, \citenamefont {v.~Zimmermann}, \citenamefont {Hayden},\ and\ \citenamefont {Chang}}]{Huecker_2014}%
  \BibitemOpen
  \bibfield  {author} {\bibinfo {author} {\bibfnamefont {M.}~\bibnamefont {H\"ucker}}, \bibinfo {author} {\bibfnamefont {N.~B.}\ \bibnamefont {Christensen}}, \bibinfo {author} {\bibfnamefont {A.~T.}\ \bibnamefont {Holmes}}, \bibinfo {author} {\bibfnamefont {E.}~\bibnamefont {Blackburn}}, \bibinfo {author} {\bibfnamefont {E.~M.}\ \bibnamefont {Forgan}}, \bibinfo {author} {\bibfnamefont {R.}~\bibnamefont {Liang}}, \bibinfo {author} {\bibfnamefont {D.~A.}\ \bibnamefont {Bonn}}, \bibinfo {author} {\bibfnamefont {W.~N.}\ \bibnamefont {Hardy}}, \bibinfo {author} {\bibfnamefont {O.}~\bibnamefont {Gutowski}}, \bibinfo {author} {\bibfnamefont {M.}~\bibnamefont {v.~Zimmermann}}, \bibinfo {author} {\bibfnamefont {S.~M.}\ \bibnamefont {Hayden}}, \ and\ \bibinfo {author} {\bibfnamefont {J.}~\bibnamefont {Chang}},\ }\href {\doibase 10.1103/physrevb.90.054514} {\bibfield  {journal} {\bibinfo  {journal} {Physical Review B}\ }\textbf {\bibinfo {volume} {90}} (\bibinfo {year} {2014}),\ 10.1103/physrevb.90.054514}\BibitemShut
  {NoStop}%
\bibitem [{\citenamefont {Wu}\ \emph {et~al.}(2015)\citenamefont {Wu}, \citenamefont {Mayaffre}, \citenamefont {Kr{\"a}mer}, \citenamefont {Horvati{\'{c}}}, \citenamefont {Berthier}, \citenamefont {Hardy}, \citenamefont {Liang}, \citenamefont {Bonn},\ and\ \citenamefont {Julien}}]{Wu2015}%
  \BibitemOpen
  \bibfield  {author} {\bibinfo {author} {\bibfnamefont {T.}~\bibnamefont {Wu}}, \bibinfo {author} {\bibfnamefont {H.}~\bibnamefont {Mayaffre}}, \bibinfo {author} {\bibfnamefont {S.}~\bibnamefont {Kr{\"a}mer}}, \bibinfo {author} {\bibfnamefont {M.}~\bibnamefont {Horvati{\'{c}}}}, \bibinfo {author} {\bibfnamefont {C.}~\bibnamefont {Berthier}}, \bibinfo {author} {\bibfnamefont {W.~N.}\ \bibnamefont {Hardy}}, \bibinfo {author} {\bibfnamefont {R.}~\bibnamefont {Liang}}, \bibinfo {author} {\bibfnamefont {D.~A.}\ \bibnamefont {Bonn}}, \ and\ \bibinfo {author} {\bibfnamefont {M.-H.}\ \bibnamefont {Julien}},\ }\href {\doibase 10.1038/ncomms7438} {\bibfield  {journal} {\bibinfo  {journal} {Nature Communications}\ }\textbf {\bibinfo {volume} {6}},\ \bibinfo {pages} {6438} (\bibinfo {year} {2015})}\BibitemShut {NoStop}%
\bibitem [{\citenamefont {Cyr-Choini{\`{e}}re}\ \emph {et~al.}(2018)\citenamefont {Cyr-Choini{\`{e}}re}, \citenamefont {LeBoeuf}, \citenamefont {Badoux}, \citenamefont {Dufour-Beaus{\'{e}}jour}, \citenamefont {Bonn}, \citenamefont {Hardy}, \citenamefont {Liang}, \citenamefont {Graf}, \citenamefont {Doiron-Leyraud},\ and\ \citenamefont {Taillefer}}]{Cyr_Choiniere_2018}%
  \BibitemOpen
  \bibfield  {author} {\bibinfo {author} {\bibfnamefont {O.}~\bibnamefont {Cyr-Choini{\`{e}}re}}, \bibinfo {author} {\bibfnamefont {D.}~\bibnamefont {LeBoeuf}}, \bibinfo {author} {\bibfnamefont {S.}~\bibnamefont {Badoux}}, \bibinfo {author} {\bibfnamefont {S.}~\bibnamefont {Dufour-Beaus{\'{e}}jour}}, \bibinfo {author} {\bibfnamefont {D.~A.}\ \bibnamefont {Bonn}}, \bibinfo {author} {\bibfnamefont {W.~N.}\ \bibnamefont {Hardy}}, \bibinfo {author} {\bibfnamefont {R.}~\bibnamefont {Liang}}, \bibinfo {author} {\bibfnamefont {D.}~\bibnamefont {Graf}}, \bibinfo {author} {\bibfnamefont {N.}~\bibnamefont {Doiron-Leyraud}}, \ and\ \bibinfo {author} {\bibfnamefont {L.}~\bibnamefont {Taillefer}},\ }\href {\doibase 10.1103/physrevb.98.064513} {\bibfield  {journal} {\bibinfo  {journal} {Physical Review B}\ }\textbf {\bibinfo {volume} {98}} (\bibinfo {year} {2018}),\ 10.1103/physrevb.98.064513}\BibitemShut {NoStop}%
\bibitem [{\citenamefont {Ka\ifmmode \check{c}\else \v{c}\fi{}mar\ifmmode~\check{c}\else \v{c}\fi{}\'{\i}k}\ \emph {et~al.}(2018)\citenamefont {Ka\ifmmode \check{c}\else \v{c}\fi{}mar\ifmmode~\check{c}\else \v{c}\fi{}\'{\i}k}, \citenamefont {Vinograd}, \citenamefont {Michon}, \citenamefont {Rydh}, \citenamefont {Demuer}, \citenamefont {Zhou}, \citenamefont {Mayaffre}, \citenamefont {Liang}, \citenamefont {Hardy}, \citenamefont {Bonn}, \citenamefont {Doiron-Leyraud}, \citenamefont {Taillefer}, \citenamefont {Julien}, \citenamefont {Marcenat},\ and\ \citenamefont {Klein}}]{PhysRevLett.121.167002}%
  \BibitemOpen
  \bibfield  {author} {\bibinfo {author} {\bibfnamefont {J.}~\bibnamefont {Ka\ifmmode \check{c}\else \v{c}\fi{}mar\ifmmode~\check{c}\else \v{c}\fi{}\'{\i}k}}, \bibinfo {author} {\bibfnamefont {I.}~\bibnamefont {Vinograd}}, \bibinfo {author} {\bibfnamefont {B.}~\bibnamefont {Michon}}, \bibinfo {author} {\bibfnamefont {A.}~\bibnamefont {Rydh}}, \bibinfo {author} {\bibfnamefont {A.}~\bibnamefont {Demuer}}, \bibinfo {author} {\bibfnamefont {R.}~\bibnamefont {Zhou}}, \bibinfo {author} {\bibfnamefont {H.}~\bibnamefont {Mayaffre}}, \bibinfo {author} {\bibfnamefont {R.}~\bibnamefont {Liang}}, \bibinfo {author} {\bibfnamefont {W.~N.}\ \bibnamefont {Hardy}}, \bibinfo {author} {\bibfnamefont {D.~A.}\ \bibnamefont {Bonn}}, \bibinfo {author} {\bibfnamefont {N.}~\bibnamefont {Doiron-Leyraud}}, \bibinfo {author} {\bibfnamefont {L.}~\bibnamefont {Taillefer}}, \bibinfo {author} {\bibfnamefont {M.-H.}\ \bibnamefont {Julien}}, \bibinfo {author} {\bibfnamefont {C.}~\bibnamefont {Marcenat}}, \ and\ \bibinfo {author}
  {\bibfnamefont {T.}~\bibnamefont {Klein}},\ }\href {\doibase 10.1103/PhysRevLett.121.167002} {\bibfield  {journal} {\bibinfo  {journal} {Phys. Rev. Lett.}\ }\textbf {\bibinfo {volume} {121}},\ \bibinfo {pages} {167002} (\bibinfo {year} {2018})}\BibitemShut {NoStop}%
\bibitem [{\citenamefont {Vinograd}\ \emph {et~al.}(2019)\citenamefont {Vinograd}, \citenamefont {Zhou}, \citenamefont {Mayaffre}, \citenamefont {Kr\"amer}, \citenamefont {Liang}, \citenamefont {Hardy}, \citenamefont {Bonn},\ and\ \citenamefont {Julien}}]{PhysRevB.100.094502}%
  \BibitemOpen
  \bibfield  {author} {\bibinfo {author} {\bibfnamefont {I.}~\bibnamefont {Vinograd}}, \bibinfo {author} {\bibfnamefont {R.}~\bibnamefont {Zhou}}, \bibinfo {author} {\bibfnamefont {H.}~\bibnamefont {Mayaffre}}, \bibinfo {author} {\bibfnamefont {S.}~\bibnamefont {Kr\"amer}}, \bibinfo {author} {\bibfnamefont {R.}~\bibnamefont {Liang}}, \bibinfo {author} {\bibfnamefont {W.~N.}\ \bibnamefont {Hardy}}, \bibinfo {author} {\bibfnamefont {D.~A.}\ \bibnamefont {Bonn}}, \ and\ \bibinfo {author} {\bibfnamefont {M.-H.}\ \bibnamefont {Julien}},\ }\href {\doibase 10.1103/PhysRevB.100.094502} {\bibfield  {journal} {\bibinfo  {journal} {Phys. Rev. B}\ }\textbf {\bibinfo {volume} {100}},\ \bibinfo {pages} {094502} (\bibinfo {year} {2019})}\BibitemShut {NoStop}%
\bibitem [{\citenamefont {Dash}\ and\ \citenamefont {S\'en\'echal}(2021)}]{Senechal}%
  \BibitemOpen
  \bibfield  {author} {\bibinfo {author} {\bibfnamefont {S.~S.}\ \bibnamefont {Dash}}\ and\ \bibinfo {author} {\bibfnamefont {D.}~\bibnamefont {S\'en\'echal}},\ }\href {\doibase 10.1103/PhysRevB.103.045142} {\bibfield  {journal} {\bibinfo  {journal} {Phys. Rev. B}\ }\textbf {\bibinfo {volume} {103}},\ \bibinfo {pages} {045142} (\bibinfo {year} {2021})}\BibitemShut {NoStop}%
\bibitem [{\citenamefont {Loret}\ \emph {et~al.}(2019)\citenamefont {Loret}, \citenamefont {Auvray}, \citenamefont {Gallais}, \citenamefont {Cazayous}, \citenamefont {Forget}, \citenamefont {Colson}, \citenamefont {Julien}, \citenamefont {Paul}, \citenamefont {Civelli},\ and\ \citenamefont {Sacuto}}]{Alain}%
  \BibitemOpen
  \bibfield  {author} {\bibinfo {author} {\bibfnamefont {B.}~\bibnamefont {Loret}}, \bibinfo {author} {\bibfnamefont {N.}~\bibnamefont {Auvray}}, \bibinfo {author} {\bibfnamefont {Y.}~\bibnamefont {Gallais}}, \bibinfo {author} {\bibfnamefont {M.}~\bibnamefont {Cazayous}}, \bibinfo {author} {\bibfnamefont {A.}~\bibnamefont {Forget}}, \bibinfo {author} {\bibfnamefont {D.}~\bibnamefont {Colson}}, \bibinfo {author} {\bibfnamefont {M.-H.}\ \bibnamefont {Julien}}, \bibinfo {author} {\bibfnamefont {I.}~\bibnamefont {Paul}}, \bibinfo {author} {\bibfnamefont {M.}~\bibnamefont {Civelli}}, \ and\ \bibinfo {author} {\bibfnamefont {A.}~\bibnamefont {Sacuto}},\ }\href {\doibase 10.1038/s41567-019-0509-5} {\bibfield  {journal} {\bibinfo  {journal} {Nature Physics}\ }\textbf {\bibinfo {volume} {15}},\ \bibinfo {pages} {771} (\bibinfo {year} {2019})}\BibitemShut {NoStop}%
\bibitem [{\citenamefont {Efetov}\ \emph {et~al.}(2013)\citenamefont {Efetov}, \citenamefont {Meier},\ and\ \citenamefont {P{\'{e}}pin}}]{efetov}%
  \BibitemOpen
  \bibfield  {author} {\bibinfo {author} {\bibfnamefont {K.~B.}\ \bibnamefont {Efetov}}, \bibinfo {author} {\bibfnamefont {H.}~\bibnamefont {Meier}}, \ and\ \bibinfo {author} {\bibfnamefont {C.}~\bibnamefont {P{\'{e}}pin}},\ }\href {\doibase 10.1038/nphys2641} {\bibfield  {journal} {\bibinfo  {journal} {Nature Physics}\ }\textbf {\bibinfo {volume} {9}},\ \bibinfo {pages} {442} (\bibinfo {year} {2013})}\BibitemShut {NoStop}%
\bibitem [{\citenamefont {Fradkin}\ \emph {et~al.}(2015)\citenamefont {Fradkin}, \citenamefont {Kivelson},\ and\ \citenamefont {Tranquada}}]{Fradkin_2015}%
  \BibitemOpen
  \bibfield  {author} {\bibinfo {author} {\bibfnamefont {E.}~\bibnamefont {Fradkin}}, \bibinfo {author} {\bibfnamefont {S.~A.}\ \bibnamefont {Kivelson}}, \ and\ \bibinfo {author} {\bibfnamefont {J.~M.}\ \bibnamefont {Tranquada}},\ }\href {\doibase 10.1103/revmodphys.87.457} {\bibfield  {journal} {\bibinfo  {journal} {Reviews of Modern Physics}\ }\textbf {\bibinfo {volume} {87}},\ \bibinfo {pages} {457} (\bibinfo {year} {2015})}\BibitemShut {NoStop}%
\bibitem [{\citenamefont {Chakraborty}\ \emph {et~al.}(2019)\citenamefont {Chakraborty}, \citenamefont {Grandadam}, \citenamefont {Hamidian}, \citenamefont {Davis}, \citenamefont {Sidis},\ and\ \citenamefont {P{\'{e}}pin}}]{pepin}%
  \BibitemOpen
  \bibfield  {author} {\bibinfo {author} {\bibfnamefont {D.}~\bibnamefont {Chakraborty}}, \bibinfo {author} {\bibfnamefont {M.}~\bibnamefont {Grandadam}}, \bibinfo {author} {\bibfnamefont {M.~H.}\ \bibnamefont {Hamidian}}, \bibinfo {author} {\bibfnamefont {J.~C.~S.}\ \bibnamefont {Davis}}, \bibinfo {author} {\bibfnamefont {Y.}~\bibnamefont {Sidis}}, \ and\ \bibinfo {author} {\bibfnamefont {C.}~\bibnamefont {P{\'{e}}pin}},\ }\href {\doibase 10.1103/physrevb.100.224511} {\bibfield  {journal} {\bibinfo  {journal} {Physical Review B}\ }\textbf {\bibinfo {volume} {100}} (\bibinfo {year} {2019}),\ 10.1103/physrevb.100.224511}\BibitemShut {NoStop}%
\bibitem [{\citenamefont {Yang}\ \emph {et~al.}(2006)\citenamefont {Yang}, \citenamefont {Rice},\ and\ \citenamefont {Zhang}}]{YZR1}%
  \BibitemOpen
  \bibfield  {author} {\bibinfo {author} {\bibfnamefont {K.-Y.}\ \bibnamefont {Yang}}, \bibinfo {author} {\bibfnamefont {T.~M.}\ \bibnamefont {Rice}}, \ and\ \bibinfo {author} {\bibfnamefont {F.-C.}\ \bibnamefont {Zhang}},\ }\href {\doibase 10.1103/PhysRevB.73.174501} {\bibfield  {journal} {\bibinfo  {journal} {Phys. Rev. B}\ }\textbf {\bibinfo {volume} {73}},\ \bibinfo {pages} {174501} (\bibinfo {year} {2006})}\BibitemShut {NoStop}%
\bibitem [{\citenamefont {Rice}\ \emph {et~al.}(2011)\citenamefont {Rice}, \citenamefont {Yang},\ and\ \citenamefont {Zhang}}]{YZR2}%
  \BibitemOpen
  \bibfield  {author} {\bibinfo {author} {\bibfnamefont {T.~M.}\ \bibnamefont {Rice}}, \bibinfo {author} {\bibfnamefont {K.-Y.}\ \bibnamefont {Yang}}, \ and\ \bibinfo {author} {\bibfnamefont {F.~C.}\ \bibnamefont {Zhang}},\ }\href {\doibase 10.1088/0034-4885/75/1/016502} {\bibfield  {journal} {\bibinfo  {journal} {Reports on Progress in Physics}\ }\textbf {\bibinfo {volume} {75}},\ \bibinfo {pages} {016502} (\bibinfo {year} {2011})}\BibitemShut {NoStop}%
\bibitem [{\citenamefont {Lee}\ \emph {et~al.}(2021)\citenamefont {Lee}, \citenamefont {Lanat\`a}, \citenamefont {Kim},\ and\ \citenamefont {Kotliar}}]{slave_cite}%
  \BibitemOpen
  \bibfield  {author} {\bibinfo {author} {\bibfnamefont {T.-H.}\ \bibnamefont {Lee}}, \bibinfo {author} {\bibfnamefont {N.}~\bibnamefont {Lanat\`a}}, \bibinfo {author} {\bibfnamefont {M.}~\bibnamefont {Kim}}, \ and\ \bibinfo {author} {\bibfnamefont {G.}~\bibnamefont {Kotliar}},\ }\href {\doibase 10.1103/PhysRevX.11.041040} {\bibfield  {journal} {\bibinfo  {journal} {Phys. Rev. X}\ }\textbf {\bibinfo {volume} {11}},\ \bibinfo {pages} {041040} (\bibinfo {year} {2021})}\BibitemShut {NoStop}%
\bibitem [{\citenamefont {Kotliar}\ \emph {et~al.}(2006)\citenamefont {Kotliar}, \citenamefont {Savrasov}, \citenamefont {Haule}, \citenamefont {Oudovenko}, \citenamefont {Parcollet},\ and\ \citenamefont {Marianetti}}]{DMFT_cite}%
  \BibitemOpen
  \bibfield  {author} {\bibinfo {author} {\bibfnamefont {G.}~\bibnamefont {Kotliar}}, \bibinfo {author} {\bibfnamefont {S.~Y.}\ \bibnamefont {Savrasov}}, \bibinfo {author} {\bibfnamefont {K.}~\bibnamefont {Haule}}, \bibinfo {author} {\bibfnamefont {V.~S.}\ \bibnamefont {Oudovenko}}, \bibinfo {author} {\bibfnamefont {O.}~\bibnamefont {Parcollet}}, \ and\ \bibinfo {author} {\bibfnamefont {C.~A.}\ \bibnamefont {Marianetti}},\ }\href {\doibase 10.1103/RevModPhys.78.865} {\bibfield  {journal} {\bibinfo  {journal} {Rev. Mod. Phys.}\ }\textbf {\bibinfo {volume} {78}},\ \bibinfo {pages} {865} (\bibinfo {year} {2006})}\BibitemShut {NoStop}%
\bibitem [{\citenamefont {Sau}\ and\ \citenamefont {Sachdev}(2014)}]{Sachdev}%
  \BibitemOpen
  \bibfield  {author} {\bibinfo {author} {\bibfnamefont {J.~D.}\ \bibnamefont {Sau}}\ and\ \bibinfo {author} {\bibfnamefont {S.}~\bibnamefont {Sachdev}},\ }\href {\doibase 10.1103/PhysRevB.89.075129} {\bibfield  {journal} {\bibinfo  {journal} {Phys. Rev. B}\ }\textbf {\bibinfo {volume} {89}},\ \bibinfo {pages} {https://pt.overleaf.com/project/62d811693bc755f5fa5c3877 {075129}} (\bibinfo {year} {2014})}\BibitemShut {NoStop}%
\bibitem [{\citenamefont {He}\ \emph {et~al.}(2011)\citenamefont {He}, \citenamefont {Hashimoto}, \citenamefont {Karapetyan}, \citenamefont {Koralek}, \citenamefont {Hinton}, \citenamefont {Testaud}, \citenamefont {Nathan}, \citenamefont {Yoshida}, \citenamefont {Yao}, \citenamefont {Tanaka}, \citenamefont {Meevasana}, \citenamefont {Moore}, \citenamefont {Lu}, \citenamefont {Mo}, \citenamefont {Ishikado}, \citenamefont {Eisaki}, \citenamefont {Hussain}, \citenamefont {Devereaux}, \citenamefont {Kivelson}, \citenamefont {Orenstein}, \citenamefont {Kapitulnik},\ and\ \citenamefont {Shen}}]{He_2011}%
  \BibitemOpen
  \bibfield  {author} {\bibinfo {author} {\bibfnamefont {R.-H.}\ \bibnamefont {He}}, \bibinfo {author} {\bibfnamefont {M.}~\bibnamefont {Hashimoto}}, \bibinfo {author} {\bibfnamefont {H.}~\bibnamefont {Karapetyan}}, \bibinfo {author} {\bibfnamefont {J.~D.}\ \bibnamefont {Koralek}}, \bibinfo {author} {\bibfnamefont {J.~P.}\ \bibnamefont {Hinton}}, \bibinfo {author} {\bibfnamefont {J.~P.}\ \bibnamefont {Testaud}}, \bibinfo {author} {\bibfnamefont {V.}~\bibnamefont {Nathan}}, \bibinfo {author} {\bibfnamefont {Y.}~\bibnamefont {Yoshida}}, \bibinfo {author} {\bibfnamefont {H.}~\bibnamefont {Yao}}, \bibinfo {author} {\bibfnamefont {K.}~\bibnamefont {Tanaka}}, \bibinfo {author} {\bibfnamefont {W.}~\bibnamefont {Meevasana}}, \bibinfo {author} {\bibfnamefont {R.~G.}\ \bibnamefont {Moore}}, \bibinfo {author} {\bibfnamefont {D.~H.}\ \bibnamefont {Lu}}, \bibinfo {author} {\bibfnamefont {S.-K.}\ \bibnamefont {Mo}}, \bibinfo {author} {\bibfnamefont {M.}~\bibnamefont {Ishikado}}, \bibinfo {author} {\bibfnamefont
  {H.}~\bibnamefont {Eisaki}}, \bibinfo {author} {\bibfnamefont {Z.}~\bibnamefont {Hussain}}, \bibinfo {author} {\bibfnamefont {T.~P.}\ \bibnamefont {Devereaux}}, \bibinfo {author} {\bibfnamefont {S.~A.}\ \bibnamefont {Kivelson}}, \bibinfo {author} {\bibfnamefont {J.}~\bibnamefont {Orenstein}}, \bibinfo {author} {\bibfnamefont {A.}~\bibnamefont {Kapitulnik}}, \ and\ \bibinfo {author} {\bibfnamefont {Z.-X.}\ \bibnamefont {Shen}},\ }\href {\doibase 10.1126/science.1198415} {\bibfield  {journal} {\bibinfo  {journal} {Science}\ }\textbf {\bibinfo {volume} {331}},\ \bibinfo {pages} {1579} (\bibinfo {year} {2011})}\BibitemShut {NoStop}%
\bibitem [{\citenamefont {Fujita}\ \emph {et~al.}(2014)\citenamefont {Fujita}, \citenamefont {Hamidian}, \citenamefont {Edkins}, \citenamefont {Kim}, \citenamefont {Kohsaka}, \citenamefont {Azuma}, \citenamefont {Takano}, \citenamefont {Takagi}, \citenamefont {Eisaki}, \citenamefont {ichi Uchida}, \citenamefont {Allais}, \citenamefont {Lawler}, \citenamefont {Kim}, \citenamefont {Sachdev},\ and\ \citenamefont {Davis}}]{d_form}%
  \BibitemOpen
  \bibfield  {author} {\bibinfo {author} {\bibfnamefont {K.}~\bibnamefont {Fujita}}, \bibinfo {author} {\bibfnamefont {M.~H.}\ \bibnamefont {Hamidian}}, \bibinfo {author} {\bibfnamefont {S.~D.}\ \bibnamefont {Edkins}}, \bibinfo {author} {\bibfnamefont {C.~K.}\ \bibnamefont {Kim}}, \bibinfo {author} {\bibfnamefont {Y.}~\bibnamefont {Kohsaka}}, \bibinfo {author} {\bibfnamefont {M.}~\bibnamefont {Azuma}}, \bibinfo {author} {\bibfnamefont {M.}~\bibnamefont {Takano}}, \bibinfo {author} {\bibfnamefont {H.}~\bibnamefont {Takagi}}, \bibinfo {author} {\bibfnamefont {H.}~\bibnamefont {Eisaki}}, \bibinfo {author} {\bibfnamefont {S.}~\bibnamefont {ichi Uchida}}, \bibinfo {author} {\bibfnamefont {A.}~\bibnamefont {Allais}}, \bibinfo {author} {\bibfnamefont {M.~J.}\ \bibnamefont {Lawler}}, \bibinfo {author} {\bibfnamefont {E.-A.}\ \bibnamefont {Kim}}, \bibinfo {author} {\bibfnamefont {S.}~\bibnamefont {Sachdev}}, \ and\ \bibinfo {author} {\bibfnamefont {J.~C.~S.}\ \bibnamefont {Davis}},\ }\href {\doibase
  10.1073/pnas.1406297111} {\bibfield  {journal} {\bibinfo  {journal} {Proceedings of the National Academy of Sciences}\ }\textbf {\bibinfo {volume} {111}},\ \bibinfo {pages} {E3026} (\bibinfo {year} {2014})},\ \Eprint {http://arxiv.org/abs/https://www.pnas.org/doi/pdf/10.1073/pnas.1406297111} {https://www.pnas.org/doi/pdf/10.1073/pnas.1406297111} \BibitemShut {NoStop}%
\bibitem [{\citenamefont {Allais}\ \emph {et~al.}(2014)\citenamefont {Allais}, \citenamefont {Chowdhury},\ and\ \citenamefont {Sachdev}}]{Allais}%
  \BibitemOpen
  \bibfield  {author} {\bibinfo {author} {\bibfnamefont {A.}~\bibnamefont {Allais}}, \bibinfo {author} {\bibfnamefont {D.}~\bibnamefont {Chowdhury}}, \ and\ \bibinfo {author} {\bibfnamefont {S.}~\bibnamefont {Sachdev}},\ }\href {\doibase 10.1038/ncomms6771} {\bibfield  {journal} {\bibinfo  {journal} {Nature Communications}\ }\textbf {\bibinfo {volume} {5}} (\bibinfo {year} {2014}),\ 10.1038/ncomms6771}\BibitemShut {NoStop}%
\bibitem [{\citenamefont {Liechtenstein}\ \emph {et~al.}(1996)\citenamefont {Liechtenstein}, \citenamefont {Gunnarsson}, \citenamefont {Andersen},\ and\ \citenamefont {Martin}}]{Liech}%
  \BibitemOpen
  \bibfield  {author} {\bibinfo {author} {\bibfnamefont {A.~I.}\ \bibnamefont {Liechtenstein}}, \bibinfo {author} {\bibfnamefont {O.}~\bibnamefont {Gunnarsson}}, \bibinfo {author} {\bibfnamefont {O.~K.}\ \bibnamefont {Andersen}}, \ and\ \bibinfo {author} {\bibfnamefont {R.~M.}\ \bibnamefont {Martin}},\ }\href {\doibase 10.1103/PhysRevB.54.12505} {\bibfield  {journal} {\bibinfo  {journal} {Phys. Rev. B}\ }\textbf {\bibinfo {volume} {54}},\ \bibinfo {pages} {12505} (\bibinfo {year} {1996})}\BibitemShut {NoStop}%
\bibitem [{\citenamefont {Mattheiss}(1990)}]{Mattheiss}%
  \BibitemOpen
  \bibfield  {author} {\bibinfo {author} {\bibfnamefont {L.~F.}\ \bibnamefont {Mattheiss}},\ }\href {\doibase 10.1103/PhysRevB.42.354} {\bibfield  {journal} {\bibinfo  {journal} {Phys. Rev. B}\ }\textbf {\bibinfo {volume} {42}},\ \bibinfo {pages} {354} (\bibinfo {year} {1990})}\BibitemShut {NoStop}%
\bibitem [{\citenamefont {Schabel}\ \emph {et~al.}(1998)\citenamefont {Schabel}, \citenamefont {Park}, \citenamefont {Matsuura}, \citenamefont {Shen}, \citenamefont {Bonn}, \citenamefont {Liang},\ and\ \citenamefont {Hardy}}]{Schabel}%
  \BibitemOpen
  \bibfield  {author} {\bibinfo {author} {\bibfnamefont {M.~C.}\ \bibnamefont {Schabel}}, \bibinfo {author} {\bibfnamefont {C.-H.}\ \bibnamefont {Park}}, \bibinfo {author} {\bibfnamefont {A.}~\bibnamefont {Matsuura}}, \bibinfo {author} {\bibfnamefont {Z.-X.}\ \bibnamefont {Shen}}, \bibinfo {author} {\bibfnamefont {D.~A.}\ \bibnamefont {Bonn}}, \bibinfo {author} {\bibfnamefont {R.}~\bibnamefont {Liang}}, \ and\ \bibinfo {author} {\bibfnamefont {W.~N.}\ \bibnamefont {Hardy}},\ }\href {\doibase 10.1103/PhysRevB.57.6090} {\bibfield  {journal} {\bibinfo  {journal} {Phys. Rev. B}\ }\textbf {\bibinfo {volume} {57}},\ \bibinfo {pages} {6090} (\bibinfo {year} {1998})}\BibitemShut {NoStop}%
\bibitem [{\citenamefont {Verret}\ \emph {et~al.}(2017)\citenamefont {Verret}, \citenamefont {Charlebois}, \citenamefont {S\'en\'echal},\ and\ \citenamefont {Tremblay}}]{Verret}%
  \BibitemOpen
  \bibfield  {author} {\bibinfo {author} {\bibfnamefont {S.}~\bibnamefont {Verret}}, \bibinfo {author} {\bibfnamefont {M.}~\bibnamefont {Charlebois}}, \bibinfo {author} {\bibfnamefont {D.}~\bibnamefont {S\'en\'echal}}, \ and\ \bibinfo {author} {\bibfnamefont {A.-M.~S.}\ \bibnamefont {Tremblay}},\ }\href {\doibase 10.1103/PhysRevB.95.054518} {\bibfield  {journal} {\bibinfo  {journal} {Phys. Rev. B}\ }\textbf {\bibinfo {volume} {95}},\ \bibinfo {pages} {054518} (\bibinfo {year} {2017})}\BibitemShut {NoStop}%
\bibitem [{\citenamefont {Comin}\ \emph {et~al.}(2014)\citenamefont {Comin}, \citenamefont {Frano}, \citenamefont {Yee}, \citenamefont {Yoshida}, \citenamefont {Eisaki}, \citenamefont {Schierle}, \citenamefont {Weschke}, \citenamefont {Sutarto}, \citenamefont {He}, \citenamefont {Soumyanarayanan}, \citenamefont {He}, \citenamefont {Tacon}, \citenamefont {Elfimov}, \citenamefont {Hoffman}, \citenamefont {Sawatzky}, \citenamefont {Keimer},\ and\ \citenamefont {Damascelli}}]{Comin_2014}%
  \BibitemOpen
  \bibfield  {author} {\bibinfo {author} {\bibfnamefont {R.}~\bibnamefont {Comin}}, \bibinfo {author} {\bibfnamefont {A.}~\bibnamefont {Frano}}, \bibinfo {author} {\bibfnamefont {M.~M.}\ \bibnamefont {Yee}}, \bibinfo {author} {\bibfnamefont {Y.}~\bibnamefont {Yoshida}}, \bibinfo {author} {\bibfnamefont {H.}~\bibnamefont {Eisaki}}, \bibinfo {author} {\bibfnamefont {E.}~\bibnamefont {Schierle}}, \bibinfo {author} {\bibfnamefont {E.}~\bibnamefont {Weschke}}, \bibinfo {author} {\bibfnamefont {R.}~\bibnamefont {Sutarto}}, \bibinfo {author} {\bibfnamefont {F.}~\bibnamefont {He}}, \bibinfo {author} {\bibfnamefont {A.}~\bibnamefont {Soumyanarayanan}}, \bibinfo {author} {\bibfnamefont {Y.}~\bibnamefont {He}}, \bibinfo {author} {\bibfnamefont {M.~L.}\ \bibnamefont {Tacon}}, \bibinfo {author} {\bibfnamefont {I.~S.}\ \bibnamefont {Elfimov}}, \bibinfo {author} {\bibfnamefont {J.~E.}\ \bibnamefont {Hoffman}}, \bibinfo {author} {\bibfnamefont {G.~A.}\ \bibnamefont {Sawatzky}}, \bibinfo {author} {\bibfnamefont
  {B.}~\bibnamefont {Keimer}}, \ and\ \bibinfo {author} {\bibfnamefont {A.}~\bibnamefont {Damascelli}},\ }\href {\doibase 10.1126/science.1242996} {\bibfield  {journal} {\bibinfo  {journal} {Science}\ }\textbf {\bibinfo {volume} {343}},\ \bibinfo {pages} {390} (\bibinfo {year} {2014})}\BibitemShut {NoStop}%
\bibitem [{\citenamefont {Devereaux}\ and\ \citenamefont {Hackl}(2007)}]{Devereaux}%
  \BibitemOpen
  \bibfield  {author} {\bibinfo {author} {\bibfnamefont {T.~P.}\ \bibnamefont {Devereaux}}\ and\ \bibinfo {author} {\bibfnamefont {R.}~\bibnamefont {Hackl}},\ }\href {\doibase 10.1103/RevModPhys.79.175} {\bibfield  {journal} {\bibinfo  {journal} {Rev. Mod. Phys.}\ }\textbf {\bibinfo {volume} {79}},\ \bibinfo {pages} {175} (\bibinfo {year} {2007})}\BibitemShut {NoStop}%
\bibitem [{\citenamefont {Georges}\ \emph {et~al.}(1996)\citenamefont {Georges}, \citenamefont {Kotliar}, \citenamefont {Krauth},\ and\ \citenamefont {Rozenberg}}]{DMFT}%
  \BibitemOpen
  \bibfield  {author} {\bibinfo {author} {\bibfnamefont {A.}~\bibnamefont {Georges}}, \bibinfo {author} {\bibfnamefont {G.}~\bibnamefont {Kotliar}}, \bibinfo {author} {\bibfnamefont {W.}~\bibnamefont {Krauth}}, \ and\ \bibinfo {author} {\bibfnamefont {M.~J.}\ \bibnamefont {Rozenberg}},\ }\href {\doibase 10.1103/RevModPhys.68.13} {\bibfield  {journal} {\bibinfo  {journal} {Rev. Mod. Phys.}\ }\textbf {\bibinfo {volume} {68}},\ \bibinfo {pages} {13} (\bibinfo {year} {1996})}\BibitemShut {NoStop}%
\bibitem [{\citenamefont {Sacuto}\ \emph {et~al.}(2011)\citenamefont {Sacuto}, \citenamefont {Gallais}, \citenamefont {Cazayous}, \citenamefont {Blanc}, \citenamefont {M\'easson}, \citenamefont {Wen}, \citenamefont {Xu}, \citenamefont {Gu},\ and\ \citenamefont {Colson}}]{Alain2}%
  \BibitemOpen
  \bibfield  {author} {\bibinfo {author} {\bibfnamefont {A.}~\bibnamefont {Sacuto}}, \bibinfo {author} {\bibfnamefont {Y.}~\bibnamefont {Gallais}}, \bibinfo {author} {\bibfnamefont {M.}~\bibnamefont {Cazayous}}, \bibinfo {author} {\bibfnamefont {S.}~\bibnamefont {Blanc}}, \bibinfo {author} {\bibfnamefont {M.-A.}\ \bibnamefont {M\'easson}}, \bibinfo {author} {\bibfnamefont {J.}~\bibnamefont {Wen}}, \bibinfo {author} {\bibfnamefont {Z.}~\bibnamefont {Xu}}, \bibinfo {author} {\bibfnamefont {G.}~\bibnamefont {Gu}}, \ and\ \bibinfo {author} {\bibfnamefont {D.}~\bibnamefont {Colson}},\ }\href {\doibase https://doi.org/10.1016/j.crhy.2011.04.001} {\bibfield  {journal} {\bibinfo  {journal} {Comptes Rendus Physique}\ }\textbf {\bibinfo {volume} {12}},\ \bibinfo {pages} {480} (\bibinfo {year} {2011})},\ \bibinfo {note} {superconductivity of strongly correlated systems}\BibitemShut {NoStop}%
\bibitem [{\citenamefont {Li}\ and\ \citenamefont {Comin}(2019)}]{Li2019_it}%
  \BibitemOpen
  \bibfield  {author} {\bibinfo {author} {\bibfnamefont {J.}~\bibnamefont {Li}}\ and\ \bibinfo {author} {\bibfnamefont {R.}~\bibnamefont {Comin}},\ }\href@noop {} {\bibfield  {journal} {\bibinfo  {journal} {Nature Physics}\ }\textbf {\bibinfo {volume} {15}},\ \bibinfo {pages} {736} (\bibinfo {year} {2019})}\BibitemShut {NoStop}%
\bibitem [{\citenamefont {Chang}\ \emph {et~al.}(2012)\citenamefont {Chang}, \citenamefont {Blackburn}, \citenamefont {Holmes}, \citenamefont {Christensen}, \citenamefont {Larsen}, \citenamefont {Mesot}, \citenamefont {Liang}, \citenamefont {Bonn}, \citenamefont {Hardy}, \citenamefont {Watenphul}, \citenamefont {v.~Zimmermann}, \citenamefont {Forgan},\ and\ \citenamefont {Hayden}}]{Chang_2012}%
  \BibitemOpen
  \bibfield  {author} {\bibinfo {author} {\bibfnamefont {J.}~\bibnamefont {Chang}}, \bibinfo {author} {\bibfnamefont {E.}~\bibnamefont {Blackburn}}, \bibinfo {author} {\bibfnamefont {A.~T.}\ \bibnamefont {Holmes}}, \bibinfo {author} {\bibfnamefont {N.~B.}\ \bibnamefont {Christensen}}, \bibinfo {author} {\bibfnamefont {J.}~\bibnamefont {Larsen}}, \bibinfo {author} {\bibfnamefont {J.}~\bibnamefont {Mesot}}, \bibinfo {author} {\bibfnamefont {R.}~\bibnamefont {Liang}}, \bibinfo {author} {\bibfnamefont {D.~A.}\ \bibnamefont {Bonn}}, \bibinfo {author} {\bibfnamefont {W.~N.}\ \bibnamefont {Hardy}}, \bibinfo {author} {\bibfnamefont {A.}~\bibnamefont {Watenphul}}, \bibinfo {author} {\bibfnamefont {M.}~\bibnamefont {v.~Zimmermann}}, \bibinfo {author} {\bibfnamefont {E.~M.}\ \bibnamefont {Forgan}}, \ and\ \bibinfo {author} {\bibfnamefont {S.~M.}\ \bibnamefont {Hayden}},\ }\href {\doibase 10.1038/nphys2456} {\bibfield  {journal} {\bibinfo  {journal} {Nature Physics}\ }\textbf {\bibinfo {volume} {8}},\ \bibinfo {pages}
  {871} (\bibinfo {year} {2012})}\BibitemShut {NoStop}%
\bibitem [{\citenamefont {Blanco-Canosa}\ \emph {et~al.}(2014)\citenamefont {Blanco-Canosa}, \citenamefont {Frano}, \citenamefont {Schierle}, \citenamefont {Porras}, \citenamefont {Loew}, \citenamefont {Minola}, \citenamefont {Bluschke}, \citenamefont {Weschke}, \citenamefont {Keimer},\ and\ \citenamefont {Le~Tacon}}]{Incom1}%
  \BibitemOpen
  \bibfield  {author} {\bibinfo {author} {\bibfnamefont {S.}~\bibnamefont {Blanco-Canosa}}, \bibinfo {author} {\bibfnamefont {A.}~\bibnamefont {Frano}}, \bibinfo {author} {\bibfnamefont {E.}~\bibnamefont {Schierle}}, \bibinfo {author} {\bibfnamefont {J.}~\bibnamefont {Porras}}, \bibinfo {author} {\bibfnamefont {T.}~\bibnamefont {Loew}}, \bibinfo {author} {\bibfnamefont {M.}~\bibnamefont {Minola}}, \bibinfo {author} {\bibfnamefont {M.}~\bibnamefont {Bluschke}}, \bibinfo {author} {\bibfnamefont {E.}~\bibnamefont {Weschke}}, \bibinfo {author} {\bibfnamefont {B.}~\bibnamefont {Keimer}}, \ and\ \bibinfo {author} {\bibfnamefont {M.}~\bibnamefont {Le~Tacon}},\ }\href {\doibase 10.1103/PhysRevB.90.054513} {\bibfield  {journal} {\bibinfo  {journal} {Phys. Rev. B}\ }\textbf {\bibinfo {volume} {90}},\ \bibinfo {pages} {054513} (\bibinfo {year} {2014})}\BibitemShut {NoStop}%
\bibitem [{\citenamefont {Mesaros}\ \emph {et~al.}(2016)\citenamefont {Mesaros}, \citenamefont {Fujita}, \citenamefont {Edkins}, \citenamefont {Hamidian}, \citenamefont {Eisaki}, \citenamefont {ichi Uchida}, \citenamefont {Davis}, \citenamefont {Lawler},\ and\ \citenamefont {Kim}}]{doi:10.1073/pnas.1614247113}%
  \BibitemOpen
  \bibfield  {author} {\bibinfo {author} {\bibfnamefont {A.}~\bibnamefont {Mesaros}}, \bibinfo {author} {\bibfnamefont {K.}~\bibnamefont {Fujita}}, \bibinfo {author} {\bibfnamefont {S.~D.}\ \bibnamefont {Edkins}}, \bibinfo {author} {\bibfnamefont {M.~H.}\ \bibnamefont {Hamidian}}, \bibinfo {author} {\bibfnamefont {H.}~\bibnamefont {Eisaki}}, \bibinfo {author} {\bibfnamefont {S.}~\bibnamefont {ichi Uchida}}, \bibinfo {author} {\bibfnamefont {J.~C.~S.}\ \bibnamefont {Davis}}, \bibinfo {author} {\bibfnamefont {M.~J.}\ \bibnamefont {Lawler}}, \ and\ \bibinfo {author} {\bibfnamefont {E.-A.}\ \bibnamefont {Kim}},\ }\href {\doibase 10.1073/pnas.1614247113} {\bibfield  {journal} {\bibinfo  {journal} {Proceedings of the National Academy of Sciences}\ }\textbf {\bibinfo {volume} {113}},\ \bibinfo {pages} {12661} (\bibinfo {year} {2016})},\ \Eprint {http://arxiv.org/abs/https://www.pnas.org/doi/pdf/10.1073/pnas.1614247113} {https://www.pnas.org/doi/pdf/10.1073/pnas.1614247113} \BibitemShut {NoStop}%
\bibitem [{\citenamefont {Vinograd}\ \emph {et~al.}(2021)\citenamefont {Vinograd}, \citenamefont {Zhou}, \citenamefont {Hirata}, \citenamefont {Wu}, \citenamefont {Mayaffre}, \citenamefont {Kr{\"a}mer}, \citenamefont {Liang}, \citenamefont {Hardy}, \citenamefont {Bonn},\ and\ \citenamefont {Julien}}]{Vinograd2021}%
  \BibitemOpen
  \bibfield  {author} {\bibinfo {author} {\bibfnamefont {I.}~\bibnamefont {Vinograd}}, \bibinfo {author} {\bibfnamefont {R.}~\bibnamefont {Zhou}}, \bibinfo {author} {\bibfnamefont {M.}~\bibnamefont {Hirata}}, \bibinfo {author} {\bibfnamefont {T.}~\bibnamefont {Wu}}, \bibinfo {author} {\bibfnamefont {H.}~\bibnamefont {Mayaffre}}, \bibinfo {author} {\bibfnamefont {S.}~\bibnamefont {Kr{\"a}mer}}, \bibinfo {author} {\bibfnamefont {R.}~\bibnamefont {Liang}}, \bibinfo {author} {\bibfnamefont {W.~N.}\ \bibnamefont {Hardy}}, \bibinfo {author} {\bibfnamefont {D.~A.}\ \bibnamefont {Bonn}}, \ and\ \bibinfo {author} {\bibfnamefont {M.-H.}\ \bibnamefont {Julien}},\ }\href {\doibase 10.1038/s41467-021-23140-w} {\bibfield  {journal} {\bibinfo  {journal} {Nature Communications}\ }\textbf {\bibinfo {volume} {12}},\ \bibinfo {pages} {3274} (\bibinfo {year} {2021})}\BibitemShut {NoStop}%
\bibitem [{\citenamefont {Kyung}\ \emph {et~al.}(2006)\citenamefont {Kyung}, \citenamefont {Kancharla}, \citenamefont {S\'en\'echal}, \citenamefont {Tremblay}, \citenamefont {Civelli},\ and\ \citenamefont {Kotliar}}]{PhysRevB.73.165114}%
  \BibitemOpen
  \bibfield  {author} {\bibinfo {author} {\bibfnamefont {B.}~\bibnamefont {Kyung}}, \bibinfo {author} {\bibfnamefont {S.~S.}\ \bibnamefont {Kancharla}}, \bibinfo {author} {\bibfnamefont {D.}~\bibnamefont {S\'en\'echal}}, \bibinfo {author} {\bibfnamefont {A.-M.~S.}\ \bibnamefont {Tremblay}}, \bibinfo {author} {\bibfnamefont {M.}~\bibnamefont {Civelli}}, \ and\ \bibinfo {author} {\bibfnamefont {G.}~\bibnamefont {Kotliar}},\ }\href {\doibase 10.1103/PhysRevB.73.165114} {\bibfield  {journal} {\bibinfo  {journal} {Phys. Rev. B}\ }\textbf {\bibinfo {volume} {73}},\ \bibinfo {pages} {165114} (\bibinfo {year} {2006})}\BibitemShut {NoStop}%
\bibitem [{\citenamefont {Stanescu}\ and\ \citenamefont {Kotliar}(2006)}]{PhysRevB.74.125110}%
  \BibitemOpen
  \bibfield  {author} {\bibinfo {author} {\bibfnamefont {T.~D.}\ \bibnamefont {Stanescu}}\ and\ \bibinfo {author} {\bibfnamefont {G.}~\bibnamefont {Kotliar}},\ }\href {\doibase 10.1103/PhysRevB.74.125110} {\bibfield  {journal} {\bibinfo  {journal} {Phys. Rev. B}\ }\textbf {\bibinfo {volume} {74}},\ \bibinfo {pages} {125110} (\bibinfo {year} {2006})}\BibitemShut {NoStop}%
\bibitem [{\citenamefont {Sakai}\ \emph {et~al.}(2009)\citenamefont {Sakai}, \citenamefont {Motome},\ and\ \citenamefont {Imada}}]{PhysRevLett.102.056404}%
  \BibitemOpen
  \bibfield  {author} {\bibinfo {author} {\bibfnamefont {S.}~\bibnamefont {Sakai}}, \bibinfo {author} {\bibfnamefont {Y.}~\bibnamefont {Motome}}, \ and\ \bibinfo {author} {\bibfnamefont {M.}~\bibnamefont {Imada}},\ }\href {\doibase 10.1103/PhysRevLett.102.056404} {\bibfield  {journal} {\bibinfo  {journal} {Phys. Rev. Lett.}\ }\textbf {\bibinfo {volume} {102}},\ \bibinfo {pages} {056404} (\bibinfo {year} {2009})}\BibitemShut {NoStop}%
\bibitem [{\citenamefont {Sakai}\ \emph {et~al.}(2016)\citenamefont {Sakai}, \citenamefont {Civelli},\ and\ \citenamefont {Imada}}]{hidden}%
  \BibitemOpen
  \bibfield  {author} {\bibinfo {author} {\bibfnamefont {S.}~\bibnamefont {Sakai}}, \bibinfo {author} {\bibfnamefont {M.}~\bibnamefont {Civelli}}, \ and\ \bibinfo {author} {\bibfnamefont {M.}~\bibnamefont {Imada}},\ }\href {\doibase 10.1103/PhysRevB.94.115130} {\bibfield  {journal} {\bibinfo  {journal} {Phys. Rev. B}\ }\textbf {\bibinfo {volume} {94}},\ \bibinfo {pages} {115130} (\bibinfo {year} {2016})}\BibitemShut {NoStop}%
\bibitem [{\citenamefont {Coleman}(2015)}]{coleman2015introduction}%
  \BibitemOpen
  \bibfield  {author} {\bibinfo {author} {\bibfnamefont {P.}~\bibnamefont {Coleman}},\ }\href {https://books.google.fr/books?id=ESB0CwAAQBAJ} {\emph {\bibinfo {title} {Introduction to Many-Body Physics}}}\ (\bibinfo  {publisher} {Cambridge University Press},\ \bibinfo {year} {2015})\BibitemShut {NoStop}%
\bibitem [{\citenamefont {Fischer}\ \emph {et~al.}(2007)\citenamefont {Fischer}, \citenamefont {Kugler}, \citenamefont {Maggio-Aprile}, \citenamefont {Berthod},\ and\ \citenamefont {Renner}}]{RevModPhys.79.353}%
  \BibitemOpen
  \bibfield  {author} {\bibinfo {author} {\bibfnamefont {O.}~\bibnamefont {Fischer}}, \bibinfo {author} {\bibfnamefont {M.}~\bibnamefont {Kugler}}, \bibinfo {author} {\bibfnamefont {I.}~\bibnamefont {Maggio-Aprile}}, \bibinfo {author} {\bibfnamefont {C.}~\bibnamefont {Berthod}}, \ and\ \bibinfo {author} {\bibfnamefont {C.}~\bibnamefont {Renner}},\ }\href {\doibase 10.1103/RevModPhys.79.353} {\bibfield  {journal} {\bibinfo  {journal} {Rev. Mod. Phys.}\ }\textbf {\bibinfo {volume} {79}},\ \bibinfo {pages} {353} (\bibinfo {year} {2007})}\BibitemShut {NoStop}%
\bibitem [{\citenamefont {Fujita}\ \emph {et~al.}(2012)\citenamefont {Fujita}, \citenamefont {R.~Schmidt}, \citenamefont {Kim}, \citenamefont {J.~Lawler}, \citenamefont {Hai~Lee}, \citenamefont {Davis}, \citenamefont {Eisaki},\ and\ \citenamefont {Uchida}}]{JPSJ.81.011005}%
  \BibitemOpen
  \bibfield  {author} {\bibinfo {author} {\bibfnamefont {K.}~\bibnamefont {Fujita}}, \bibinfo {author} {\bibfnamefont {A.}~\bibnamefont {R.~Schmidt}}, \bibinfo {author} {\bibfnamefont {E.-A.}\ \bibnamefont {Kim}}, \bibinfo {author} {\bibfnamefont {M.}~\bibnamefont {J.~Lawler}}, \bibinfo {author} {\bibfnamefont {D.}~\bibnamefont {Hai~Lee}}, \bibinfo {author} {\bibfnamefont {J.}~\bibnamefont {Davis}}, \bibinfo {author} {\bibfnamefont {H.}~\bibnamefont {Eisaki}}, \ and\ \bibinfo {author} {\bibfnamefont {S.-i.}\ \bibnamefont {Uchida}},\ }\href {\doibase 10.1143/JPSJ.81.011005} {\bibfield  {journal} {\bibinfo  {journal} {Journal of the Physical Society of Japan}\ }\textbf {\bibinfo {volume} {81}},\ \bibinfo {pages} {011005} (\bibinfo {year} {2012})},\ \Eprint {http://arxiv.org/abs/https://doi.org/10.1143/JPSJ.81.011005} {https://doi.org/10.1143/JPSJ.81.011005} \BibitemShut {NoStop}%
\bibitem [{\citenamefont {Hoffman}\ \emph {et~al.}(2002)\citenamefont {Hoffman}, \citenamefont {Hudson}, \citenamefont {Lang}, \citenamefont {Madhavan}, \citenamefont {Eisaki}, \citenamefont {Uchida},\ and\ \citenamefont {Davis}}]{science.1066974}%
  \BibitemOpen
  \bibfield  {author} {\bibinfo {author} {\bibfnamefont {J.~E.}\ \bibnamefont {Hoffman}}, \bibinfo {author} {\bibfnamefont {E.~W.}\ \bibnamefont {Hudson}}, \bibinfo {author} {\bibfnamefont {K.~M.}\ \bibnamefont {Lang}}, \bibinfo {author} {\bibfnamefont {V.}~\bibnamefont {Madhavan}}, \bibinfo {author} {\bibfnamefont {H.}~\bibnamefont {Eisaki}}, \bibinfo {author} {\bibfnamefont {S.}~\bibnamefont {Uchida}}, \ and\ \bibinfo {author} {\bibfnamefont {J.~C.}\ \bibnamefont {Davis}},\ }\href {\doibase 10.1126/science.1066974} {\bibfield  {journal} {\bibinfo  {journal} {Science}\ }\textbf {\bibinfo {volume} {295}},\ \bibinfo {pages} {466} (\bibinfo {year} {2002})},\ \Eprint {http://arxiv.org/abs/https://www.science.org/doi/pdf/10.1126/science.1066974} {https://www.science.org/doi/pdf/10.1126/science.1066974} \BibitemShut {NoStop}%
\bibitem [{\citenamefont {Lanzara}\ \emph {et~al.}(2001)\citenamefont {Lanzara}, \citenamefont {Bogdanov}, \citenamefont {Zhou}, \citenamefont {Kellar}, \citenamefont {Feng}, \citenamefont {Lu}, \citenamefont {Yoshida}, \citenamefont {Eisaki}, \citenamefont {Fujimori}, \citenamefont {Kishio}, \citenamefont {Shimoyama}, \citenamefont {Noda}, \citenamefont {Uchida}, \citenamefont {Hussain},\ and\ \citenamefont {Shen}}]{Lanzara2001-he}%
  \BibitemOpen
  \bibfield  {author} {\bibinfo {author} {\bibfnamefont {A.}~\bibnamefont {Lanzara}}, \bibinfo {author} {\bibfnamefont {P.~V.}\ \bibnamefont {Bogdanov}}, \bibinfo {author} {\bibfnamefont {X.~J.}\ \bibnamefont {Zhou}}, \bibinfo {author} {\bibfnamefont {S.~A.}\ \bibnamefont {Kellar}}, \bibinfo {author} {\bibfnamefont {D.~L.}\ \bibnamefont {Feng}}, \bibinfo {author} {\bibfnamefont {E.~D.}\ \bibnamefont {Lu}}, \bibinfo {author} {\bibfnamefont {T.}~\bibnamefont {Yoshida}}, \bibinfo {author} {\bibfnamefont {H.}~\bibnamefont {Eisaki}}, \bibinfo {author} {\bibfnamefont {A.}~\bibnamefont {Fujimori}}, \bibinfo {author} {\bibfnamefont {K.}~\bibnamefont {Kishio}}, \bibinfo {author} {\bibfnamefont {J.~I.}\ \bibnamefont {Shimoyama}}, \bibinfo {author} {\bibfnamefont {T.}~\bibnamefont {Noda}}, \bibinfo {author} {\bibfnamefont {S.}~\bibnamefont {Uchida}}, \bibinfo {author} {\bibfnamefont {Z.}~\bibnamefont {Hussain}}, \ and\ \bibinfo {author} {\bibfnamefont {Z.~X.}\ \bibnamefont {Shen}},\ }\href@noop {} {\bibfield  {journal}
  {\bibinfo  {journal} {Nature}\ }\textbf {\bibinfo {volume} {412}},\ \bibinfo {pages} {510} (\bibinfo {year} {2001})}\BibitemShut {NoStop}%
\bibitem [{\citenamefont {Kordyuk}\ \emph {et~al.}(2005)\citenamefont {Kordyuk}, \citenamefont {Borisenko}, \citenamefont {Koitzsch}, \citenamefont {Fink}, \citenamefont {Knupfer},\ and\ \citenamefont {Berger}}]{PhysRevB.71.214513}%
  \BibitemOpen
  \bibfield  {author} {\bibinfo {author} {\bibfnamefont {A.~A.}\ \bibnamefont {Kordyuk}}, \bibinfo {author} {\bibfnamefont {S.~V.}\ \bibnamefont {Borisenko}}, \bibinfo {author} {\bibfnamefont {A.}~\bibnamefont {Koitzsch}}, \bibinfo {author} {\bibfnamefont {J.}~\bibnamefont {Fink}}, \bibinfo {author} {\bibfnamefont {M.}~\bibnamefont {Knupfer}}, \ and\ \bibinfo {author} {\bibfnamefont {H.}~\bibnamefont {Berger}},\ }\href {\doibase 10.1103/PhysRevB.71.214513} {\bibfield  {journal} {\bibinfo  {journal} {Phys. Rev. B}\ }\textbf {\bibinfo {volume} {71}},\ \bibinfo {pages} {214513} (\bibinfo {year} {2005})}\BibitemShut {NoStop}%
\bibitem [{\citenamefont {Graf}\ \emph {et~al.}(2007)\citenamefont {Graf}, \citenamefont {Gweon}, \citenamefont {McElroy}, \citenamefont {Zhou}, \citenamefont {Jozwiak}, \citenamefont {Rotenberg}, \citenamefont {Bill}, \citenamefont {Sasagawa}, \citenamefont {Eisaki}, \citenamefont {Uchida}, \citenamefont {Takagi}, \citenamefont {Lee},\ and\ \citenamefont {Lanzara}}]{PhysRevLett.98.067004}%
  \BibitemOpen
  \bibfield  {author} {\bibinfo {author} {\bibfnamefont {J.}~\bibnamefont {Graf}}, \bibinfo {author} {\bibfnamefont {G.-H.}\ \bibnamefont {Gweon}}, \bibinfo {author} {\bibfnamefont {K.}~\bibnamefont {McElroy}}, \bibinfo {author} {\bibfnamefont {S.~Y.}\ \bibnamefont {Zhou}}, \bibinfo {author} {\bibfnamefont {C.}~\bibnamefont {Jozwiak}}, \bibinfo {author} {\bibfnamefont {E.}~\bibnamefont {Rotenberg}}, \bibinfo {author} {\bibfnamefont {A.}~\bibnamefont {Bill}}, \bibinfo {author} {\bibfnamefont {T.}~\bibnamefont {Sasagawa}}, \bibinfo {author} {\bibfnamefont {H.}~\bibnamefont {Eisaki}}, \bibinfo {author} {\bibfnamefont {S.}~\bibnamefont {Uchida}}, \bibinfo {author} {\bibfnamefont {H.}~\bibnamefont {Takagi}}, \bibinfo {author} {\bibfnamefont {D.-H.}\ \bibnamefont {Lee}}, \ and\ \bibinfo {author} {\bibfnamefont {A.}~\bibnamefont {Lanzara}},\ }\href {\doibase 10.1103/PhysRevLett.98.067004} {\bibfield  {journal} {\bibinfo  {journal} {Phys. Rev. Lett.}\ }\textbf {\bibinfo {volume} {98}},\ \bibinfo {pages} {067004}
  (\bibinfo {year} {2007})}\BibitemShut {NoStop}%
\bibitem [{\citenamefont {Valla}\ \emph {et~al.}(2007)\citenamefont {Valla}, \citenamefont {Kidd}, \citenamefont {Yin}, \citenamefont {Gu}, \citenamefont {Johnson}, \citenamefont {Pan},\ and\ \citenamefont {Fedorov}}]{PhysRevLett.98.167003}%
  \BibitemOpen
  \bibfield  {author} {\bibinfo {author} {\bibfnamefont {T.}~\bibnamefont {Valla}}, \bibinfo {author} {\bibfnamefont {T.~E.}\ \bibnamefont {Kidd}}, \bibinfo {author} {\bibfnamefont {W.-G.}\ \bibnamefont {Yin}}, \bibinfo {author} {\bibfnamefont {G.~D.}\ \bibnamefont {Gu}}, \bibinfo {author} {\bibfnamefont {P.~D.}\ \bibnamefont {Johnson}}, \bibinfo {author} {\bibfnamefont {Z.-H.}\ \bibnamefont {Pan}}, \ and\ \bibinfo {author} {\bibfnamefont {A.~V.}\ \bibnamefont {Fedorov}},\ }\href {\doibase 10.1103/PhysRevLett.98.167003} {\bibfield  {journal} {\bibinfo  {journal} {Phys. Rev. Lett.}\ }\textbf {\bibinfo {volume} {98}},\ \bibinfo {pages} {167003} (\bibinfo {year} {2007})}\BibitemShut {NoStop}%
\bibitem [{\citenamefont {Xie}\ \emph {et~al.}(2007)\citenamefont {Xie}, \citenamefont {Yang}, \citenamefont {Shen}, \citenamefont {Zhao}, \citenamefont {Ou}, \citenamefont {Wei}, \citenamefont {Gu}, \citenamefont {Arita}, \citenamefont {Qiao}, \citenamefont {Namatame}, \citenamefont {Taniguchi}, \citenamefont {Kaneko}, \citenamefont {Eisaki}, \citenamefont {Tsuei}, \citenamefont {Cheng}, \citenamefont {Vobornik}, \citenamefont {Fujii}, \citenamefont {Rossi}, \citenamefont {Yang},\ and\ \citenamefont {Feng}}]{PhysRevLett.98.147001}%
  \BibitemOpen
  \bibfield  {author} {\bibinfo {author} {\bibfnamefont {B.~P.}\ \bibnamefont {Xie}}, \bibinfo {author} {\bibfnamefont {K.}~\bibnamefont {Yang}}, \bibinfo {author} {\bibfnamefont {D.~W.}\ \bibnamefont {Shen}}, \bibinfo {author} {\bibfnamefont {J.~F.}\ \bibnamefont {Zhao}}, \bibinfo {author} {\bibfnamefont {H.~W.}\ \bibnamefont {Ou}}, \bibinfo {author} {\bibfnamefont {J.}~\bibnamefont {Wei}}, \bibinfo {author} {\bibfnamefont {S.~Y.}\ \bibnamefont {Gu}}, \bibinfo {author} {\bibfnamefont {M.}~\bibnamefont {Arita}}, \bibinfo {author} {\bibfnamefont {S.}~\bibnamefont {Qiao}}, \bibinfo {author} {\bibfnamefont {H.}~\bibnamefont {Namatame}}, \bibinfo {author} {\bibfnamefont {M.}~\bibnamefont {Taniguchi}}, \bibinfo {author} {\bibfnamefont {N.}~\bibnamefont {Kaneko}}, \bibinfo {author} {\bibfnamefont {H.}~\bibnamefont {Eisaki}}, \bibinfo {author} {\bibfnamefont {K.~D.}\ \bibnamefont {Tsuei}}, \bibinfo {author} {\bibfnamefont {C.~M.}\ \bibnamefont {Cheng}}, \bibinfo {author} {\bibfnamefont {I.}~\bibnamefont {Vobornik}},
  \bibinfo {author} {\bibfnamefont {J.}~\bibnamefont {Fujii}}, \bibinfo {author} {\bibfnamefont {G.}~\bibnamefont {Rossi}}, \bibinfo {author} {\bibfnamefont {Z.~Q.}\ \bibnamefont {Yang}}, \ and\ \bibinfo {author} {\bibfnamefont {D.~L.}\ \bibnamefont {Feng}},\ }\href {\doibase 10.1103/PhysRevLett.98.147001} {\bibfield  {journal} {\bibinfo  {journal} {Phys. Rev. Lett.}\ }\textbf {\bibinfo {volume} {98}},\ \bibinfo {pages} {147001} (\bibinfo {year} {2007})}\BibitemShut {NoStop}%
\bibitem [{\citenamefont {Meevasana}\ \emph {et~al.}(2007)\citenamefont {Meevasana}, \citenamefont {Zhou}, \citenamefont {Sahrakorpi}, \citenamefont {Lee}, \citenamefont {Yang}, \citenamefont {Tanaka}, \citenamefont {Mannella}, \citenamefont {Yoshida}, \citenamefont {Lu}, \citenamefont {Chen}, \citenamefont {He}, \citenamefont {Lin}, \citenamefont {Komiya}, \citenamefont {Ando}, \citenamefont {Zhou}, \citenamefont {Ti}, \citenamefont {Xiong}, \citenamefont {Zhao}, \citenamefont {Sasagawa}, \citenamefont {Kakeshita}, \citenamefont {Fujita}, \citenamefont {Uchida}, \citenamefont {Eisaki}, \citenamefont {Fujimori}, \citenamefont {Hussain}, \citenamefont {Markiewicz}, \citenamefont {Bansil}, \citenamefont {Nagaosa}, \citenamefont {Zaanen}, \citenamefont {Devereaux},\ and\ \citenamefont {Shen}}]{PhysRevB.75.174506}%
  \BibitemOpen
  \bibfield  {author} {\bibinfo {author} {\bibfnamefont {W.}~\bibnamefont {Meevasana}}, \bibinfo {author} {\bibfnamefont {X.~J.}\ \bibnamefont {Zhou}}, \bibinfo {author} {\bibfnamefont {S.}~\bibnamefont {Sahrakorpi}}, \bibinfo {author} {\bibfnamefont {W.~S.}\ \bibnamefont {Lee}}, \bibinfo {author} {\bibfnamefont {W.~L.}\ \bibnamefont {Yang}}, \bibinfo {author} {\bibfnamefont {K.}~\bibnamefont {Tanaka}}, \bibinfo {author} {\bibfnamefont {N.}~\bibnamefont {Mannella}}, \bibinfo {author} {\bibfnamefont {T.}~\bibnamefont {Yoshida}}, \bibinfo {author} {\bibfnamefont {D.~H.}\ \bibnamefont {Lu}}, \bibinfo {author} {\bibfnamefont {Y.~L.}\ \bibnamefont {Chen}}, \bibinfo {author} {\bibfnamefont {R.~H.}\ \bibnamefont {He}}, \bibinfo {author} {\bibfnamefont {H.}~\bibnamefont {Lin}}, \bibinfo {author} {\bibfnamefont {S.}~\bibnamefont {Komiya}}, \bibinfo {author} {\bibfnamefont {Y.}~\bibnamefont {Ando}}, \bibinfo {author} {\bibfnamefont {F.}~\bibnamefont {Zhou}}, \bibinfo {author} {\bibfnamefont {W.~X.}\ \bibnamefont {Ti}},
  \bibinfo {author} {\bibfnamefont {J.~W.}\ \bibnamefont {Xiong}}, \bibinfo {author} {\bibfnamefont {Z.~X.}\ \bibnamefont {Zhao}}, \bibinfo {author} {\bibfnamefont {T.}~\bibnamefont {Sasagawa}}, \bibinfo {author} {\bibfnamefont {T.}~\bibnamefont {Kakeshita}}, \bibinfo {author} {\bibfnamefont {K.}~\bibnamefont {Fujita}}, \bibinfo {author} {\bibfnamefont {S.}~\bibnamefont {Uchida}}, \bibinfo {author} {\bibfnamefont {H.}~\bibnamefont {Eisaki}}, \bibinfo {author} {\bibfnamefont {A.}~\bibnamefont {Fujimori}}, \bibinfo {author} {\bibfnamefont {Z.}~\bibnamefont {Hussain}}, \bibinfo {author} {\bibfnamefont {R.~S.}\ \bibnamefont {Markiewicz}}, \bibinfo {author} {\bibfnamefont {A.}~\bibnamefont {Bansil}}, \bibinfo {author} {\bibfnamefont {N.}~\bibnamefont {Nagaosa}}, \bibinfo {author} {\bibfnamefont {J.}~\bibnamefont {Zaanen}}, \bibinfo {author} {\bibfnamefont {T.~P.}\ \bibnamefont {Devereaux}}, \ and\ \bibinfo {author} {\bibfnamefont {Z.-X.}\ \bibnamefont {Shen}},\ }\href {\doibase 10.1103/PhysRevB.75.174506}
  {\bibfield  {journal} {\bibinfo  {journal} {Phys. Rev. B}\ }\textbf {\bibinfo {volume} {75}},\ \bibinfo {pages} {174506} (\bibinfo {year} {2007})}\BibitemShut {NoStop}%
\bibitem [{\citenamefont {Hashimoto}\ \emph {et~al.}(2010)\citenamefont {Hashimoto}, \citenamefont {He}, \citenamefont {Tanaka}, \citenamefont {Testaud}, \citenamefont {Meevasana}, \citenamefont {Moore}, \citenamefont {Lu}, \citenamefont {Yao}, \citenamefont {Yoshida}, \citenamefont {Eisaki}, \citenamefont {Devereaux}, \citenamefont {Hussain},\ and\ \citenamefont {Shen}}]{Hashimoto2010-qa}%
  \BibitemOpen
  \bibfield  {author} {\bibinfo {author} {\bibfnamefont {M.}~\bibnamefont {Hashimoto}}, \bibinfo {author} {\bibfnamefont {R.-H.}\ \bibnamefont {He}}, \bibinfo {author} {\bibfnamefont {K.}~\bibnamefont {Tanaka}}, \bibinfo {author} {\bibfnamefont {J.-P.}\ \bibnamefont {Testaud}}, \bibinfo {author} {\bibfnamefont {W.}~\bibnamefont {Meevasana}}, \bibinfo {author} {\bibfnamefont {R.~G.}\ \bibnamefont {Moore}}, \bibinfo {author} {\bibfnamefont {D.}~\bibnamefont {Lu}}, \bibinfo {author} {\bibfnamefont {H.}~\bibnamefont {Yao}}, \bibinfo {author} {\bibfnamefont {Y.}~\bibnamefont {Yoshida}}, \bibinfo {author} {\bibfnamefont {H.}~\bibnamefont {Eisaki}}, \bibinfo {author} {\bibfnamefont {T.~P.}\ \bibnamefont {Devereaux}}, \bibinfo {author} {\bibfnamefont {Z.}~\bibnamefont {Hussain}}, \ and\ \bibinfo {author} {\bibfnamefont {Z.-X.}\ \bibnamefont {Shen}},\ }\href@noop {} {\bibfield  {journal} {\bibinfo  {journal} {Nature Physics}\ }\textbf {\bibinfo {volume} {6}},\ \bibinfo {pages} {414} (\bibinfo {year}
  {2010})}\BibitemShut {NoStop}%
\bibitem [{\citenamefont {Smallwood}\ \emph {et~al.}(2012)\citenamefont {Smallwood}, \citenamefont {Hinton}, \citenamefont {Jozwiak}, \citenamefont {Zhang}, \citenamefont {Koralek}, \citenamefont {Eisaki}, \citenamefont {Lee}, \citenamefont {Orenstein},\ and\ \citenamefont {Lanzara}}]{science.1217423}%
  \BibitemOpen
  \bibfield  {author} {\bibinfo {author} {\bibfnamefont {C.~L.}\ \bibnamefont {Smallwood}}, \bibinfo {author} {\bibfnamefont {J.~P.}\ \bibnamefont {Hinton}}, \bibinfo {author} {\bibfnamefont {C.}~\bibnamefont {Jozwiak}}, \bibinfo {author} {\bibfnamefont {W.}~\bibnamefont {Zhang}}, \bibinfo {author} {\bibfnamefont {J.~D.}\ \bibnamefont {Koralek}}, \bibinfo {author} {\bibfnamefont {H.}~\bibnamefont {Eisaki}}, \bibinfo {author} {\bibfnamefont {D.-H.}\ \bibnamefont {Lee}}, \bibinfo {author} {\bibfnamefont {J.}~\bibnamefont {Orenstein}}, \ and\ \bibinfo {author} {\bibfnamefont {A.}~\bibnamefont {Lanzara}},\ }\href {\doibase 10.1126/science.1217423} {\bibfield  {journal} {\bibinfo  {journal} {Science}\ }\textbf {\bibinfo {volume} {336}},\ \bibinfo {pages} {1137} (\bibinfo {year} {2012})},\ \Eprint {http://arxiv.org/abs/https://www.science.org/doi/pdf/10.1126/science.1217423} {https://www.science.org/doi/pdf/10.1126/science.1217423} \BibitemShut {NoStop}%
\bibitem [{\citenamefont {Yang}\ \emph {et~al.}(2008)\citenamefont {Yang}, \citenamefont {Rameau}, \citenamefont {Johnson}, \citenamefont {Valla}, \citenamefont {Tsvelik},\ and\ \citenamefont {Gu}}]{Yang2008-zv}%
  \BibitemOpen
  \bibfield  {author} {\bibinfo {author} {\bibfnamefont {H.-B.}\ \bibnamefont {Yang}}, \bibinfo {author} {\bibfnamefont {J.~D.}\ \bibnamefont {Rameau}}, \bibinfo {author} {\bibfnamefont {P.~D.}\ \bibnamefont {Johnson}}, \bibinfo {author} {\bibfnamefont {T.}~\bibnamefont {Valla}}, \bibinfo {author} {\bibfnamefont {A.}~\bibnamefont {Tsvelik}}, \ and\ \bibinfo {author} {\bibfnamefont {G.~D.}\ \bibnamefont {Gu}},\ }\href@noop {} {\bibfield  {journal} {\bibinfo  {journal} {Nature}\ }\textbf {\bibinfo {volume} {456}},\ \bibinfo {pages} {77} (\bibinfo {year} {2008})}\BibitemShut {NoStop}%
\bibitem [{\citenamefont {Sobota}\ \emph {et~al.}(2021)\citenamefont {Sobota}, \citenamefont {He},\ and\ \citenamefont {Shen}}]{RevModPhys.93.025006}%
  \BibitemOpen
  \bibfield  {author} {\bibinfo {author} {\bibfnamefont {J.~A.}\ \bibnamefont {Sobota}}, \bibinfo {author} {\bibfnamefont {Y.}~\bibnamefont {He}}, \ and\ \bibinfo {author} {\bibfnamefont {Z.-X.}\ \bibnamefont {Shen}},\ }\href {\doibase 10.1103/RevModPhys.93.025006} {\bibfield  {journal} {\bibinfo  {journal} {Rev. Mod. Phys.}\ }\textbf {\bibinfo {volume} {93}},\ \bibinfo {pages} {025006} (\bibinfo {year} {2021})}\BibitemShut {NoStop}%
\bibitem [{\citenamefont {Benhabib}\ \emph {et~al.}(2015)\citenamefont {Benhabib}, \citenamefont {Sacuto}, \citenamefont {Civelli}, \citenamefont {Paul}, \citenamefont {Cazayous}, \citenamefont {Gallais}, \citenamefont {M\'easson}, \citenamefont {Zhong}, \citenamefont {Schneeloch}, \citenamefont {Gu}, \citenamefont {Colson},\ and\ \citenamefont {Forget}}]{Collapse}%
  \BibitemOpen
  \bibfield  {author} {\bibinfo {author} {\bibfnamefont {S.}~\bibnamefont {Benhabib}}, \bibinfo {author} {\bibfnamefont {A.}~\bibnamefont {Sacuto}}, \bibinfo {author} {\bibfnamefont {M.}~\bibnamefont {Civelli}}, \bibinfo {author} {\bibfnamefont {I.}~\bibnamefont {Paul}}, \bibinfo {author} {\bibfnamefont {M.}~\bibnamefont {Cazayous}}, \bibinfo {author} {\bibfnamefont {Y.}~\bibnamefont {Gallais}}, \bibinfo {author} {\bibfnamefont {M.-A.}\ \bibnamefont {M\'easson}}, \bibinfo {author} {\bibfnamefont {R.~D.}\ \bibnamefont {Zhong}}, \bibinfo {author} {\bibfnamefont {J.}~\bibnamefont {Schneeloch}}, \bibinfo {author} {\bibfnamefont {G.~D.}\ \bibnamefont {Gu}}, \bibinfo {author} {\bibfnamefont {D.}~\bibnamefont {Colson}}, \ and\ \bibinfo {author} {\bibfnamefont {A.}~\bibnamefont {Forget}},\ }\href {\doibase 10.1103/PhysRevLett.114.147001} {\bibfield  {journal} {\bibinfo  {journal} {Phys. Rev. Lett.}\ }\textbf {\bibinfo {volume} {114}},\ \bibinfo {pages} {147001} (\bibinfo {year} {2015})}\BibitemShut {NoStop}%
\bibitem [{\citenamefont {Loret}\ \emph {et~al.}(2017)\citenamefont {Loret}, \citenamefont {Sakai}, \citenamefont {Benhabib}, \citenamefont {Gallais}, \citenamefont {Cazayous}, \citenamefont {M\'easson}, \citenamefont {Zhong}, \citenamefont {Schneeloch}, \citenamefont {Gu}, \citenamefont {Forget}, \citenamefont {Colson}, \citenamefont {Paul}, \citenamefont {Civelli},\ and\ \citenamefont {Sacuto}}]{PhysRevB.96.094525}%
  \BibitemOpen
  \bibfield  {author} {\bibinfo {author} {\bibfnamefont {B.}~\bibnamefont {Loret}}, \bibinfo {author} {\bibfnamefont {S.}~\bibnamefont {Sakai}}, \bibinfo {author} {\bibfnamefont {S.}~\bibnamefont {Benhabib}}, \bibinfo {author} {\bibfnamefont {Y.}~\bibnamefont {Gallais}}, \bibinfo {author} {\bibfnamefont {M.}~\bibnamefont {Cazayous}}, \bibinfo {author} {\bibfnamefont {M.~A.}\ \bibnamefont {M\'easson}}, \bibinfo {author} {\bibfnamefont {R.~D.}\ \bibnamefont {Zhong}}, \bibinfo {author} {\bibfnamefont {J.}~\bibnamefont {Schneeloch}}, \bibinfo {author} {\bibfnamefont {G.~D.}\ \bibnamefont {Gu}}, \bibinfo {author} {\bibfnamefont {A.}~\bibnamefont {Forget}}, \bibinfo {author} {\bibfnamefont {D.}~\bibnamefont {Colson}}, \bibinfo {author} {\bibfnamefont {I.}~\bibnamefont {Paul}}, \bibinfo {author} {\bibfnamefont {M.}~\bibnamefont {Civelli}}, \ and\ \bibinfo {author} {\bibfnamefont {A.}~\bibnamefont {Sacuto}},\ }\href {\doibase 10.1103/PhysRevB.96.094525} {\bibfield  {journal} {\bibinfo  {journal} {Phys. Rev. B}\
  }\textbf {\bibinfo {volume} {96}},\ \bibinfo {pages} {094525} (\bibinfo {year} {2017})}\BibitemShut {NoStop}%
\bibitem [{\citenamefont {Agterberg}\ \emph {et~al.}(2020)\citenamefont {Agterberg}, \citenamefont {Davis}, \citenamefont {Edkins}, \citenamefont {Fradkin}, \citenamefont {Harlingen}, \citenamefont {Kivelson}, \citenamefont {Lee}, \citenamefont {Radzihovsky}, \citenamefont {Tranquada},\ and\ \citenamefont {Wang}}]{PDW}%
  \BibitemOpen
  \bibfield  {author} {\bibinfo {author} {\bibfnamefont {D.~F.}\ \bibnamefont {Agterberg}}, \bibinfo {author} {\bibfnamefont {J.~S.}\ \bibnamefont {Davis}}, \bibinfo {author} {\bibfnamefont {S.~D.}\ \bibnamefont {Edkins}}, \bibinfo {author} {\bibfnamefont {E.}~\bibnamefont {Fradkin}}, \bibinfo {author} {\bibfnamefont {D.~J.~V.}\ \bibnamefont {Harlingen}}, \bibinfo {author} {\bibfnamefont {S.~A.}\ \bibnamefont {Kivelson}}, \bibinfo {author} {\bibfnamefont {P.~A.}\ \bibnamefont {Lee}}, \bibinfo {author} {\bibfnamefont {L.}~\bibnamefont {Radzihovsky}}, \bibinfo {author} {\bibfnamefont {J.~M.}\ \bibnamefont {Tranquada}}, \ and\ \bibinfo {author} {\bibfnamefont {Y.}~\bibnamefont {Wang}},\ }\href@noop {} {\bibfield  {journal} {\bibinfo  {journal} {Annual Review of Condensed Matter Physics}\ }\textbf {\bibinfo {volume} {11}},\ \bibinfo {pages} {231} (\bibinfo {year} {2020})}\BibitemShut {NoStop}%
\end{thebibliography}%

\end{document}